\documentclass[12pt]{article}
\pdfoutput=1

\usepackage[applemac]{inputenc}

\usepackage{jcappub}
\usepackage{ifpdf}

\usepackage{amssymb}
\usepackage{amsfonts}
\usepackage{color}
\usepackage{graphicx,psfrag, subfigure}
\usepackage{amsmath}
\usepackage{hyperref}
\usepackage[applemac]{inputenc}
\usepackage{dsfont}
\usepackage{comment} 

\usepackage{wrapfig}

\usepackage{soul}
\definecolor{dm}{cmyk}{.20, 0, .30, 0}
\sethlcolor{dm}

                 \usepackage{cleveref}

\renewcommand{\d}{\mathrm{d}}

\newcommand{\vect}[1]{
}

\newcommand\be{\begin{equation}}
\newcommand\ee{\end{equation}}
\newcommand{\bea}{\begin{eqnarray}}
\newcommand{\eea}{\end{eqnarray}}
\newcommand\ini{\varphi^{(i)}}
\newcommand\fin{\varphi^{(f)}}
\newcommand\dd{{\rm d}}

\DeclareMathOperator{\PathOrder}{\mathsf{P}}

\renewcommand{\d}{\textrm{d}}

\numberwithin{equation}{section}

\def\be{\begin{equation}}
\def\ee{\end{equation}}
\def\bea{\begin{eqnarray}}
\def\eea{\end{eqnarray}}

\def\Mpl{M_{\rm Pl}}

\def\Lh{\Lambda_{\rm h}}
\def\Lv{\Lambda_{\rm v}}
\def\eV{\epsilon_{\rm V}}
\def\etaV{\eta_{\rm V}}
\def\mmin{m^2_{\rm min}}

\begin{document}

\begin{titlepage}

\begin{center}
\rightline{\small DESY-17-082}
\vskip -1.3cm
\end{center}

\setcounter{page}{1} \baselineskip=15.5pt \thispagestyle{empty}

\bigskip\
\begin{center}
{\Large \bf  Seven Lessons from
\\
\vskip 5pt
 Manyfield Inflation  in Random Potentials}
\vskip 5pt

\vskip 15pt
\end{center}
\vspace{0.3cm}
\begin{center}
{
\large
Mafalda Dias,$^1$ Jonathan Frazer,$^1$ M.C.~David Marsh,$^2$}
\end{center}

\vspace{0.1cm}

\begin{center}
\begingroup
    \fontsize{11pt}{12pt}\selectfont

\vskip 4pt
\textsl{
$^1$ Deutsches Elektronen-Synchrotron DESY, Theory Group, 22603 Hamburg, Germany \\
$^2$ Department of Applied Mathematics and Theoretical Physics, University of Cambridge, Cambridge, CB3 0WA, United Kingdom}\\
\emailAdd{mafalda.dias@desy.de}
\emailAdd{jonathan.frazer@desy.de}
\emailAdd{m.c.d.marsh@damtp.cam.ac.uk}


\endgroup
\end{center}

{\small  \noindent  \\[0.2cm]
\noindent
We study inflation in models with many interacting fields subject to randomly generated scalar potentials. 
We use methods from 
non-equilibrium random matrix theory to construct the potentials and an adaption of the `transport method' to evolve the two-point correlators during inflation.  
This construction allows, for the first time, for an 
explicit 
study of models  with up to 100 interacting fields supporting a period of  `approximately saddle-point' inflation. We determine the statistical 
predictions for observables 
by generating over 30,000 
models
with 2--100 fields supporting at least 60 efolds of inflation. 
These studies lead us to seven lessons: 
\emph{i)} Manyfield inflation is not single-field inflation, 
\emph{ii)} The larger the number of fields, the simpler and sharper the predictions,
\emph{iii)} Planck compatibility is not rare, but future experiments may rule out this class of models,
\emph{iv)} The smoother the potentials, the sharper the predictions, 
\emph{v)} Hyperparameters can transition from stiff to sloppy, 
\emph{vi)} Despite tachyons, isocurvature can decay, 
\emph{vii)} Eigenvalue repulsion drives the predictions.
We conclude that many of the `generic predictions' of single-field inflation can be emergent features of complex inflation models. }

\vspace{0.3cm}

\vspace{0.6cm}

\vfil
\begin{flushleft}
\small \today
\end{flushleft}
\end{titlepage}

\newpage
\tableofcontents

\section{Introduction}

Inflation is the leading paradigm for explaining the origin of  the observed large-scale structure (LSS) of the universe and the anisotropies of the Cosmic Microwave Background (CMB). According to inflation, the 
primordial seeds of structure 
arose from
quantum scalar fluctuations during a period of accelerated expansion. Inflation can be achieved with rather simple ingredients, and the observationally inferred properties of the CMB anisotropies are in excellent agreement with the generic predictions of some of the simplest models of inflation involving only a single scalar field. 

However, there 
are good reasons to consider more general models involving several fields. 
 First, observations of distinctive signals of multifield effects during inflation can 
 rule out the single-field paradigm. 
 For example, it is well known that single-field inflation generates only highly suppressed contributions to non-Gaussianities of the local type \cite{astro-ph/0210603, astro-ph/0306122, astro-ph/0503692, 0707.3378}, but multiple-field effects can lead to enhanced levels \cite{hep-ph/0207295, astro-ph/0504045, astro-ph/0506056, astro-ph/0506704, astro-ph/0509719}.  
  Second, there is no theoretical reason to expect only a single scalar field to be  important in the early universe. Inflation may have probed energy scales far above those accessible by terrestrial experiments and is thereby sensitive to the 
 degrees of freedom of the ultraviolet completion of the Standard Model. Both particle physics models and string theory compactifications commonly give rise to effective field theories involving many weakly coupled fields with masses between the weak scale and the Planck scale. 
  In such theories, single-field inflation can only be realised if   
   hierarchies in the mass spectrum of the fields are induced, which may come at the cost of additional tuning.  
 
This motivates the development of a systematic understanding of the observational predictions of multiple-field inflation. 
Unfortunately, 
the study of
generic inflationary models with several interacting fields have been hampered by the significant computational difficulty 
associated with constructing them and deriving their predictions. As a result, most work in this area has focussed on deriving  the particular effects that can arise in special multifield scenarios (see e.g.~\cite{1002.3110, Vennin:2015egh, 1612.05248} and references therein), typically focussing on the computationally less taxing two-field case. These works reveal the range of observational possibilities offered by multifield effects, but do not 
provide an insight into
 the  characteristic (or typical) signatures of inflationary  models with many interacting fields.

In this paper, we take steps to address these issues by computing the statistical  predictions of models of inflation involving many ($N_f \gg 1$) interacting fields, and we identify the parameters most relevant for determining the distributions of observables. 
Our approach is to sample a large class of randomly generated inflationary models  following the Random Matrix theory (RMT) method first developed in \cite{1307.3559} (for other applications of RMT in early universe cosmology, see for example \cite{Aazami:2005jf,hep-th/0512102, 1404.7496, 1604.05970}).   
  This method  draws on
  two novel techniques
  to
    overcome the computational difficulties associated with studying large, coupled inflationary systems.
    First, the randomly generated scalar potential, $V(\phi_1, \ldots , \phi_{N_f})$, is constructed \emph{only locally}, in a patch-by-patch manner,  around the dynamically determined field trajectory, rather than globally over the entire field space. 
 As the inflationary evolution 
depends only on the properties of  the scalar potential in the immediate vicinity of the inflaton trajectory, this approach  drastically 
simplifies the task of constructing the model and obtaining its observational predictions.
Second, the random scalar potential is generated by postulating that the Hessian matrix, 
\be
{\cal H}_{ab} \equiv \partial^2_{\phi^a\, \phi^b} V(\phi_1, \ldots , \phi_{N_f}) \, ,
\ee
evolves from one patch to the next along the field trajectory according to `Dyson Brownian Motion' (DBM) \cite{Dyson}, a 
non-equilibrium extension of Random Matrix Theory.\footnote{In contrast Refs.~\cite{Aazami:2005jf,hep-th/0512102, 1404.7496} draw on results from equilibrium Random Matrix Theory} In the continuum limit of  closely spaced patches, this prescription gives a smooth evolution of 
the 
local Taylor coefficients of the potential up to quadratic order along the field trajectory. Cubic and higher order interaction terms are implicitly captured through the non-trivial evolution of the lower order Taylor coefficients. 
DBM has the property that, regardless of the matrix configuration chosen in the initial patch, ${\cal H}_{ab}$ eventually evolves into 
a random and statistically isotropic matrix distributed according to the Gaussian Orthogonal Ensemble (GOE). 
We will refer to potentials constructed this way as `DBM potentials'. 

The  random DBM potentials  
may  elucidate also
other classes of multifield potentials.  
Key to this is the principle of universality:  
for complicated systems with  many degrees of freedom, one may expect that large-$N_f$ central limit behaviour will lead to results that 
only depend on some qualitative properties of the potentials, while being independent to the details of how these are constructed. 
 By identifying the mechanism  that drives the inflationary predictions, one may also gain some insight into the predictions of other classes of potentials in 
 which the same mechanism is present. 
 In this paper,  we  concretise these statements  by showing that `eigenvalue repulsion' in the Hessian matrix determines many of the inflationary predictions.
 Eigenvalue repulsion is a generic feature of systems with interacting fields, and common to broad classes of potentials constructed by methods very different from  the DBM prescription.

In this paper, we compute the  two-point correlation function of the inflationary perturbations in
`manyfield' inflation close to approximate saddle points of otherwise random DBM potentials. 
We adopt the `transport method' \cite{1609.00379, 1302.3842, 1203.2635, 1502.03125}  to efficiently compute the perturbations: by taking advantage of the simple form of the potential in each local patch along the field trajectory, the differential equations governing the evolution of the perturbations can be solved analytically, and the perturbations are easily transported from one local patch to the next. The efficiency of this method allows us, for the first time, to explore the observational predictions of models of inflation with a large number of interacting fields.  

We sample more than 100,000 
inflationary realisations with $N_f$ from 2 to 100 
and
compute observables for over 30,000 models that support at least $60$ efolds of inflation. 
We find the prevalence of many active fields throughout inflation, leading to significant superhorizon evolution of the curvature perturbation. Our results can become independent of hyperparameters of the construction leading to universal behaviour and sharp observable predictions. We also find a large region of hyperparameter space to be compatible with constraints from CMB surveys. We interpret these results in the light of RMT and therefore expect this universality to be present in a wider class of multifield potentials.

Our results provide a first view into the observational predictions of manyfield inflation; confirm and extend results on universality in the large $N_f$ limit; and serve as a sharp reality check for speculations about multifield inflation based on single-field or few-field models.

Throughout this paper, we set the reduced Planck mass  $\Mpl = 2.48\times 10^{18}\, {\rm GeV} =1$, but we occasionally reinsert it for clarity of presentation.  

\section{Inflation in random potentials}

In this section we discuss the challenges involved in studying inflation with many interacting fields, and we review the construction of DBM potentials. We emphasise the range of applicability of this class of potentials, and critically review the present literature on the subject.

Models of inflation involve a scalar potential, $V( \phi_i)$, which is a function of the fields $\phi_i$ for $i = 1, \ldots, N_{f}$.  Unless the fields $\phi_i$ are subject to additional symmetries, 
 the scalar potential in general involves interactions between all the fields. For  $N_{f}>1$, the potential is then a complicated function of many parameters, and extracting the observational predictions of these models can be challenging for several reasons: first, both the homogeneous background equations of motion and the equations governing the perturbations become non-linear coupled differential equations, which are challenging to solve for  $N_f\gg 1$. Second, multiple-field effects allow for novel phenomena that do not exist in single-field models, such as entropic modes that may source superhorizon evolution of the curvature perturbation $\zeta$. To compute observables, one should then follow the perturbations until the entropic modes have all decayed and $\zeta$ has `frozen' (\text{i.e.} to the `adiabatic limit'), or, alternatively, all through the reheating phase after the end of inflation. 
Third, in addition to the very large number of parameters appearing in the potential, multifield inflation can exhibit strong sensitivity to initial conditions. 
A good example of this difficulty is simply finding sustained inflation in a random potential --- in many classes of random potentials, inflation is an exponentially rare event implying that, without a prescription for initial conditions, studying inflation with large numbers of fields is numerically prohibitive.

Our approach in this paper is to search for universality that may arise in fully interacting models in the limit where $N_f$ is large. To do so, we randomly generate ensembles of scalar potentials with many interacting fields and use these to study the generation of observables during inflation. 
Key to overcoming the challenges associated with multiple-field effects is the local procedure first introduced in \cite{1307.3559}, which we now review.

\subsection{Local potentials, generated from Random Matrix Theory }
\label{sec:DBMreview}
The DBM prescription for constructing random potentials takes as its starting point the potential defined locally, to quadratic order in the fields $\phi^a$, in a patch around a point $p_0$ in field space,
\be
V = \Lv^4 \sqrt{N_f} \left[ v_0 + v_a \frac{\phi^a}{\Lh} + \frac{1}{2} v_{ab} \frac{\phi^a}{\Lh}\frac{\phi^b}{\Lh}\right] \, ,
\label{eq:V}
\ee   
where $\Lv$ and $\Lh$ respectively denote the (globally defined) `vertical' and `horizontal' scales of the potential.\footnote{
In this paper we restrict our attention to a flat field space with a Euclidean metric. In particular, this implies that we do not distinguish between upper and lower tensor indices.  
We expect our results to be directly applicable 
also to curved field spaces as long as the inflationary trajectory does not pass too close to field-space singularities or regions of very high curvature. 
 An interesting example of 
 random models of inflation in
 the presence of curvature is the case of the `$\alpha$-attractors', where the kinetic term presents a pole (cf.~Ref.~\cite{1612.04505} for a specific multifield example).}  
We have here followed the convention of \cite{1307.3559} and introduced an explicit $\sqrt{N_f}$ prefactor in the potential. 
 The
  potential in a nearby patch centred at the point $p_1$ which 
is related by the small displacement $||\delta \phi^a/\Lh||\ll 1$ from $p_0$,
 is then to leading order determined by the expansion coefficients,
\bea
v_0 \big|_{p_1} &=& v_0 \big|_{p_0} + v_a \big|_{p_0} \frac{\delta \phi^a}{\Lh} \, , \nonumber \\
v_a \big|_{p_1} &=& v_a \big|_{p_0} + v_{ab} \big|_{p_0} \frac{\delta \phi^b}{\Lh} \, , \nonumber \\
v_{ab} \big|_{p_1} &=& v_{ab} \big|_{p_0} + \delta v_{ab} \big|_{p_0 \to p_1} \, , 
\label{eq:DBMpot}
\eea 
where $\delta v_{ab}$ is a stochastic matrix that we will soon define. By repeating this procedure 
for a large set of points along the path $L$, the random scalar potential is glued together in a patch-by-patch manner.\footnote{
As we review in \S\ref{sec:subtleties},  the path $L$ is not completely arbitrary. 
} 

The properties of the potentials defined in this way are determined by the definition of the  stochastic matrices $\delta v_{ab}$. 
To specify the distribution of $\delta v_{ab}$, we postulate that $v_{ab}$ evolves, at large distances,
 into a random 
matrix in the ensemble of symmetric $N_f \times N_f$  matrices with independent and identically distributed (i.i.d.) Gaussian elements, i.e.~the Gaussian Orthogonal Ensemble (GOE). This ensemble is invariant under orthogonal transformations, so this choice ensures our potential is 
statistically isotropic. 
While this particular choice is made for simplicity and generality, many of the properties of the GOE extend to more general classes of random matrices. For example, Wigner famously showed that,  for $N_f \gg 1$, the spectral density of 
a symmetric matrix with i.i.d.~entries is 
 well described by a semi-circle \cite{wigner},
\be
\rho(\lambda) = \frac{1}{\pi N_f \sigma^2} \sqrt{2N_f \sigma^2 - \lambda^2} \, .
\label{eq:semicirc}
\ee  
Here, $\sigma$ denotes the  standard deviation of the randomly distributed matrix elements, 
which we will take to be $\sigma = \sqrt{2/N_f}$ so that the width of the semi-circle is independent of $N_f$, with end-points located at $\pm2$. 
Moreover, the i.i.d.~assumption can be further relaxed, and Eq.~\eqref{eq:semicirc} remains the limiting eigenvalue distribution for large classes of matrix ensembles with correlations and non-identical distributions \cite{Mehta, SchenkerShulzBaldes, HofmannCredner,Deift, Kuijlaars}.
  The persistent relevance of the Wigner semi-circle is one of the most well-known examples of large-$N$ universality in random matrix theory.

To make $v_{ab}$ evolve from a given initial condition into a random sample of the GOE, we use Dyson Brownian Motion: from one patch to the next, the elements of $\delta v_{ab}$ undergo independent Brownian motions subject to a restoring, harmonic force,
\bea
\langle \delta v_{ab} \big|_{p_i \to p_{i+1}} \rangle &=& - v_{ab}\big|_{p_i} \frac{||\delta \phi^a||}{\Lh} \, , \nonumber \\
\langle \delta v_{ab}^2 \big|_{p_i \to p_{i+1}} \rangle &=& \sigma^2 \left( 1+\delta_{ab} \right) \frac{||\delta \phi^a||}{\Lh} \, .
\label{eq:DBM}
\eea
In the continuum limit where ${\rm d}s =||\delta \phi^a||/\Lh$ is infinitesimal, Eq.~\eqref{eq:DBM} leads to a Smoluchowski-Fokker-Planck equation for the probability distribution of $v_{ab}$ \cite{Dyson, Uhlenbeck}.  
This equation has the solution,
\be
P(v_{ab}(s)) \sim {\rm exp}\left[ -\frac{{\rm tr}\left((v_{ab}(s) - q v_{ab}(0))^2\right)}{2 \sigma^2(1-q^2)}\right] \xrightarrow[s \gg 1~~]{}
 {\rm exp}\left[ -\frac{{\rm tr}(v_{ab}(s)^2 )}{2 \sigma^2}\right] 
 \, ,
 \label{eq:P(vab)}
\ee
%
%
where $q = {\rm exp}(-s)$, and $s$ is the path length in units of $\Lh$ from $p_0$ where $v_{ab} = v_{ab}(0)$. The stationary limiting distribution for $s \gg 1$ is  the Gaussian Orthogonal Ensemble, as desired, and the eigenvalue distribution at large $N_f$ is given by Eq.~\eqref{eq:semicirc}. 

According to Eq.~\eqref{eq:P(vab)} and the definition of $s$, the parameter $\Lh$ sets the scale in field space over which 
the initial condition is `forgotten' and 
the potential randomises. However, the parametrisation \eqref{eq:V} is clearly degenerate: a rescaling of either $\Lh$ or $\Lv$ can be absorbed by a rescaling of the Taylor coefficients. More precisely, the two scalings are given by \cite{1608.00041},
\bea
\Lv &\to& \lambda_1 \Lv\, :~~~(v_0, v_a, v_{ab}) \to (\frac{1}{\lambda^4_1}v_0, \frac{1}{\lambda^4_1} v_a, \frac{1}{\lambda^4_1}v_{ab}) \, , 
\label{eq:scaling1}\\
\Lh &\to& \lambda_2 \Lh \, :~~~(v_0, v_a, v_{ab}) \to (v_0, \lambda_2 v_a, \lambda_2^2v_{ab})
\label{eq:scaling2}
 \, .  
\eea
The physical parameters, $V,  \partial_a V =V_a$ and $\partial^2_{ab} V = V_{ab}$ do not transform under these rescalings. 
Equation  \eqref{eq:scaling2}
can be recognised as the  degeneracy between 
the ultra-violet cut-off scale and the Wilson coefficients common to  bottom-up effective field theories. If this degeneracy is fixed so that the typical values of $(v_0,\,  ||v_a||,\,  {\rm Eig}(v_{ab}))$  are ${\cal O}(1)$, 
 $\Lh$ can be interpreted as the ultraviolet cut-off scale of the theory. 
 
 In sum, DBM potentials have the global `hyperparameters' $(N_f, \Lh, \Lv, \sigma)$, and, in addition, parameters that are locally fixed in the first patch but stochastically determined in subsequent patches: $(v_0, v_a, v_{ab})\big|_{p_0}$.

\subsection{Inflation in DBM potentials}
\label{sec:DBMinflation}
To apply the DBM construction to cosmology, we may choose the path $L$ to be the dynamically determined trajectory in field space. That is, given some set of initial conditions for the potential and the field velocity at the point $p_0$ in field space, 
the field trajectory in the local patch around $p_0$ determines a segment of $L$. After evolving the field to some $p_1$ located a sufficiently small distance $\delta s \ll 1$ from $p_0$, we  update the potential according to Eq.~\eqref{eq:DBMpot} and thereby generate the local potential in a new patch, centred at $p_1$. By solving the equations of motion in this new patch, another segment of $L$ is mapped out. This iterative procedure can be used to map out the potential along the entire inflationary trajectory, while it remains undetermined far from the field space regions probed during inflation. As DBM is stochastic, this prescription allows us to efficiently generate large sets of inflationary models of many fields from a given deterministic set of initial conditions.

\subsubsection{Multiple field slow-roll inflation}
\label{sec:multiSR}


The fact that the typical, or equilibrium, spectral density of $v_{ab}$ is well described the semi-circle law, with equal probability of positive and tachyonic masses, is a statement of how unlikely inflation is in a random potential with Hessians belonging to the GOE.
A significant benefit of the DBM construction is the fact that we can chose to start our trajectory with rare initial conditions, as will be described in \S\ref{sec:regimes}. 
To study inflation in DBM potentials, we take the initial Taylor coefficients of the potential and the field velocity at $p_0$ to be suitable for 
slow-roll inflation. 
The equations of motion of $N_f$ scalar fields $\phi^a$ minimally coupled to gravity are given by,
\bea
\ddot{\phi}^a + 3H \dot{\phi}^a +  V_a = 0 \, ~ \, , \\
3 H^2 = \frac{1}{2} \sum_{a=1}^{N_f} (\dot{\phi}^a)^2 + V(\phi^a) \, .
\eea
Here, $V_a = \partial_a V$. 
 Inflation occurs whenever,
\bea
\epsilon = - \dot{H}/H^2 
=
\frac{1}{2  H^2} \dot{\phi}^a \dot{\phi}_a
< 1\, , 
\eea
and inflation is sustained for many efolds if also, 
\bea
 |\eta| &=&\frac{1}{H} \Big|\frac{{\rm d} \ln \epsilon}{d t}\Big| \ll 1\, .
 \eea
In single-field slow-roll inflation these conditions are equivalent to, 
\be
\epsilon_{\rm V} = \frac{1}{2} \frac{ V_{\phi}^2}{V^2} < 1~~~{\rm and}~~~|\eta_{\rm V}|  = \left| \frac{ V_{\phi \phi}}{V}\right| \ll 1\, .
\label{eq:SRsingle}
\ee
For the multifield case,  a convenient generalisation of the `potential' slow-roll parameters of Eq.~\eqref{eq:SRsingle} is,
\bea
\eV &\equiv& 
\frac{1}{2 } \frac{ V_a V_a}{V^2}
 \, , \\
\etaV &\equiv&  \frac{\mmin}{V} \, ,
\label{eq:etaV}
\eea 
where $\mmin ={\rm Min}\left( {\rm Eig}\left(V_{ab} \right) \right)$ and repeated indices are summed over. In slow-roll inflation, $\eV = \epsilon <1$. To see the relevance of Eq.~\eqref{eq:etaV}, let us consider the tangent vector of the gradient flow,
\be
n^a = - \frac{V_a}{\sqrt{ V_b  V_b}} \,, 
\ee
whose rate of change satisfies,
\be
n'^a =  - \frac{ V_{ab}\, n^b}{V}+ n^a \frac{n^c\, V_{cd}\, n^d}{V} \,.
\label{eq:tprime}
\ee
Here, we denote derivatives with respect to the number of efolds by $'  ={\rm d}/H{\rm d}t= {\rm d}/{\rm d}N$. 
The second term of Eq.~\eqref{eq:tprime} enforces the preservation of the unit norm of $n^a$, so that
$n_a n'^a =0$. By considering Eq.~\eqref{eq:tprime} in an eigenbasis of the Hessian matrix (so that $ V_{ab} = \delta_{ab}\, m^2_a$, with no sum on $a$), we see that the non-vanishing components of $n^a$ evolve as,
\be
\label{eq:mosttachy}
(\ln n^a)' = - \frac{m^2_a}{V} + \sum_{c=1}^{N_f} (n^c)^2 \frac{ m^2_c}{V} \, , 
\ee
indicating  that the gradient direction tends to align with the eigendirection of the most tachyonic eigenvalue.

The slow-roll Klein Gordon equation can be written as,
\be
\phi'^{\, a} \equiv \frac{1}{H} \dot{\phi}^a \approx - \frac{ V_a}{V} = \sqrt{2 \eV}\, n^a \, ,
\label{eq:SReqn}
\ee  
and the acceleration of the field is given by,
\be
\phi''^a = \sqrt{2 \eV}\, \left( 
- \frac{n^b  V_{ab}}{V} + 2 \eV\, n^a 
\right) \, .
\ee
We then find that, 
\be
\eta= \frac{\epsilon'}{\epsilon} = \frac{1}{\epsilon} \phi'' \cdot \phi' = 2 \left(-\frac{n^a\,  V_{ab}\, n^b}{V} + 2 \eV \right) \, ,
\ee
which implies that if $n^a$ is aligned with the eigenvector of the smallest eigenvalue,  
\be
\eta= 2 \left(-\etaV + 2 \eV \right) \, .
\ee
Motivated by this we choose the initial configuration at $p_0$ to have $\eV < 1$ and $|\etaV| \ll 1$.\footnote{Note however that our numerical studies include a full non-slow-roll analysis in addition to the extensive slow-roll analysis. In both cases however, we start in a field configuration compatible with slow-roll at $p_0$.} 
 
\subsubsection{Approximate saddle-point inflation in DBM potentials}
\label{sec:regimes}
We now discuss how inflation can be achieved in DBM potentials, focussing in particular on the case of `approximate saddle-point' inflation. 

DBM potentials can be used to study various regimes of  inflation.
Inflation may be supported by taking $\Lh \gg \Mpl$ so that the potentials are slowly varying over Planckian distances.
In this case, the field displacement becomes super-Planckian during inflation, $\Delta \phi > \Mpl$, and moreover, a substantial fraction of the efolds of inflation may arise close to the minimum of the potential.
 For $\Lh <\Mpl$, inflation can be supported in two regimes: 
for
 very large values of $v_0$
  inflation may proceed `off a high slope' with typical configurations of the gradient vector and the Hessian  matrix.
 The total field displacement during inflation may be greater than $\Lh$, but not necessarily larger than $\Mpl$. 
 However, reference \cite{1307.3559} argued that for $N_f \lesssim 100$, this type of inflation is more rare than  `approximate saddle-point inflation' occurring as the field passes close to  an approximate critical point of the potential. 
 %
   %
The Hessian matrix 
  \begin{wrapfigure}{l}{0.45\textwidth}
  \vspace{-20pt}
  \begin{center}
    \includegraphics[width=0.30\textwidth]{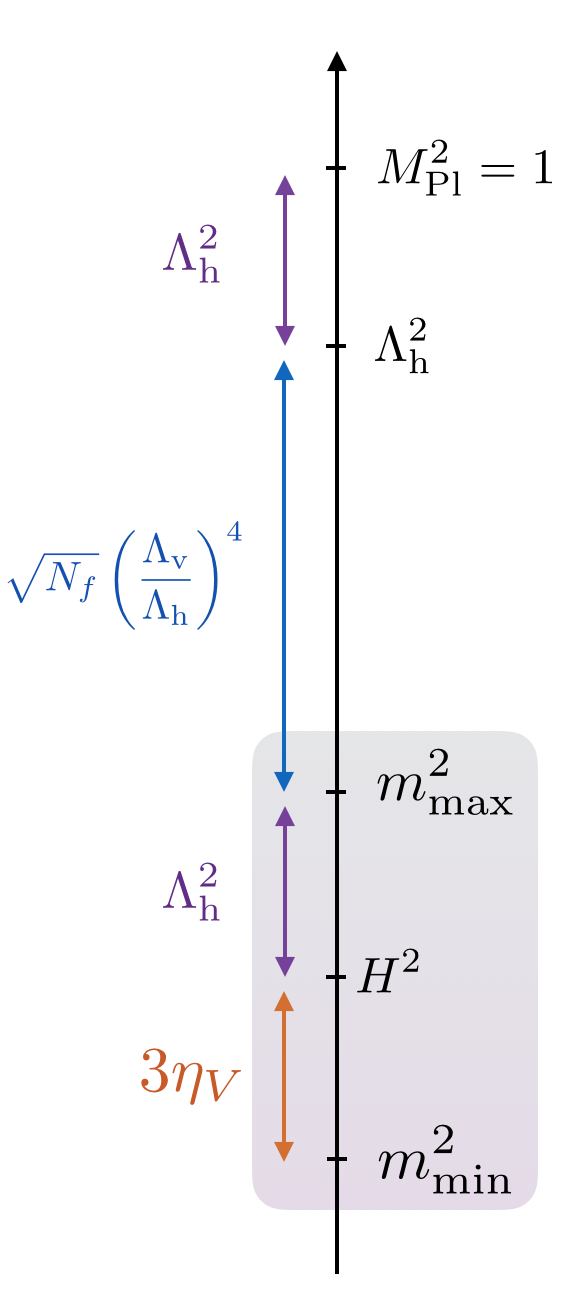}
  \end{center}
  \vspace{-20pt}
  \caption{Energy scales of the DBM potentials. 
  }
    \vspace{-15pt}
  \label{fig:Escales}
\end{wrapfigure}
at a typical critical point has both positive and negative eigenvalues, so that the approximate critical point is a saddle point.  
  A small $\eta_V$ parameter can be obtained if the minimum eigenvalue of the Hessian, $m^2_{\rm min}$, is suppressed relative to $V$.  In this paper, 
  we focus solely on this type of  `approximate saddle point inflation'.\footnote{The techniques we develop in this case extend straightforwardly to inflation `off a high slope', while inflation with $\Lh \gg 1$ would require a detailed modelling of the final approach to the minimum of the potential.}

  The relevant (squared) energy scales of the DBM models include $\Lh^2 < \Mpl^2$, which we interpret as the UV cut-off of the theory; 
the inflationary Hubble parameter, $H^2 = V/3 $; 
  and finally, the smallest and largest eigenvalues of the Hessian matrix, 
  $m^2_{\rm min}$ and $m^2_{\rm max}$. As we will explain in \S\ref{sec:prescription}, 
  we take $v_0=1$ at the critical point $p_0$ so that 
  %
  $v_0 \approx 1$ during inflation. Moreover, in all models  that we study $\Lv\ll \Lh$. 
  We  take $\sigma=\sqrt{2/N_f}$ so that the positive endpoint of the Wigner semicircle are located at  ${\rm Max(}\lambda_a) \approx 2$,
   where $\lambda_a$ denotes an eigenvalue of $v_{ab}$.

    Figure \ref{fig:Escales}  schematically illustrates the various hierarchies of scales. 
  In all cases that we study, the squared masses are substantially below the `UV cut-off' of the theory, cf.
  \be
 \frac{m^2_{\rm max}}{\Lh^2} = \frac{\Lv^4 \sqrt{N_f}}{\Lh^4} {\rm Max(}\lambda_a) \approx 
2  \frac{\Lv^4 \sqrt{N_f}}{\Lh^4} \ll 1
  \, .
  \ee
  The dynamically relevant energy scale during inflation is the Hubble parameter,
\be
H^2 = \frac{1}{3} \Lv^4 \sqrt{N_f} v_0 \approx \frac{1}{3} \Lv^4 \sqrt{N_f} \, \, ,
\ee
in relation to which 
the eigenvalues of the Hessian are given by,
\be
\frac{m^2_a}{H^2} = \frac{3 \lambda_a}{\Lh^2} \, .
\label{eq:mvH}
\ee 
The spectrum of DBM models includes both `heavy' modes with $m^2_a > H^2$, and `light' modes with $m_a^2 < H^2$.\footnote{Clearly, with $\sigma$ fixed and taking $\Lh \ll 1$, the natural mass scale of the fields becomes much larger than the Hubble parameter and the corresponding models do not exhibit interesting multifield dynamics. Since we are interested in multifield inflation, we will not discuss this regime.  } We note however that the eigenvalues of the Hessian matrix are not fixed during inflation, and modes can evolve from being heavy to light, and possibly further to being massless or even tachyonic.

\begin{figure}[t]
 \centering
 \includegraphics[width=0.6\textwidth]{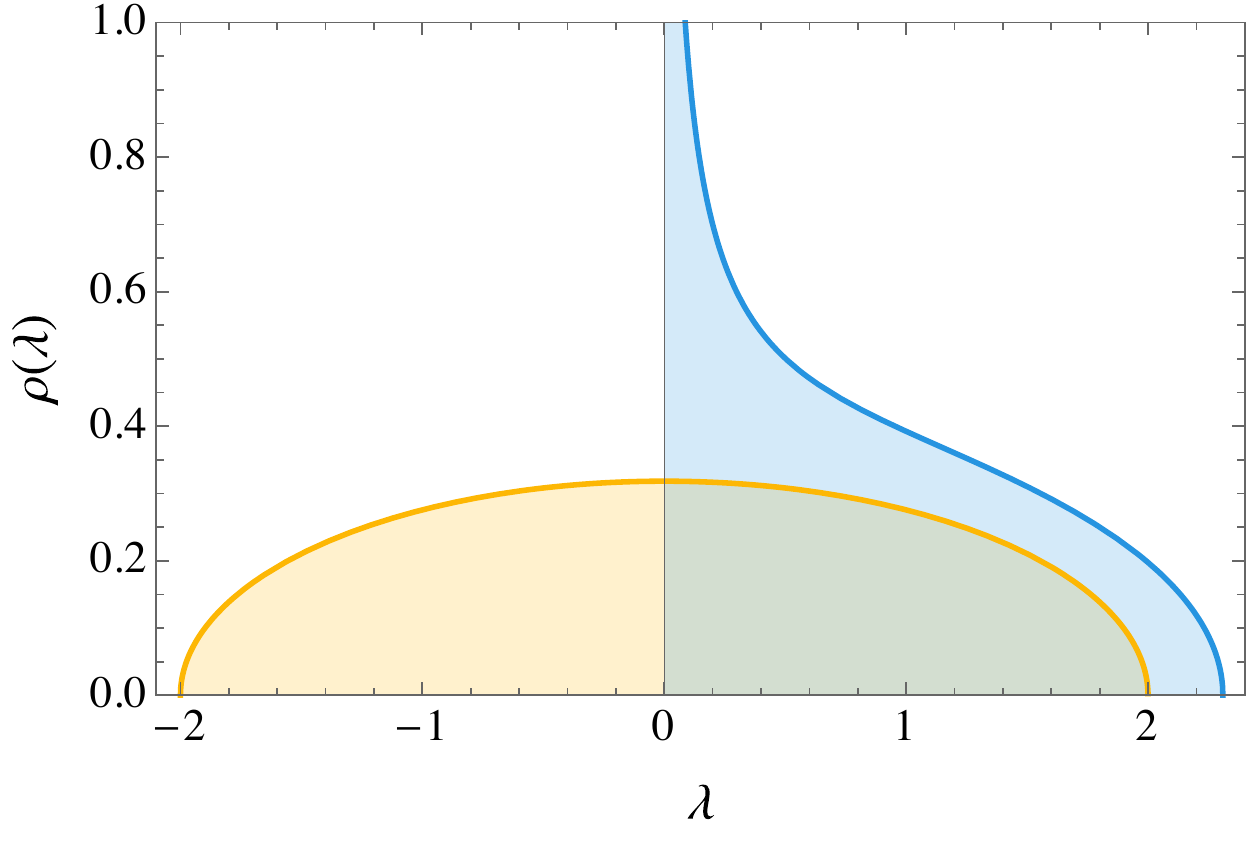}
  \caption{The Wigner semi-circle and the `fluctuated' probability distribution used as initial condition for the Hessian matrix.}
   \label{fig:wigner}
\end{figure}

Sustained 
approximate saddle point inflation is possible if $|\eta_V| <1$ at the critical point, $p_0$. Thus, the spectrum of the Hessian matrix cannot simply be determined by the Wigner semi-circle, cf.~Eq.~\eqref{eq:semicirc}. To set the initial configuration for $v_{ab}$, we consider the spectrum of the \emph{subset} of the GOE that has  no eigenvalue smaller than a given lower bound, say $\xi_{\rm min}$. Such `fluctuated' spectra were computed in Ref.~\cite{cond-mat/0609651}, and
  Fig.~\ref{fig:wigner}
  illustrates the 
fluctuated eigenvalue density    for  $\xi_{\rm min}=0$. To set the initial condition of the Hessian matrix at $p_0$, we take the initial spectrum to be a fluctuated configuration and use 
the initial  $\eta_V$ parameter as the hyperparameter controlling the spectrum at $p_0$,\footnote{The eigenvalues of the Hessian are drawn from the fluctuated spectrum, by taking all eigenvalues to be equally spaced in the cumulative probability function.}
\be
\eta_{V0} = \frac{m^2_{\rm min}}{V}  = \frac{1}{\Lh^2} \xi_{\rm min} \, .
\label{eq:etaV0}
\ee

\subsubsection{Recipe for DBM inflation and our statistical ensemble}
\label{sec:prescription}
For clarity and reproducibility, we here summarise the step-by-step procedure for studying inflation in DBM potentials. 
\\
\\
\noindent {\bf Global hyperparameters:}
We recall the global parameters of the model are $N_f,~\Lh$, $\Lv$ and $\sigma$. We refer to $N_f,~\Lh$,  and $\sigma$ as hyperparameters since they 
are
the dials controlling 
different ensembles of DBM potentials,
and
may moreover  be regarded as the parameters of the prior distribution of the model (see Ref.~\cite{Price:2015qqb} for more discussion of hyperparameters in the context of inflation).
We construct ensembles of DBM potentials for certain fixed choices of $N_f$ and $\Lh$, as we explain below.
The value of $\Lv$ does not affect the slow-roll dynamics, and is fixed individually for each model supporting at least 60  efolds  of inflation by `COBE normalising' the amplitude of the scalar perturbations at the pivot scale, which we take to lie 55 efolds before the end of inflation.
  
  Throughout our studies, we take
  $\sigma = \sqrt{2/N_f}$ so that the width of the spectrum \eqref{eq:semicirc} is independent of $N_f$, with end-points at $\pm2$. 
 This choice is motivated by Eq.~\eqref{eq:mvH}: for this choice of $\sigma$, an increase in the number of fields does not lead to a broadening of the spectrum in units of $H$, but rather a direct increase in the number of fields within the relevant mass range.  This choice  fixes the redundancy of Eq.~\eqref{eq:scaling2}. 
 \\
\\
\noindent {\bf Initial Conditions:}
In addition to the global parameters described above, there are in principle $1+N_f+(N_{f}^2+N_{f})/2$ parameters that characterise the quadratic expansion of the first patch: these are $(v_0, v_a, v_{ab})\big|_{p_0}$. 
 In practice we reduce this number to 3.

We first fix the degeneracy of Eq.~\eqref{eq:scaling1} by taking $v_0\big|_{p_0} =1$.\footnote{In comparison, Ref.~\cite{1307.3559} chose $v_0\big|_{p_0} = 1/\sqrt{N_f}$ which gives a sharp reduction in the number of efolds generated as $N_f$ is increased (with all else constant).} The gradient vector $v_a\big|_{p_0}$ has an orientation and a norm. 
As we have seen in \S\ref{sec:multiSR} (see also 
 Ref.~\cite{1307.3559}), 
$v_a$ quickly aligns with the eigendirection of 
the smallest eigenvalue of $v_{ab}$  during slow-roll inflation.
We therefore choose 
these two vectors to be aligned from the outset.
The norm of $v_a\big|_{p_0}$ is conveniently parametrised by $\epsilon_{V0}$, which we retain as a hyperparameter. 
The spectrum of $v_{ab}\big|_{p_0}$ is taken to be a `fluctuated' configuration, cf.~Fig.~\ref{fig:wigner}, with $\eta_{V0}$ controlling the value of the smallest eigenvalue, cf.~Eq.~\eqref{eq:etaV0}. 

The complete list of hyperparameters that we choose to study is then $(N_f, \Lh, \epsilon_{V0}, \eta_{V0})$. 
 \\
 \\
\noindent {\bf Step-by-step procedure:}
\begin{enumerate}
\item We first (non-randomly)  fix the hyperparameters $(N_f, \Lh, \epsilon_{V0}, \eta_{V0})$ of the ensemble we wish to study. As described above, this fixes the potential in the the initial patch around $p_0$. 
\item We then solve the equations of motion for gravity coupled to the homogeneous, slowly-rolling multifield system for a small displacement $\delta s \ll 1$ to the point $p_1$. The displacement is chosen to be sufficiently small to  map out the potential and its derivatives with high accuracy. 
\item At the new point, we update the Taylor coefficients of the potential according to the stochastic description Eq.~\eqref{eq:DBM}. 
\item In the new potential, we again solve the equations of motion for the system.
\item We repeat points 3 and 4 for many patches, until $\eV=1$ and inflation ends. 
\item If the random realisation has resulted in at least 60 efolds of inflation, we continue to study the evolution of the perturbations in this background. If the models yields fewer than 60 efolds, we discard it. 
\item Given the homogeneous background solutions for the field and the randomly generated potential,  we compute the transport matrices of non-homogeneous field perturbations, using the analytical  expressions derived in \S\ref{sec:perts}. As will be discussed in that section, we do this both in full non-slow roll and simplified slow roll descriptions.
\end{enumerate}
We repeat the above prescription a large number of times, keeping the  hyperparameters fixed, to generate an ensemble of DBM models corresponding to a given set of hyperparameters. We then use these ensembles to 
study the statistics of the resulting predictions. 
 \\
\\
\noindent {\bf The statistical ensemble of models:}
For each choice of hyperparameters that we study, we generate $\gtrsim1000$ inflationary realisations with at least 60 efolds of inflation. 
For the ease of analysis (and to save computational cost), we do not study a grid of the four-dimensional hyperparameter space, but instead choose illustrative lines through it.
Specifically, we generate ensembles for:
\begin{itemize}
\item  $N_f$ between 2 and 100, keeping $\Lh=0.4 \Mpl$, $\epsilon_{V0}=10^{-11}$ and $\eta_{V0}= -10^{-4}$,
\item $\Lh$ between $0.2 \Mpl$ and $\Mpl$,  keeping $N_f =20$, $\epsilon_{V0}=10^{-11}$ and $\eta_{V0}= -10^{-4}$,
\item $\epsilon_{V0}$ in the range $10^{-12}$ to $10^{-9}$,  keeping $N_f=20$, $\Lh=0.4 \Mpl$,  and $\eta_{V0}= -10^{-4}$,
\item $\eta_{V0}$ between $-10^{-5}$ and $-0.1$,  keeping $N_f=20$, $\Lh=0.4 \Mpl$,  and  $\epsilon_{V0}=10^{-11}$.
\end{itemize}
In total, we generate more than 100,000 models of inflation, and study the observables of approximately 30,000 inflationary realisations yielding at least 60 efolds of inflation. Throughout our analysis, we take the pivot scale at which CMB observables are estimated to be the scale which left the horizon 55 efolds before the end of inflation.

\subsection{Subtleties of the DBM potentials}
\label{sec:subtleties}
The DBM prescription results in a large 
class of random potentials with many  fields, and, for the first time, makes it possible to study inflation in models with dozens or even hundreds of interacting fields.\footnote{Models with ${ \cal O} (100)$ non-interacting fields have been studied in the context of N-flation \cite{1312.4035}. Symmetries protecting interactions amongst fields make these potentials and subsequent phenomenology much simpler than the models we wish to explore with this work.} But, owing to their Brownian origin, the DBM potentials differ in some important ways from more conventional constructions of scalar potentials, and cannot be used to address all possible questions of interest. 
In this subsection, we review the subtleties of the DBM construction. 
 
The DBM prescription leads to a scalar potential defined locally to quadratic order along a path $L$ in field space. 
The path $L$ is not completely arbitrary in order to avoid self-intersecting paths which do not lead to single valued potentials.  For
 models of inflation with $N_f \gg 1$ fields, this is not a severe restriction, as self-intersecting trajectories are very rare in a high-dimensional field space.

 Moreover, 
 as the potential is mapped out in a stochastic fashion along the path, it does not automatically evolve to  a vacuum configuration with a small or vanishing vacuum energy (should the path be continued  past the end of inflation).
 Thus, a DBM prescription that extends beyond inflation must also incorporate a vacuum generation mechanism (see also \cite{1611.07059}).

In general, the DBM prescription is not well-suited to address questions relating to the global nature of random scalar potentials. For example, Morse theory inequalities are not easily verifiable given the potential in only a small subset of the field space \cite{1307.3559}. Moreover, for very long paths ($s \gg1$),  the energy scale of the potential and the Hubble parameter may exceed the putative ultraviolet cut-off of the theory, $\Lh$,  indicating a break-down of the effective field theory interpretation of the model. This is related to the fact that,  while the Hessian matrix evolves into a stationary statistical distribution for $s\gg 1$, the variance of the potential and its gradient grow with $s$ \cite{MarshPajer}. For inflationary models with $s \lesssim {\cal O}({\rm few})$, this concern is rarely of practical relevance. 

A more serious concern that has caused some confusion in the literature relates to the intrinsically Brownian nature of Dyson Brownian Motion. It is well-known that a particle undergoing Brownian motion traces out a path that is (`almost surely') continuous, but also (`almost surely') nowhere differentiable. This follows from the Markovian nature of the random steps. 
Consequently, the evolution of $v_{ab}$ is (`almost surely') continuous along the path $L$, but also (`almost surely') not differentiable in the continuous limit. That is, DBM potentials are twice differentiable (by construction), but higher derivatives are not well-defined.\footnote{This should be contrasted by claims that DBM potentials are only once differentiable \cite{1409.5135}.} 
Fortunately, to compute the two-point statistics of the perturbations in models of multiple-field inflation, only the second derivatives along the field trajectory are needed, not higher orders. Moreover, the values of $v_{ab}$ along the path are only relevant up to a certain numerical accuracy, which effectively regulates the unknown higher-order terms. Hence, for the purpose of computing the power spectrum of the field perturbations, the curvature power spectrum and the isocurvature, the DBM construction suffices. Computing higher-order correlations and non-Gaussianities however, requires significant amendments to the construction, going beyond the Markovian definition of the stochastic evolution. 

A promising alternative method to explore inflation in high-dimensional field spaces invokes multi-dimensional Gaussian Random Fields (GRFs) as the field-dependent scalar potential. Early work using this method include \cite{1111.6646}, in which inflation in potentials with $\Lh > \Mpl$ and up to six fields was studied. Subsequent work
 includes \cite{Battefeld:2012qx, Bachlechner:2014rqa, 1612.03960} (see also \cite{astro-ph/0410281} for earlier studies of single-field inflation in random potentials). The Hessian matrix  of GRFs is simply related to the GOE, and eigenvalue repulsion is critically important for the evolution of fluctuated eigenvalue configurations also in GRFs.\footnote{In particular, the Hessian of an $N_f$-dimensional, mean zero GRF with Gaussian covariance function is an element of the GOE plus a shift, 
 proportional to minus the value of the potential times the unit matrix. For the set of points at which the potential is zero, the spectrum of the Hessian is precisely the GOE semi-circle law.} Hence, we expect that the lessons derived from studies of DBM potentials should apply also to potentials modelled by Gaussian Random Fields.

\section{Method: Computing Observables}
\label{sec:perts}
In this section we describe the methods for computing inflationary observables used in our analysis. Computing observables for a large number of fields is in general a numerically heavy task, even if we limit our study to the power spectrum of curvature, isocurvature and tensor perturbations.\footnote{We leave the study of non-gaussianities to a further work. For details on how to compute three-point statistics of inflationary observables see Ref.~\cite{1609.00379}.} The heaviest computational cost comes from the evaluation of the two-point correlation function of 
the field perturbations, $\langle \delta \phi^a_{-{\bf k} }\delta \phi^b_{\bf k}\rangle$, 
which involves solving $\mathcal{O}(N_f^2)$ coupled ordinary differential equations for a given scale $k$. This task then needs to be repeated for a range of scales in order to construct the full power spectrum of the curvature perturbations, $P_\zeta(k)$. Given a complicated  inflationary potential, we have have no reason to expect $P_\zeta(k)$ to have a simple functional form and hence a large number of modes must be computed in order to guarantee a good characterisation of $P_\zeta(k)$. 
For these reasons, the study of 
complex multifield models 
crucially requires a
highly  efficient method for computing observables.

The  section is organised as follows: in \S \ref{sec:generalmethod} we 
closely follow Ref.~\cite{1609.00379, 1302.3842, 1203.2635, 1502.03125} and 
present our most general approach to computing two-point statistics of scalar perturbations in the flat gauge. 
In \S\ref{sec:srcalc}, we show that by assuming slow-roll, and by only considering superhorizon scales, the calculation of observables can be sped up  dramatically.
 This slow-roll method is sufficiently general to capture a very broad class of models, 
 including the complicated models of manyfield inflation constructed with the DBM method. 
 As far as we are aware, this is the first example of a general method for computing observables that is feasible for models with $\mathcal{O}(100)$ fields and a complicated potential. In \S \ref{sec:gauge}, we briefly summarise how to use the separate universe assumption to obtain the gauge transformation required to move from the flat gauge to the constant density gauge.
 In \S\ref{sec:Pzeta} and \S\ref{sec:ns}
 we obtain compact expressions for the power spectrum $P_{\zeta}(k)$ and related quantities such as the spectral index, running and isocurvature power spectrum. In \S \ref{sec:methodcomparison} we 
 compare the results of the efficient slow-roll method with the more general, slower method. 
 
\subsection{The general method}
\label{sec:generalmethod}

A common 
approach to computing the statistics of $\zeta$ is to break the calculation into two steps: the computation of correlation functions of fluctuations evaluated on flat surfaces, and the subsequent evaluation of the gauge transformation which relates field perturbations to the curvature perturbation evaluated on constant density surfaces. The first step encompasses the bulk of the computational effort as it corresponds to evolving field perturbations through non-linear evolution until the desired time of evaluation. The second step is a simple analytical expression which depends only on background quantities. There are a number of methods for computing correlation functions in the flat gauge (see Ref.~\cite{1609.00379} for a review). The approach taken here is to compute the propagator ${\Gamma^{\alpha}}_{\beta}(N,N_{0})$, defined such that \cite{1203.2635},
\be\label{eq:deltaphi}
X^{\alpha}(N)={\Gamma^{\alpha}}_{\beta}(N,N_{0})X^{\beta}(N_{0}) \, ,
\ee
where we have collected the fields and momenta $X^{\alpha} \equiv  \{\delta\phi^{a} , \delta\pi^{a}\}$. Here and in what follows, greek indices run over the $2N_{f}$ fields and momenta and latin indices run over the corresponding $N_{f}$ fields or momenta. Defining the dimensionless two-point correlation function as,
\begin{equation}
    \label{eq:generic-2pf}
    \langle X^\alpha({\bf k}_1) X^\beta({\bf k}_2) \rangle
  \equiv
    (2\pi)^3 \delta({\bf k}_1 + {\bf k}_2) \frac{\Sigma^{\alpha\beta}(k)}{k^3} \, ,
\end{equation}
it simply evolves according to two copies of the propagator,
\begin{equation}\label{eq:sigmasol}
\Sigma^{\alpha\beta}(N)={\Gamma^{\alpha}}_{\gamma}(N,N_{0}){\Gamma^{\beta}}_{\delta}(N,N_{0})\Sigma^{\gamma\delta}(N_{0})\, .
\end{equation}
Here we drop the label $k$ for simplicity; this expression describes how $\Sigma^{\alpha\beta}$ for a single $k$-mode evolves in time.

To compute the propagator, we consider the action governing small fluctuations $\delta\phi_{\bf k}^a(t)$ around a homogeneous background $\phi^a(t)$. In the flat gauge, to second order in amplitude, this is given by \cite{astro-ph/9507001}:
\begin{equation}
    \label{pertaction}
    S \supseteq \frac{1}{2} \int \frac{\d^3 k}{(2\pi)^3} \, \d t \; a^3
    \bigg\{
        \big[ \partial_{t} \delta\phi_a({{\bf k}}) \big]
        \big[ \partial_{t} \delta\phi^a(-{{\bf k}}) \big]
        - \bigg(
            \frac{k^2}{a^2}\delta_{ab}+ M_{ab}
        \bigg)
        \delta\phi^a({{\bf k}}) \delta\phi^b(-{{\bf k}})
    \bigg\}
    ,
\end{equation}
where the effective mass matrix $M_{ab}$ is given by,
\begin{equation} 
\label{masseq}
    M_{ab}
    \equiv 
        V_{ab}
        - \frac{1}{a^3} \partial_{t}
        \bigg(
            a^3 \frac{\dot{\phi}_\alpha \dot{\phi}_\beta}{H}
        \bigg) \, .
\end{equation}
Using the definition $\delta\pi^{a} \equiv \d \delta\phi^a/ \d N$, the equation of motion at tree-level has the form,
\begin{equation}\label{eq:EOMperts}
\frac{\d X^{\alpha}}{\d N}={u^{\alpha}}_{\beta}X^{\beta} \, ,
\end{equation}
and hence the propagator may be obtained by solving,
\begin{equation}\label{eq:GammaEoM}
\frac{\d\Gamma^{\alpha \beta}}{\d N}={u^{\alpha}}_{\gamma}\Gamma^{\gamma \beta}\, ,
\end{equation}
where,
\begin{equation}
{u^\alpha}_\beta = \left(\begin{split}
    & 0  \	\	\	\	\	\	\	\	\	\	\	\	\	\		\	\	\	\	\	\	\	\	  	\	\	\	\	\	   {\delta^a}_b \\
    -{\delta^a}_b \frac{k^2}{a^2H^2}& - \frac{{M^a}_b}{H^2}  	\	\	\	\	\		\	\	\	\	\	\	\	\	  (\epsilon-3){\delta^a}_b 
\end{split}\right) \, .
\label{eq:u-tensor}
\end{equation}
Similar equations, although even simpler, can be obtained for the two-point correlator of tensor perturbations, cf.~Ref.~\cite{1502.03125}.

One might wonder if it would be computationally cheaper to solve Eq.~\eqref{eq:EOMperts} instead of Eq.~\eqref{eq:GammaEoM}. Even though there are only ${\cal O}(N_{f})$ equations in~\eqref{eq:EOMperts}, to compute the correlation functions we need to solve ${\cal O}(N_{f}^{2})$ equations; this was originally pointed out by Salopek, Bond \& Bardeen \cite{Salopek:1988qh}. A single solution of Eq.~\eqref{eq:EOMperts} would describe how $\delta\phi^\alpha$ and $\delta\pi^\alpha$ respond at late times to a specific initial condition. So to compute the correlation functions, for which we need to know how the perturbations respond to arbitrary initial conditions, we have to compute $2N_f$ equations for each of the independent initial conditions.\footnote{Reference \cite{Salopek:1988qh} deals with this by decomposing $X^{\alpha}$ in terms of two index mode functions. This is the approach taken in \cite{1410.0685} and is very similar to solving Eq.~\eqref{eq:GammaEoM}. 
}

Equation \eqref{eq:GammaEoM}, together with Eq.~\eqref{eq:sigmasol}, determines the evolution of the correlation functions in the flat gauge. To specify the initial conditions for the two-point correlations,  $\Sigma^{\alpha\beta}(N_{0})$,  we consider two prescriptions. First, 
sufficiently deep inside the horizon the correlation functions are kinetic dominated and the modes are approximately massless.
 In this case, the Bunch-Davies initial conditions are~\cite{1502.03125}, 
\begin{equation}
  \Sigma^{\alpha\beta} (N_0)=\left(
\begin{split}
    \frac{H^2_0 \delta^{ab}_0}{2}  &  |k\tau_0|^2  
    \	\	\	\	\	\	\	\	\	\
    - \frac{H^2_0 \delta^{ab}_0}{2}   |k\tau_0|^2\\
     - \frac{H^2_0 \delta^{ab}_0}{2}  & |k\tau_0|^2    \	\	\	\	\	\	\	\	\	\	\	\	\	\
    \frac{H^2_0 \delta^{ab}_\ast}{2}   |k\tau_0|^4  
\end{split}\right),
\label{eq:universal-ics}
\end{equation}
where $\tau=-1/aH$ is the conformal time and a subscript `0' denotes evaluation at the initial time $N_{0}$ deep inside the horizon when $|k / aH| \approx |k\tau| \gg 1$.
This initial condition is suitable for our most general analysis, but is also computationally costly as $N_0 \ll N_{\rm exit}$. The second prescription for specifying the initial conditions for \eqref{eq:sigmasol}, valid in slow-roll and if the field trajectory is slowly turning at horizon crossing, takes $N_0 = N_{\rm exit}$ and treats the eigenmodes of the effective mass matrix \eqref{masseq} as independent correlators. We will return to this initial condition in \S\ref{sec:SR}.





Equations \eqref{eq:GammaEoM} and \eqref{eq:sigmasol} should be evolved until some final time at which the observational predictions are determined.
In multifield models of inflation, the observational predictions can in general continue to evolve well after the end of inflation, at which point they will be affected by the reheating dyanamics. 
%
 In this work we will not model the reheating phase (as this would require a modification of the DBM construction of the inflationary potentials), but  focus on the inflationary era and take the final time of evaluation 
 to be the end of inflation. 
We expect that the general method for computing the perturbations presented here will be applicable also for the post-inflationary evolution.

The set of ODEs Eq.~\eqref{eq:GammaEoM} is well suited for numerical implementation, making this method a highly versatile 
recipe for computing observables at tree level in multiple field models.\footnote{A very closely related method also extends in a simple way to the study of three-point statistics. See \href{https://transportmethod.wordpress.com/}{transportmethod.com} for a publicly available implementation in both mathematica and C++.}  Nevertheless, as already mentioned, the computational cost of solving this system of equations scales poorly with the number of fields, making the evaluation of $P_\zeta(k)$ quite expensive in certain regions of our hyperparameter space.

For our trajectories built using DBM, the matrix ${u^{\alpha}}_{\beta}$ is obtained from the numerical solution to the background equations of motion for each patch. 
Using the fact that the propagator satisfies composition,
\begin{equation}
{\Gamma^{\alpha}}_{\beta}(N_{3},N_{1})={\Gamma^{\alpha}}_{\gamma}(N_{3},N_{2}){\Gamma^{\gamma}}_{\beta}(N_{2},N_{1}) \, ,
\end{equation}
for times $N_{3}>N_{2}>N_{1}$, we can solve Eq.~\eqref{eq:GammaEoM} for each patch independently and obtain the result for the whole evolution from $N_0$ at the beginning of patch $p_0$ to any final time at point $p_f$ by computing the path ordered product,
\bea
\label{stringingGammas}
X^\alpha \big|_{p_f} &=& 
{\Gamma^\alpha}_{\gamma}(p_{f},p_{f-1}) \, 
\ldots {\Gamma^\rho}_{\beta}(p_{1},p_{0}) \,  X^\beta \big|_{p_{0}} 
\nonumber \\
&\equiv&
{\Gamma^\alpha}_{\beta}(p_f, p_0) \, 
X^\beta \big|_{p_0} 
 \, .
\eea
Note that the initial conditions for Eq.~\eqref{eq:GammaEoM} in each patch is always simply the identity matrix, making this approach especially easy to implement.
\footnote{We note that  one could alternatively consider solving the integration Eq.~\eqref{eq:GammaEoM} only once using an interpolating function for $u_{\alpha\beta}$ from $N_0$ to the final time of evaluation rather than performing the calculation in each patch individually. We found this method required very high numerical accuracy and was generally harder to implement in such a way that it was both efficient and robust. For all calculations of the full non-slow-roll treatment, we therefore evaluated the propagator patch by patch.
}

In what follows we refer to this method of computing the propagator as the full non-slow-roll treatment. It provides a precise result, capturing all tree-level effects.\footnote{See Ref.~\cite{1609.00379} for a more careful discussion of what is captured at tree-level.}

\subsection{Drastic   slow-roll simplifications}
\label{sec:srcalc}
\label{sec:SR}

The slow-roll conditions lead to drastic simplifications of the evolution of the perturbations. In this subsection, we show that if the slow-roll approximation holds throughout the last 60 or so efolds of inflation (which is typically the case in the examples we study), and the field perturbations at horizon exit are approximately uncorrelated (which we will check in \S\ref{sec:methodcomparison}), the propagator $\Gamma^{\alpha}_{~\beta}$ can be solved for analytically in each coordinate patch.  This greatly increases our ability to study systems with many fields explicitly, and is one of the key results of this paper.

The essential simplification stems from the local sum-separability of the inflaton potential. Vernizzi and Wands showed in Ref.~\cite{Vernizzi:2006ve} that for potentials of the form $V(\phi_{1},\phi_{2}) = V_{1}(\phi_{1})+V_{2}(\phi_{2})$, one can obtain expressions for observables purely in terms of background quantities, thereby entirely avoiding the necessity to solve the  
differential equations for the perturbations numerically. It was then shown in Ref.~\cite{Battefeld:2006sz} that the method generalises to an arbitrary number of fields.\footnote{The same approach can also be applied to product separable models \cite{Choi:2007su} and in the case of $H$-separable models one can even use the same approach without assuming slow-roll \cite{Byrnes:2009qy}.} While the potential studied here clearly is not of this form, it is constructed by stringing together a series of patches which, by a suitable field redefinition, individually can be expressed as sum-separable. This field redefinition is an orthogonal transformation of the field basis $\vec{\varphi}= {O}^{\rm T} \vec{\phi}$ which diagonalises the Hessian, 
 such that the potential is locally to quadratic order is sum-separable,
\be
\label{sumsep}
V(\varphi^{1},\ldots,\varphi^{N_f})=\sum_{a}^{N_{f}}V_a(\varphi^{a}) \, .
\ee
Note that in this subsection and only this subsection, we denote one partial derivative of the potential by $V'_a(\varphi^a)$. 
In this basis, the field perturbation in each patch evolves as,
\be
\delta \varphi^a \big|_{p_{i+1}} = \Gamma^a_{~b} (p_{i+1},p_i) \delta\varphi^b |_{p_i} \, ,
\ee
and Eq.~\eqref{stringingGammas} is substituted by,
\bea
\delta \vec{\phi} \big|_{p_f} &=& 
{O}^{\rm T}_{p_{f}}{\Gamma}(p_{f},p_{f-1}) {O}_{p_{f}}\, 
\ldots {O}^{\rm T}_{p_{1}}{\Gamma}(p_{1},p_{0}) {O}_{p_{1}}\,  \delta \vec{\phi} \big|_{p_{0}} \, .
\label{gammasumsep}
\eea
Note the usage of latin indices, as in the slow-roll limit we work purely in field-space, rather than the full phase-space. Here we are taking $p_0$ to correspond to the time of horizon exit and so the path-ordered product of propagators and orthogonal transformations describes evolution on superhorizon scales. 
Assuming slow-roll, the propagator on superhorizon scales $\Gamma^a_{~b} (p_{i+1},p_i)$ can be expressed purely in terms of background quantities, providing an analytic solution to the propagation of the perturbation along the inflationary trajectory. 
In this approach the work of numerically solving the coupled ODEs of Eq.\eqref{eq:GammaEoM} is 
replaced with 
finding the  orthogonal matrices $O_{p_i}$ and the local propagators $\Gamma(p_{i+1}, p_{i})$ in each patch --- both entirely specified given 
the solution of the classical background  
 ---
 and multiplying them together.  
We find this method to be exceedingly numerically efficient. 
Furthermore, in contrast to any calculation starting on subhorizon scales, there is no explicit $k$-dependence. Hence, the propagators computed this way have the full information about evolution of observables at any scale. This implies that one obtains observables for \emph{all} scales exiting the horizon during inflation at no additional cost, by simply starting the product of Eq.~\eqref{gammasumsep} at subsequent patches.

We now compute the expression for the propagator in each patch. In the slow-roll limit, and for large scales where $k/aH \ll 1$, the equations of motion reduce to, 
\begin{equation}\label{eq:JocobiEoM}
\frac{\d\delta\phi^{a}}{\d N}={u^{a}}_{b}\delta\phi^{b} \, , 
\end{equation}
where the expansion tensor $u_{ab}$ is now simply,
\be
u_{ab}=-\partial^2_{ab}\ln V \, .
\ee
On superhorizon scales, the field perturbations may be interpreted as Jacobi fields, describing the independent evolution of a bundle of trajectories in field space \cite{1203.2635}. One consequence of this is that Eq.~\eqref{eq:JocobiEoM} can be obtained by simply perturbing the background equations of motion, it also implies that we can take the infinitesimal limit of the perturbations to bring Eq.~\eqref{eq:deltaphi} to the form,
\begin{equation}
{\Gamma^a}_{b}(N_{(f)},N_{(i)})=\frac{\partial\phi^{a(f)}}{\partial\phi^{b(i)}} \, ,
\label{gammawewant}
\end{equation}
where a label ``$i$" indicates evaluation at some initial flat surface and ``$f$" denotes evaluation on some final flat surface corresponding to times $N_{(i)}$ and $N_{(f)}$ respectively.\footnote{References \cite{Vernizzi:2006ve,Battefeld:2006sz} compute a very similar object:
\begin{equation}
\frac{\partial \varphi^{(c)}_{a}}{\partial \varphi^{b(i)}}=-\frac{V^{(c)}}{V^{(i)}}\sqrt{\frac{\epsilon^{(c)}_{a}}{\epsilon^{(i)}_{b}}}\left(\frac{\epsilon^{(c)}_{b}}{\epsilon^{(c)}}-\delta_{ab}\right) \, .
\end{equation}
However, while similar, this is \emph{not} the quantity we are after because it takes the final surface `c' to be a constant density surface. The propagator \eqref{eq:deltaphi} applies to fluctuations in the flat gauge and this distinction is of course important for preserving the composition property of $\Gamma$. 
}

To determine the $N_f^2$ components of ${\Gamma^a}_{b}(N_{(f)},N_{(i)})$ of Eq.~\eqref{gammawewant}, we wish to compute the response in 
the final position in field space given a general infinitesimal variation in the initial position:
\be
\label{variation}
{\rm d} \fin_a = 
\frac{\partial \fin_a}{\partial \varphi^{b(i)}} 
{\rm d} \ini_b \, .
\ee
As we  now describe, this can be done by requiring the conservation of $N_f-1$  `slow-roll charges', and  that $\Gamma^a_{~b}$ propagates the perturbations to a final hypersurface with flat geometry.


The number of efolds parametrises the position of the fields along the path, $\varphi^a(N)$. In slow-roll and for 
 a sum-separable potential like Eq.~\eqref{sumsep}, the  number of efolds between the points $\ini_a$ and $\fin_a$ 
 is given by,
\be
N = - \sum_{a=1}^{N_f} \int_{\ini_a}^{\fin_a} \frac{V_a}{V'_a} d\varphi_a \, ,
\label{eq:Ne}
\ee 
where a prime denotes differentiation with respect to the appropriate field. In addition, each slow-roll path can be identified by $N_f -1$ integrals of motion, which are constant along the trajectory,
\be
\label{Cas}
C_a  = - \int_{\gamma} \frac{d\varphi_1}{V_1'} + \int_{\gamma} \frac{d\varphi_a}{V'_a} \, ,
\ee
for  $a = 2 \ ... \ N_f$.
Here $\gamma$ denotes a field-space path from a fiducial reference point to any point on the slow-roll trajectory. 

A small perturbation of the initial position results in a slightly perturbed slow-roll trajectory, and a variation of the charges $C_a$,
\be
\frac{d C_a}{d \ini_b} = - \left(\frac{1}{V_1'}\right)_i \delta_{b1} + \left(\frac{1}{V_a'}\right)_i \delta_{ab} \, ,
\ee
where we have taken the path $\gamma$ to end at $\ini_b$, $\delta_{ab}$ denotes the Kronecker delta function, and the repeated $a$ on the right hand side is not summed over.
 However, the path $\gamma$ can equally well be taken to end at $\fin_b$ so that,
\be
\frac{d C_a}{d \ini_b} =  \frac{\partial C_a}{\partial \varphi^{c(f)}} \frac{\partial \fin_c}{\partial \varphi^{b(i)}} 
=
\sum_{c=1}^{N_f} \left( 
- \left(\frac{1}{V_1'}\right)_f\delta_{c1} + \left(\frac{1}{V_a'}\right)_f \delta_{ac}
\right) \frac{\partial \fin_c}{\partial \varphi^{b(i)}} \, .
\ee
This gives $N_f(N_f-1)$ linear equations for the $N_f^2$ unknown elements of $\partial \fin_c/\partial \varphi_b^{(i)}$,
\be
 - \left(\frac{1}{V_1'}\right)_i \delta_{b1} + \left(\frac{1}{V_a'}\right)_i \delta_{ab} 
 =
\sum_{c=1}^{N_f} \left( 
- \left(\frac{1}{V_1'}\right)_f\delta_{c1} + \left(\frac{1}{V_a'}\right)_f \delta_{ac}
\right) \frac{\partial \fin_c}{\partial \varphi^{b(i)}} \, .
\label{eq:syst1}
\ee

To obtain  $N_{f}$ additional equations, we impose the constraint that the 
final hypersurface is flat, i.e.~that the curvature perturbation vanishes. To linear order, this means that, 
\be
\zeta(N_f, {\bf x}) =\frac{\dd N}{\dd \ini_a} \delta \ini_a ({\bf x}) = 0 \, ,
\ee
which for general perturbations requires that,
\be
\frac{\dd N}{\dd \ini_a} = \frac{\partial N}{\partial \varphi^{b(i)}} + \frac{\partial N}{\partial \varphi^{c(f)}} \frac{\partial \fin_c}{\partial \varphi^{b(i)}}  =0 \, .
\ee
Using the definition \eqref{eq:Ne}, this gives,
\be
\left(\frac{V_a}{V_a'}\right)_i - 
\left( \frac{V_b}{V'_b}\right)_f \frac{\partial \fin_b}{\partial \varphi^{a(i)}} = 0 \, .
\label{eq:syst2}
\ee 
and an additional $N_f$ equations for the unknown $\partial \fin_c/\partial \varphi_b^{(i)}$.

Equations \eqref{eq:syst1} and \eqref{eq:syst2} can be solved by inspection, giving, 
\be\label{eq:Gammaeq}
\frac{\partial \fin_a}{\partial \varphi^{b(i)}}  = 
\frac{(V'_a)_f}{(V'_b)_i} \left( 
\delta_{ab}
+
   \frac{ (V_b)_i -(V_b)_f }{ V_f} 
\right)
\, .
\ee
Consistently, as $\fin \to \ini$, $\Gamma_{ab} \to \delta_{ab}$.\footnote{As a further consistency check, we can contract this expression for $\Gamma$ with the gauge transformation which will be derived in the next section, Eq.\eqref{eq:gaugetran}, to obtain,
\begin{equation}
\frac{\partial N}{\partial\varphi^{a(i)}}=\left[\frac{1}{V_{a}'}\right]_{i}\left[(V_{a})_{i}+\left(\frac{VV_{a}'^{2}}{V_{b}'V^{'b}}-V_{a}\right)_{f}\right] \, ,
\end{equation}
which agrees with the expressions given in Refs.~\cite{Vernizzi:2006ve,Battefeld:2006sz}.}

Equation \eqref{eq:Gammaeq} is the key to striking reductions of computational cost for computing the power spectrum, $P_\zeta(k)$, for systems with a large number of fields. By only involving quantities accessible from the solution of the classical background alone, it allows us to completely circumvent the task of numerically solving the differential equations of the perturbations. Rather,   equations \eqref{gammasumsep} and \eqref{eq:sigmasol} shows how to obtain the field correlator $\Sigma_{ab}$ at any point during inflation by stringing together local solutions of \eqref{eq:Gammaeq} along the background trajectory.

An important aspect of the  expression \eqref{eq:Gammaeq} is that it is independent of $k$. 
However, as in virtually all inflationary models, the curvature power spectrum is not completely scale-invariant: 
different $k$-modes exit the horizon at different times during inflation, 
and the field perturbations at horizon crossing, 
 $\Sigma_{ab}(N_{\rm exit})$, reflect the slowly varying Hubble parameter during inflation. 
More precisely, assuming 
that the field trajectory is not turning significantly at the time of horizon crossing,\footnote{We know of no general, analytic expression for the correlation functions in the case of a fast-turning trajectory. In that case, a treatment like the one described in \S\ref{sec:generalmethod}, where the propagator is estimated from the deep subhorizon limit, needs to be used.
} 
the correlation function is diagonal in the basis that diagonalises
 the effective mass matrix Eq.~\eqref{masseq}.  
 In terms of the 
  independent mode functions in this basis, $\delta\tilde\varphi$,  the correlation functions   at horizon exit, $\tilde\Sigma^{ab} (N_{\rm exit})$, are given by \cite{ 0911.3380, 0911.3550},

\begin{equation}
    \label{eq:generic-2pf_a}
   \tilde\Sigma^{ab} (N_{\rm exit})=  \frac{H^2}{2} \	 \delta^{ab},\	\	\	\	\ {\rm if}\	\	\frac{M^2}{H^2} \leq \frac{9}{4}
\, ,
\end{equation}
and 
\begin{equation}
    \label{eq:generic-2pf_b}
   \tilde\Sigma^{ab} (N_{\rm exit})=  e^{- \pi \nu}\frac{H^2}{2} \	 \delta^{ab},\	\	\	\	\ {\rm if}\	\	\frac{M^2}{H^2}>\frac{9}{4},
\end{equation}
where $\nu=\sqrt{\frac{M^2}{H^2} - \frac{9}{4}}$ and $M$ are eigenvalues of the effective mass matrix Eq.~\eqref{masseq}. Bringing these expressions back to the original field basis gives rise to a non-diagonal expression for $\Sigma^{ab}(N_{\rm exit})$. 

An additional source of $k$-dependence is intrinsically multifield in nature. Modes exiting the horizon earlier during inflation will evolve with a longer string of local propagators, cf.~Eq.~\eqref{gammasumsep}, thus inducing a non-trivial $k$-dependence inherited from the superhorizon evolution of the  curvature perturbation.

\subsection{From field space to $\zeta$}
\label{sec:gauge}

In this subsection, we relate the 
field perturbations in the flat gauge 
to the curvature perturbation,  $\zeta$, in the uniform-density gauge.
%
Since the relevant scales for which we wish to perform this gauge transformation are superhorizon, a very simple way to do so is by invoking the separate universe assumption,
which gives an intuitive classical
description of
the evolving fluctuations~\cite{Starobinsky:1986fxa,Sasaki:1995aw,Wands:2000dp,
Rigopoulos:2003ak,Lyth:2004gb,Lyth:2005fi, 1410.3491}.  


In the superhorizon limit, we can interpret the evolution of field perturbations as a
bundle of non-interacting inflationary trajectories in field space. In this picture Eq.~\eqref{eq:deltaphi} and \eqref{eq:EOMperts} describe the evolution of Jacobi fields and hence characterise the evolution of the shape of the bundle~\cite{1203.2635}. In addition, $\zeta$ simply corresponds to the variation in the number of efolds that each trajectory takes to reach a constant density hypersurface from an initially flat hypersurface \cite{Starobinsky:1986fxa,Sasaki:1995aw,Wands:2000dp,
Rigopoulos:2003ak,Lyth:2004gb,Lyth:2005fi, 1410.3491}. Therefore, the gauge transformation that relates $\delta \phi^a$ and $\delta \pi^a $ to $\zeta$ must be a measure of this $\delta N$ when both hypersurfaces are evaluated at the same cosmic time. Reference \cite{1410.3491} presents a systematic way to compute this quantity. We will first describe how to compute the gauge transformation that connects the perturbations $X^\alpha = \{\delta\phi^{a} , \delta\pi^{a}\}$ to $\zeta$ and then present the slow-roll limit of the result.

Let $\Delta N$ be the number of efolds  between a point $p=\{\phi^a_*,\pi^a_*\}$ on the spatially
flat hypersurface at which the density is $\rho (\phi^a_*, \pi^a_*)$
and a nearby uniform density
hypersurface with density $\rho_c$. At first order in $\rho$,
\begin{equation}
    \Delta N =
    \left.
    \frac{\d N}{\d \rho}
    \right|_p
    (\rho_c - \rho)
    + \ldots .
\end{equation}
Expanding in terms of phase-space coordinates, we then have,
\begin{equation}
\begin{split}
    \delta ( \Delta N )
    & \approx
    {-\mbox{}}
    \left.
    \frac{\d N}{\d \rho}
    \right|_*
    \Big(
        \left. \frac{\partial \rho}{\partial \phi^a} \right|_* \delta \phi^a_*
        +
        \left. \frac{\partial \rho}{\partial \pi^a} \right|_* \delta \pi^a_*
    \Big)
    + \ldots
 \end{split}
\end{equation}
where `$\ldots$' denotes higher order terms in
$\delta \phi^a$ or $\delta \pi^a$.
The variation $\delta ( \Delta N )$ is precisely $\zeta$ and we identify the gauge transformations at linear order, ${N}_a$ and ${\tilde N}_{a}$, to be defined by,
\begin{equation}
\zeta=\delta ( \Delta N )
    \equiv
    {N}_a \delta \phi^a
    +
    {\tilde N}_{a} \delta \pi^a
    + \ldots .
    \label{eq:gauge-xfm}
\end{equation}
In terms of the potential and momenta they read:
\be
    \label{eq:N-unbarred}
 N_{\alpha} \equiv \{ {N}_a, {\tilde N}_{a}\}  = \left\{ \frac{1}{2\epsilon V}    V_a \ , \     \frac{1}{2\epsilon(3-\epsilon)} \pi_a \right\}.
\ee
Here, again, $V_a = \partial_a V$.

In the slow-roll limit, the perturbations are expressed purely in terms of field-space coordinates and the gauge transformation is simply given by, 
\begin{equation}\label{eq:gaugetran}
N_{a}= \frac{V}{  V_b  V_b}  V_a \, .
\end{equation}
This expression gives the instantaneous gauge transformation that relates 
a flat hypersurface to a \emph{nearby} uniform-density surface, and is thereby different from the commonly used `$\delta N$' expression for the variation in the total number of efolds with respect to fluctuations at an initial (and hence distant) flat hypersurface. These formalisms are obviously closely related, and we now review the computation of observables in the transport formalism. 

\subsection{Curvature and isocurvature power spectra}%
\label{sec:Pzeta}
In this paper, we compute the two-point correlation functions of field perturbations and the corresponding observables generated during inflation. Of primary interest is the  power spectrum of curvature perturbations, which is defined as,
\begin{equation}
    \langle
        \zeta({{\bf k}}_1)
        \zeta({{\bf k}}_2)
    \rangle
    =
    (2\pi)^3
    \delta^{(3)}({{\bf k}}_1+{{\bf k}}_2)
    \frac{P_{\zeta}(k)}{k^3} \, .
\end{equation}
In single-field inflation, $P_{\zeta}(k)$ evolves on subhorizon scales ($k>aH$) but stops evolving at around horizon crossing ($k=aH$), see for example Ref.~\cite{astro-ph/0411220, astro-ph/0306620, astro-ph/0405397}.
In multifield inflation, the curvature perturbation may evolve on superhorizon scales so that  
in general, $P_{\zeta}(k) = P_{\zeta}(k, N)$. For compactness of notation, we will often suppress the dependence of $k$ and let $P_{\zeta}(N)$ denote the time-dependent power of the curvature perturbation at a given scale.   
It follows from
Eqs.~\eqref{eq:gauge-xfm} and \eqref{eq:sigmasol} that
$P_\zeta$ can be written as,
\begin{equation}\label{eq:2p}
    P_{\zeta}(N)=N_{\alpha}N_{\beta}\Sigma^{\alpha \beta}(N)=  N_{\alpha}(N) N_{\beta}(N) {\Gamma^{\alpha}}_{\gamma}(N, N_0){\Gamma^{\beta}}_{\delta}(N, N_0)\Sigma^{\gamma\delta}(N_{0}) \, .
\end{equation}
%
This expression 
applies to each mode $k$, so that $P_\zeta(k, N)$ can be obtained by  
evaluating
Eq.~\eqref{eq:2p} for each mode within the range of interest. In the slow-roll formalism of \S\ref{sec:SR} and with the gauge transformation \eqref{eq:gaugetran}, the  $k$-dependence sits in the factor of $\Sigma^{\gamma\delta}(N_{0})$ (i.e.~the field two-point function at the horizon crossing of each mode, cf.~Eq's.~\eqref{eq:generic-2pf_a} and \eqref{eq:generic-2pf_b}), and in the length of the string of propagators making up ${\Gamma^{\alpha}}_{\gamma}(N, N_0)$ (cf.~Eq.~\eqref{gammasumsep}). 

The slow-roll gauge transformation, Eq.~\eqref{eq:gaugetran}, relates $\delta\phi^a(N)$ to $\zeta(N)$ at any time between horizon crossing and the end of inflation 
and can be written as, 
%
\be
N_a(N) = -\frac{1}{\sqrt{2 \epsilon_V(N)}} n_a(N) \, ,
\ee
where $n^a$ is the unit vector along the field trajectory. 
The component of the field fluctuation parallel to the slow-roll trajectory, $\delta \phi_{\parallel} = n^a \delta \phi^a$, is then responsible for the curvature fluctuation
\be
\label{eq:zeta}
\zeta = 
N_a \delta \phi^a = 
- \frac{1}{\sqrt{2 \epsilon}} \delta \phi_{\parallel} \, .
\ee
In this way, the curvature power spectrum, as evaluated at any time during inflation, is given by the projection,
\be
P_{\zeta}(N)= \frac{1}{ 2\epsilon_V} n^a\,  \Sigma^{ab} \,   n^b \, .
\ee

Field perturbation $\delta \phi_{\parallel}$ along the trajectory is referred to as `adiabatic', and perturbations along the $N_f-1$ perpendicular directions give rise to
`entropic' or
 `isocurvature' perturbations.
Denoting a generic orthonormal frame of basis vectors in the perpendicular directions by $v^a_j(N)$ for the vector index $a$ and with $j=1, \ldots, N_f-1$, the field perturbations are decomposed as,\footnote{The `kinematic basis' in which the unit vectors are formed from time derivatives of the background trajectory is one popular choice for $v^a_i$ \cite{GrootNibbelink:2001qt}. }
\be
{\delta\phi}^a \equiv \delta \phi_{\parallel}\,  n^{a}+ \delta \phi_{\perp}^j  \,v^a_j 
\, .
\ee
In analogy to $\zeta$, we define the isocurvature ${\cal S}^i$ as,
\be
\label{eq:iso}
{\cal S}^i \equiv \frac{1}{\sqrt{2 \epsilon_V}}  \delta \phi_{\perp}^i \, .
\ee
 The isocurvature correlations (suppressing the momentum-conserving delta function) are then given by,
\be
\label{eq:Piso}
P_{\rm iso}^{ij}(N)= 
\frac{k^3}{(2\pi)^3}\, 
\langle {\cal S}^i {\cal S}^j \rangle
=
\frac{1}{ 2\epsilon_V} 
v^{a}_i \,  \Sigma^{ab}\, v^{b}_j \, .
\ee
We refer to the isocurvature power spectrum (without indices) as,
\be
P_{\rm iso} = \delta_{ij} P^{ij}_{\rm iso} = 
\frac{1}{ 2\epsilon_V} 
v^{a}_i \,  \Sigma^{ab}\, v^{b}_i \, .
\ee
The curvature-isocurvature cross-power spectrum is given by,
\be
\label{eq:cross}
P^i_{\rm cross}(N) =
\frac{k^3}{(2\pi)^3}\, 
\langle \zeta\, {\cal S}^i  \rangle
= -
\frac{k^3}{(2\pi)^3}\, 
\frac{1}{ 2\epsilon_V}  
\langle \delta \phi_{\parallel} \delta \phi^i_{\perp} \rangle
= -
\frac{1}{ 2\epsilon_V}  
n^a \, \Sigma^{ab}\, v^b_i 
\, .
\ee
We define the cross-correlation (without indices) as,
\be
P_{\rm cross} \equiv \sqrt{P_{\rm cross}^i P_{\rm cross}^i} \, .
\ee

\subsection{Superhorizon evolution}
Isocurvature can source superhorizon evolution of $\zeta$. Using the slow-roll evolution of the field perturbations, Eq.~\eqref{eq:JocobiEoM},
it is easy to see that,
\be
\delta \phi_{\parallel}' = 
\left(
n^a \delta \phi^a
\right)'
=
\left(2 \epsilon_V -  n^a \frac{V_{ab}}{V} n^b\right)\delta \phi_{\parallel} - 2\, n^a \frac{V_{ab}}{V} v^b_j \; \delta \phi_{\perp}^j \, .
\ee
To compute the evolution of the entropic perturbations, $(\delta \phi_{\perp}^i)' = (v^a_i\,\delta \phi^a)'$, we  need to specify how the basis $v^a_i$ evolves during inflation. To preserve the unit norm, $v^a_i (v^a_i)' = 0$ for each fixed $i$.  We furthermore take $v^a_i (v^a_j)' = 0$ so that 
the entropic basis vectors do not rotate among themselves. To preserve orthogonality with $n^a$, each entropic basis vector satisfies,
$
(n^a\, v^a_i)' = (n^a)' v^a_i + n^a (v^a_i)' =  0$, 
so that
\be
(v^a_i)' = - (n^b)' v^b_i\; n^a = n^b \frac{V_{bc}}{V} v^c_i\, n^a \, .
\ee
With this choice, the entropic fluctuations evolve as,
\be
(\delta \phi_{\perp}^i)' = - v^a_j\, \frac{V_{ab}}{V} \, v^b_k\; \delta \phi_{\perp}^k \, .
\ee

The corresponding evolution equations for the curvature, $\zeta$, and isocurvature perturbations, ${\cal S}^i$, can be found by first noting that,
\be
\frac{d}{dN} \left( \frac{1}{\sqrt{2\epsilon_V}}\right) =  \frac{1}{\sqrt{2\epsilon_V}} \left( 
n^a \frac{V_{ab}}{V} n^b - 2 \epsilon_V
\right) \, ,
\ee 
so that,
\bea
\zeta' &=&  2 \left(n^a \frac{V_{ab}}{V} v^b_i\right) {\cal S}^i \, , 
\label{eq:zetaprime}
\\
({\cal S}^i)' &=& (n^a \frac{V_{ab}}{V} n^b - 2 \epsilon_V) {\cal S}^i - v^a_i\, \frac{V_{ab}}{V}\, v^b_k\; {\cal S}^k \, . 
\label{eq:Sprime}
\eea


The super-horizon evolution of the curvature and isocurvature correlations during slow-roll inflation are then governed by,
\bea
P_{\zeta}' &=&  4 n^a\, \frac{V_{ab}}{V}\, v^b_i \; P_{\rm cross}^i \, , \label{eq:Pzetaprime}  \\
(P_{\rm iso}^{ij})' &=& 2\left(n^a  \frac{V_{ab}}{V}  n^b - 2 \epsilon_V \right) P_{\rm iso}^{ij} - v_i^a  \frac{V_{ab}}{V}  v^b_k\, P_{\rm iso}^{kj}
 - v_j^a  \frac{V_{ab}}{V}  v^b_k\, P_{\rm iso}^{ik} \, , \label{eq:Pisoprime} \\
(P_{\rm cross}^i)' &=& \left(n^a  \frac{V_{ab}}{V}  n^b - 2 \epsilon_V \right) P_{\rm cross}^i - v^a_i\,  \frac{V_{ab}}{V} \, v^b_j\, P_{\rm cross}^j 
+ 2 n^a\,  \frac{V_{ab}}{V} \, v^b_j\; P_{\rm iso}^{ij}  \label{eq:Pcrossprime} \, .
\eea
%
%
In  \S\ref{sec:isocurvature}, we will find these expressions very intuitive in interpreting the numerical results of  explicit manyfield models.
 
\subsection{The tilt and the running}
\label{sec:ns}
It is conventional  to describe  $P_{\zeta}(k)$ as  a simple power law,
\be
P_{\zeta}(k) = A_s \left( 
\frac{k}{k_{\star}}
\right)^{n_s-1}
\, ,
\label{eq:powerlaw}
\ee
where Planck observations give a good fit for $\ln ({10^{10}} A_s )= 0.563 \pm 0.034$ \footnote{Our definition of dimensionless power spectrum differs from the one used in Ref.~\cite{1502.02114} by a factor of $4\pi$.} and $n_s = 0.9645 \pm 0.0049$ at the pivot scale $k_{\star} = 0.05 {\rm Mpc ^{-1}}$ \cite{1502.02114}. 
In many models of inflation, the power spectra are well approximated by simple power laws, and observations lead to direct constraints on the model  parameters.  
The randomly generated models studied in this paper however are not required 
%
to yield simple power law power spectra, 
but we find it is still convenient to define the scalar spectral index by,
\be\label{eq:tiltdef}
n_{s}=1+ \frac{\d \ln P_{\zeta}}{\d \ln k}(k_{\star}) \, . 
\ee
We note that constraining  heavily featured power spectra is more complicated than comparing the result of Eq.~\eqref{eq:tiltdef} with observational constraints derived assuming Eq.~\eqref{eq:powerlaw}.
In the numerical examples constructed in this paper, we compute 
\eqref{eq:tiltdef} by linearly fitting the power spectrum over a 10 efold range around the pivot scale, which we assume left the horizon 55 efolds before the end of inflation.  

Much intuition for the distribution of values of $n_s$ in models of multifield inflation can be gained from analytic expression for the spectral index obtained from the transport formalism. 
Following Ref.~\cite{Dias:2011xy} we denote,  
\be
n^{ab}=\frac{\d \Sigma^{ab}}{\d \ln k} \, ,
\ee
such that the spectral index is given by,\be\label{eq:nsofnab}
n_{s}-1=\frac{1}{P_{\zeta}}N_{a}N_{b}n^{ab} \, .
\ee
On superhorizon scales, the equation of motion for $n^{ab}$ is the same as the one for $\Sigma^{ab}$,\footnote{On subhorizon scales the transport equations for $\Sigma_{ab}$ and $n_{ab}$ differ \cite{1502.03125}. } leading to,
\be\label{eq:nsol}
n^{ab}(N)={\Gamma^{a}}_{c}{\Gamma^{b}}_{d} {n^{cd}}_* \, ,
\ee
where ${n^{cd}}_* \equiv n^{cd}(N_{0})$ is the value at horizon exit.
Reference \cite{Dias:2011xy} showed that,
\begin{equation}
\delta\Sigma^{ab}|_{*}=\left (-\epsilon\Sigma^{ab} - \frac{d \Sigma^{ab}}{dN}\right )_{*}\delta\ln k \, ,
\end{equation}
from which it  follows that,
\begin{equation}\label{eq:nHcrossing}
n^{ab}_{*}=-2\left(\epsilon\Sigma^{ab}+{u^a}_c \Sigma^{cb}\right )^{*} \, .
\end{equation}
For simplicity of discussion, we now take $\Sigma^{ab}_*= H^2/2 \	 \delta^{ab}$ and 
substitute  Eqs.~\eqref{eq:nsol}, \eqref{eq:nHcrossing} into Eq.~\eqref{eq:nsofnab} to obtain the full expression for the spectral index:
\begin{equation}\label{eq:SRns}
n_{s}-1=-2(\epsilon_{*}+ e_{a} e_{b}u^{ab}_{*}) \, ,
\end{equation}
where we have defined the unit vector,
  \be
  e_{a} = \frac{N_{b}\Gamma^{b}_{a}}{\| N_{c}\Gamma^{c}_{d}\|} \, .
  \label{eq:e_a}
  \ee
This expression is at the heart of the discussion of \S\ref{sec:lesson7}. 

The ansatz \eqref{eq:powerlaw} for the primordial power spectrum can be generalised by taking $n_s = n_s(k)$. The first derivative of $n_s$ is called the running of the spectral index, $\alpha_s$. In the transport formalism, the running can be expressed as \cite{1203.3792},
\begin{equation}\label{eq:alpha}
\alpha_{s}\equiv \frac{\d n_s}{\d \ln k}=\frac{1}{P_{\zeta}} N_a N_b {\Gamma^a}_c{\Gamma^{b}}_d \alpha^{cd}_*-(n_{s}-1)^{2} \, ,
\end{equation}
where, 
\begin{align}
\alpha^{ab}_* \equiv \frac{\d n^{ab}}{\d \ln k}\bigg|_{*} = [(2 \epsilon^{2}-\epsilon')\delta^{ab}- u'^{ab}+2\epsilon u^{ab}]_*H_*^2-2[u^{a}_{~c} n^{cb}]_*  \nonumber \, ,
\end{align}
and primes indicate differentiation with respect to efold time $N$. In terms of the unit vector $e$, the running can be expressed as 
\begin{equation}\label{eq:newalpha}
\alpha_s  = -2 \epsilon' -2 e_{a} e_{b}u'^{ab} +  4e_{a} e_{b}u^{a}_cu^{cb} -4\left(e_{a} e_{b}u^{ab}\right)^2.
\end{equation}

\subsection{Testing the slow-roll analysis}
\label{sec:methodcomparison}

As discussed above, a complete tree-level treatment of the curvature perturbation can be achieved by computing the propagator from deep inside the horizon through superhorizon scales. This approach 
allows for a  broad range of dynamical behaviours, and provides precise results at a numerical cost which, unfortunately, scales poorly with dimensionality. Conversely, imposing the slow-roll conditions and assuming a negligible turn-rate  around the time of horizon exit, the curvature perturbation can be computed entirely using background quantities.
This corresponds to a drastic improvement in numerical performance, but it is not clear from the outset if the random models of  DBM inflation 
satisfy these conditions: while the fields are typically evolving very slowly during most of the inflationary evolution, turns in field space can occur at any point during inflation.   


To test if the slow-roll method can be used in our case, we compute perturbations with both methods and compare the results. Given the stochastic nature of our background, it is not possible to compare slow-roll and non-slow-roll trajectories for each realisation individually. 
It is however possible to compare ensembles of results for fixed values of the hyperparameters.  We find that, at the level of ensembles, both methods give comparable results. 

\begin{figure}[t]
 \centering
 \includegraphics[width=0.6\textwidth]{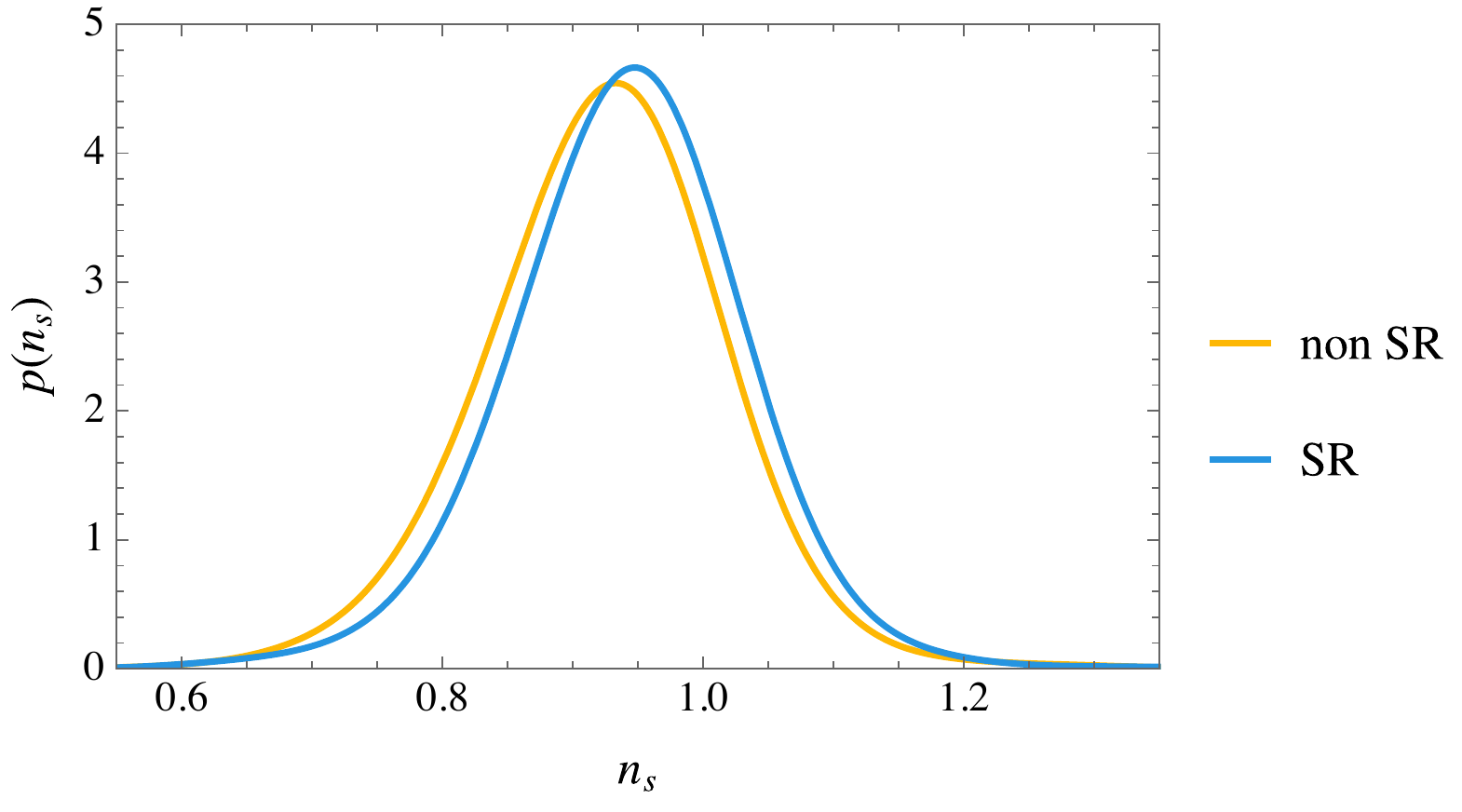}
  \caption{Probability density function of the spectral index estimated by linear fit of the power spectrum for $N_f=10$, $ \Lambda_{\rm h}=0.4$, and initial $\epsilon_{V0} = 10^{-11}$ and $\eta_{V0} = -10^{-4}$. The yellow line shows the result for the full treatment, whereas the blue line represents the result of the slow-roll method.}
   \label{fig:comparison}
\end{figure}

As an example, we present in Fig.~\ref{fig:comparison} the probability distribution of the spectral index computed with both methods, for ensembles of $1000$ realisations with 20 fields. The distributions only differ by a few per cent off-set. 
Since
our aim is to identify general lessons from manyfield behaviour rather than a detailed quantitative description and stringent constraints on hyperparameters,  we regard these results as comparable and in good agreement. 
Moreover, we have verified that 
manyfield effects such as 
the levels of 
superhorizon evolution of observables are consistent between both methods. 

\section{Lesson I: Manyfield inflation is not single-field inflation}
\label{sec:casestudy}
\label{sec:Example}
In this section, we focus on one of the several tens of thousands of inflationary models  
studied
in the subsequent sections of this paper. By considering this example in detail, we can highlight several of the properties that are characteristic of much larger classes of inflationary models. Moreover,  by considering an in many ways typical example of a randomly generated manyfield model, we are able to rule out the applicability of single-field models as faithful proxies of typical multiple-field systems. 


One immediate aspect of the multifield models that we study is that they typically contain several fields with masses of the order of the Hubble parameter. Such `light' fields cannot be integrated out during inflation, and  commonly contribute to multifield effects that impact observables. In this sense,  manyfield models of inflation are clearly not identical to single-field models. However, it has been shown that particularly engineered single-field models of inflation can capture some aspects of multifield models. For example, a simple model of a single canonically normalised inflaton field subject to the potential,
\be
V(\phi) = V_0 - V_1 \phi - \frac{1}{2} m^2(\phi) \phi^2 \, ,
\ee 
with $m^2(\phi)\sim \sqrt{1-{\rm exp}(-\phi/\Lh)}$ (so that $m^2(\phi) \sim (\phi/\Lh)^{1/2}$ for $\phi\ll \Lh$) 
provides, for certain choices of the parameters,  a rather accurate estimate of the total number of efolds of the more complicated DBM models \cite{1307.3559}. 
Even though it only involves one field, this model was constructed to reproduce the large-$N_f$ evolution of the inflaton mass and therefore should not be expected to agree with DBM models with $N_f = 1$.
This single-field model was further developed in Ref.~\cite{1608.00041}, which in particular argued that $m^2(\phi) \sim \phi^{2/3}$ better captured the properties of the multifield DBM models for sufficiently small field displacements. By assuming that the primordial power spectrum of such a single-field model 
agrees with that of the multifield DBM models, Ref.~\cite{1608.00041} went on to draw strong conclusions about the incompatibility of DBM potentials with CMB data.
 We will return to and test this assumption and its consequences in detail in \S\ref{sec:Planckcompat}. Other work on using random single-field models as proxies for more complicated multifield `landscapes' include \cite{astro-ph/0410281, 1512.02637, 1409.6698, Amin:2017wvc}, and more recently \cite{1612.03960}.   

  Thus, while single-field models cannot capture all aspects of more complicated multiple-field models, they have commonly been used to build intuition 
  for the latter. By studying one particular multifield model in detail, we here assess the importance of multifield effects on the inflationary observables, thereby allowing us to gauge the limitations of the single-field intuition. 

\begin{figure*}
\centering
\begin{minipage}{\textwidth}
\includegraphics[width=0.5\textwidth]{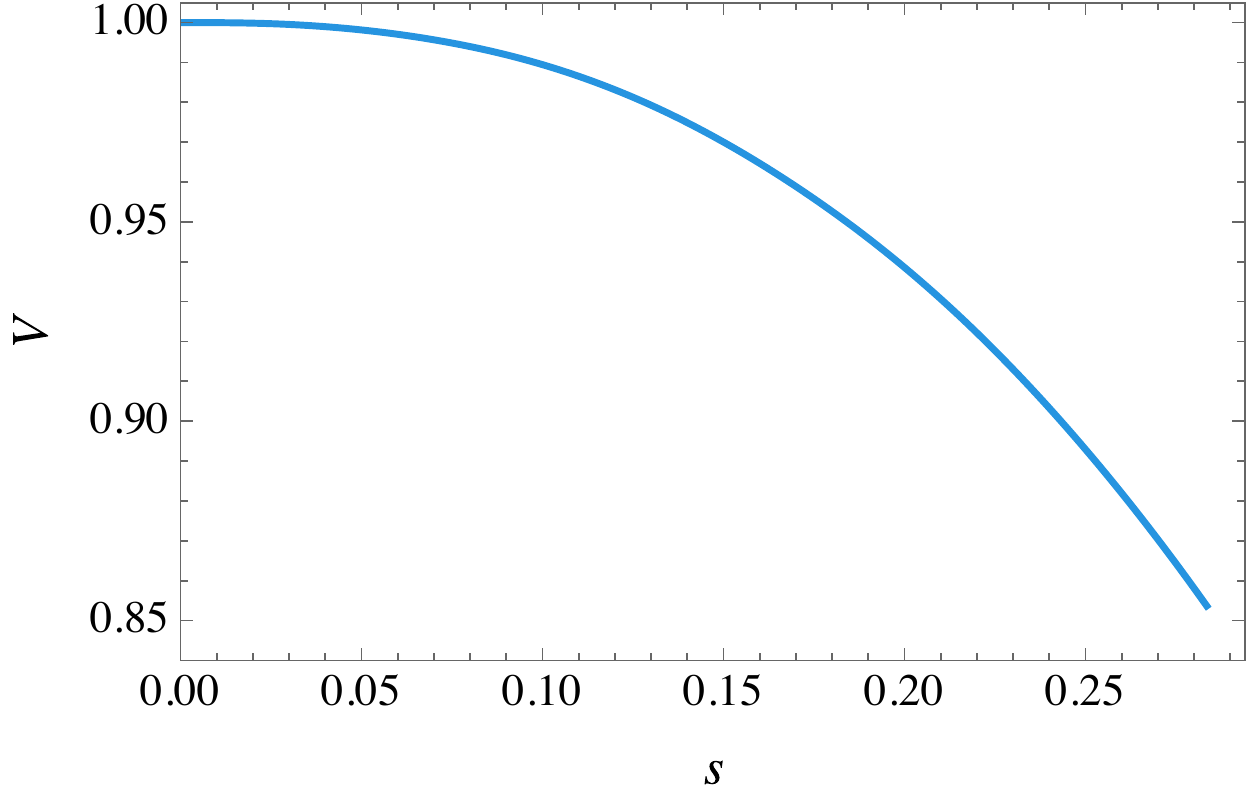}
\includegraphics[width=0.5\textwidth]{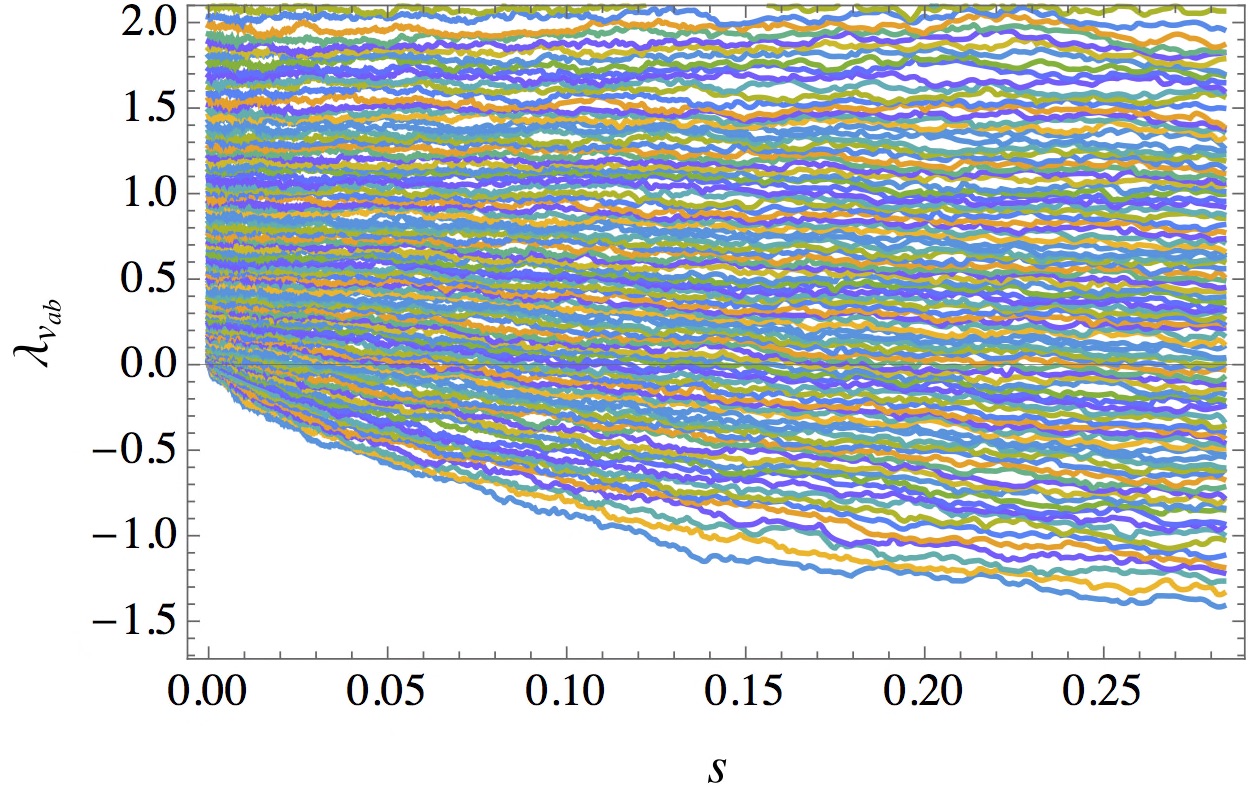}
\caption{The value of the potential (left) and the eigenvalues of $v_{ab}$ as functions of the path length, $s= \Delta \phi/\Lh$.}
 \label{fig:Expl1}
\end{minipage}
\end{figure*}

\subsection{Case study: 100-field inflation}
We consider a randomly generated model of inflation with $N_f=100$ fields and $\Lh = 0.4 \Mpl$, constructed according to the DBM prescription \S\ref{sec:DBMreview}.
The initial conditions at the approximate inflection point were taken to be  $\epsilon_{V0} = 10^{-11}$ and $\eta_{V0} = - 10^{-4}$, with the spectrum of the Hessian matrix given by the `fluctuated' spectrum discussed in \S\ref{sec:regimes}.
These initial conditions were chosen so that randomly generated 
models supporting  at least 60 efolds of inflation are not overly rare (the mean value of the total number of efolds is  71.5, and 49.9\% of the 2,600 examples tested support at least 60 efolds of inflation). This particular example yields a total of $63.2$ efolds of slow-roll inflation with a total field space displacement of $0.28 \Lh = 0.11\Mpl$. It takes 1031 patches to construct this example.

Figure \ref{fig:Expl1} 
shows the evolution of the value of the potential and the eigenvalues of the Hessian matrix as a function of the path length $s= \Delta \phi/\Lh$ along the inflaton trajectory. Two characteristic features are worth highlighting: first, 
despite the random nature of the DBM potentials, 
the evolution of the value of the scalar potential maps out a very smooth approximate saddle-point. The absence of large features in the sampled potential 
is not surprising: the `gradient flow' field evolution of the inflaton seeks out the locally steepest path away from the inflection point, making `bumps', `steps' and other large features in the sampled potential highly unlikely.  

Second, many of the   eigenvalues of the Hessian matrix rather quickly `drop' to negative values, 
thereby erasing the details of the initial, `fluctuated' spectrum. 
%
Multiple fields get tachyonic eigenvalues of the Hessian during inflation, and towards the end of inflation, very nearly half of the eigenvalues are negative. Moreover, Fig.~\ref{fig:Expl1} illustrates the continuity  of the evolution of the Hessian eigenvalues as well as the intrinsic yet regulated `raggedness' that is characteristic of Brownian motions, cf.~our discussion in \S\ref{sec:subtleties}.

\begin{figure*}
\centering
\begin{minipage}{\textwidth}
\includegraphics[width=0.5\textwidth]{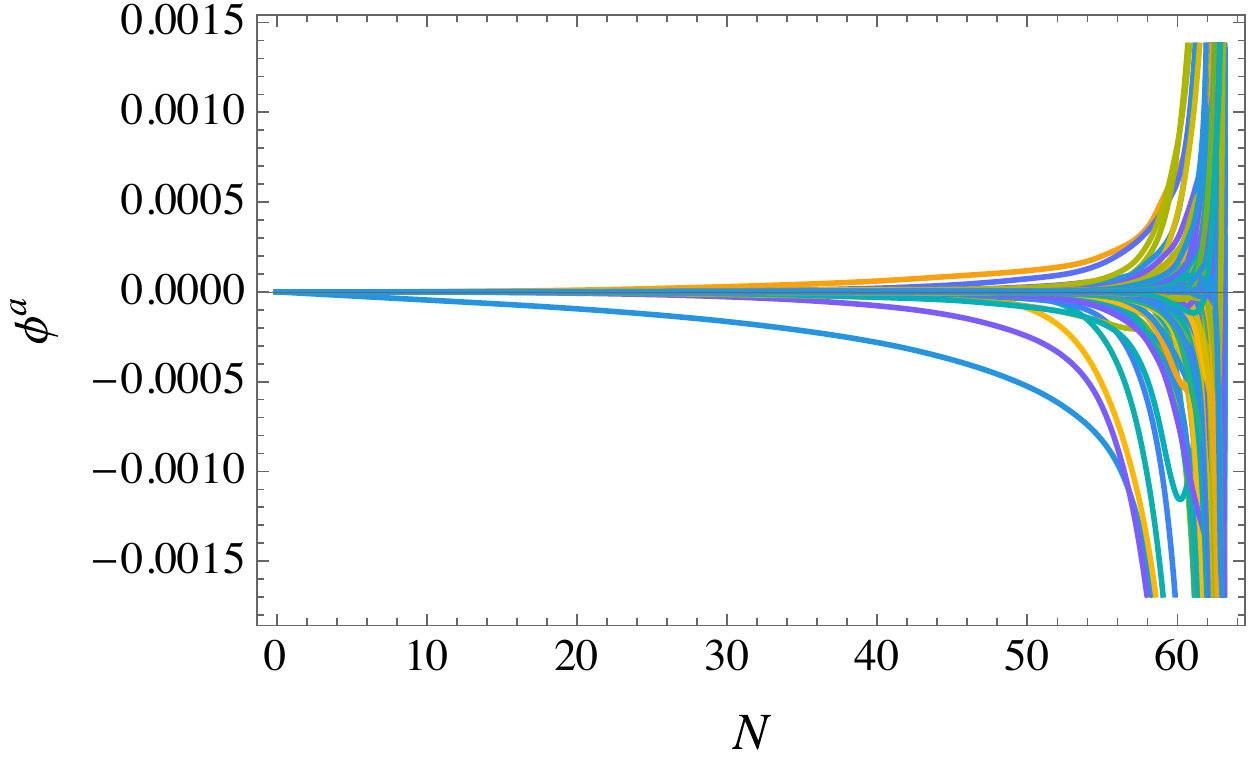}
\includegraphics[width=0.5\textwidth]{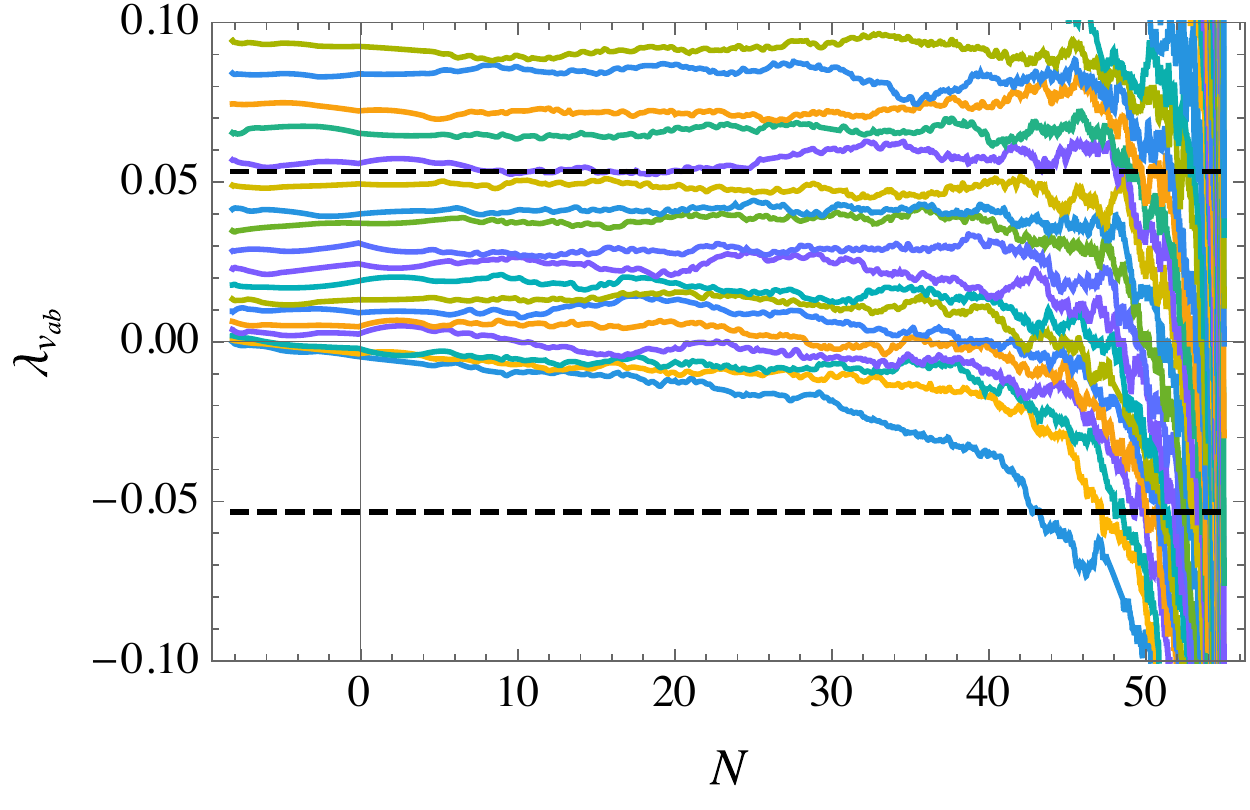}
\caption{The field evolution (left) and the eigenvalues of $v_{ab}$ (right) as functions of the number of efolds. The region between horizontal black lines correspond to modes with $-H^2_{\star}\leq m^2 \leq +H_{\star}^2$. }
 \label{fig:Expl2}
\end{minipage}
\end{figure*}

The field is `rolling' very slowly during inflation, and most the evolution evident in Fig.~\ref{fig:Expl1}  occurs during the last few efolds. 
Fig.~\ref{fig:Expl2} shows the evolution of the components of the vector $ \phi^a$  during inflation (with respect to a fixed basis in which $\dot \phi^a(0) \propto \delta^a_1$). Clearly,  the fields evolve non-trivially in the high-dimensional field space. As $\epsilon$, and hence the speed of the field, grows towards the end of inflation, much of the evolution occurs during the last few efolds. The right panel of Fig.~\ref{fig:Expl2} shows the evolution of the eigenvalues of the Hessian matrix (just as Fig.~\ref{fig:Expl1}), here as a function of the number of efolds. The number of `light' fields with $m^2 < H_{\star}^2$ is not fixed and increases during inflation, but, again, due to the slow initial motion of the field, most of the interesting dynamics occur towards the end of inflation.  An interesting aspect of the eigenvalue evolution depicted in Fig.~\ref{fig:Expl2} is that $|\eta_{\rm V}| = |m^2_{\rm min}|/3H^2$ exceeds unity for $N_e \geq 59.9$, still inflation persists until $N_e=63.2$ when $\epsilon_V =1$. 

With multiple modes becoming tachyonic, one may worry that isocurvature perturbations grow and become dominant by the end of inflation. 
The case however, is just the opposite. At horizon crossing of the pivot scale, the field perturbations are given by Eqs.~\eqref{eq:generic-2pf_a} and \eqref{eq:generic-2pf_b}. 
The adiabatic mode is the linear combination of these modes in the direction aligned with $\dot \phi^a(N_{\star})$ (cf.~Eq.~\eqref{eq:2p}), and all perpendicular entropic modes contribute to the isocurvature. As there are many light fields in this model, the power in the entropic modes exceeds the power in the adiabatic mode at horizon crossing. 

\begin{figure*}
\centering
\includegraphics[width=0.392\textwidth, height=4.08cm]{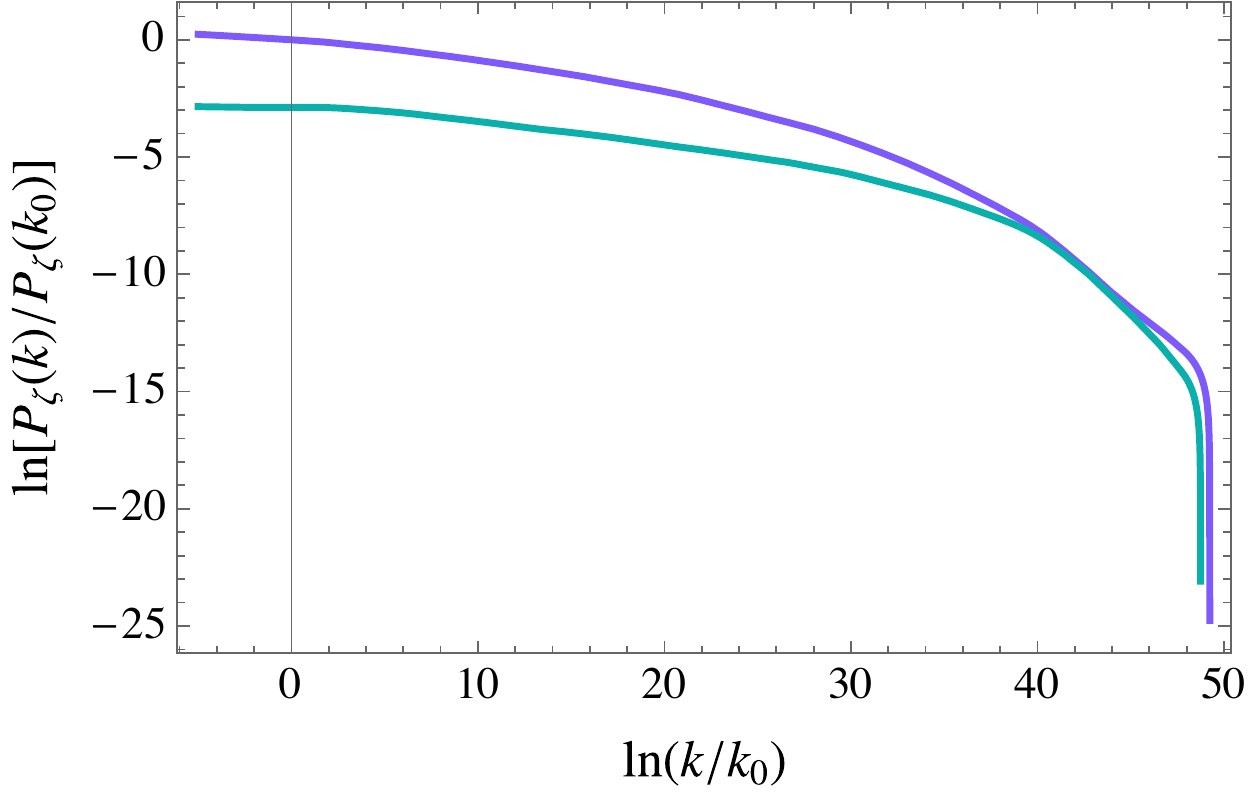}
\includegraphics[width=0.6\textwidth]{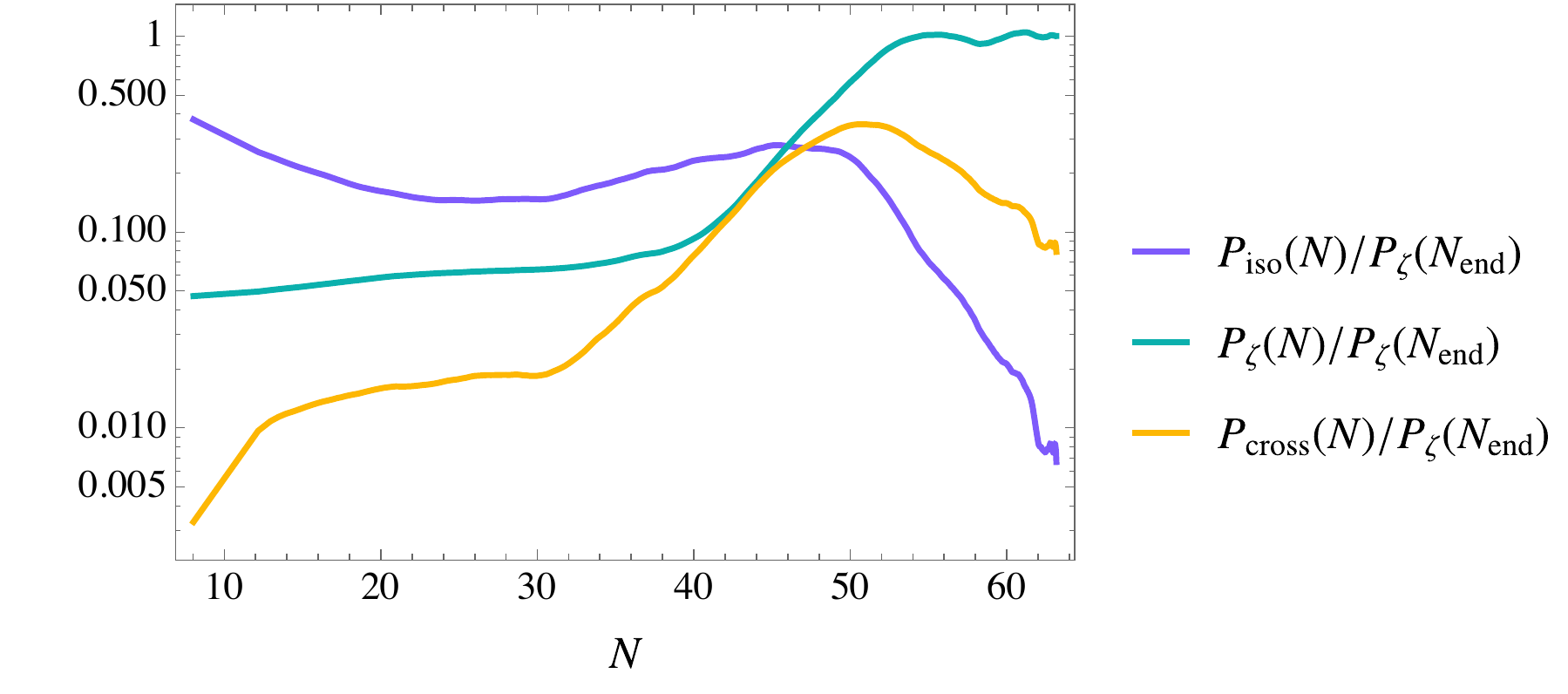}
\caption{
Left: Power spectrum of curvature perturbations 
 as evaluated at horizon exit (cyan) and at the end of inflation (purple). 
 Right: Superhorizon evolution of the power of the pivot-scale modes of curvature, isocurvature and their cross-correlation. 
}
 \label{fig:P(k)Expl}
 \label{fig:P(N)Expl}
\end{figure*}

Figure \ref{fig:P(N)Expl} shows the superhorizon evolution of the power in the curvature and isocurvature perturbations at the pivot scale,
together with  their cross-correlation. In a single-field model, the curvature perturbation freezes out on superhorizon scales and $P_{\zeta}(N)$ is constant. In multiple-field models isocurvature can source superhorizon evolution of the curvature perturbation (cf.~Eq.~\eqref{eq:Pzetaprime}). In slow-roll inflation, this always leads to a net increase in the power of the adiabatic mode: $P_{\zeta}(N_{\rm end})/P_{\zeta}(N_{\star}) \geq 1$.
  In our case-study, the superhorizon evolution of the curvature perturbation is characteristically substantial: $P_{\zeta}(N_{\rm end})/P_{\zeta}(N_{\star}) = 20.3$. This indicates that isocurvature modes, and hence genuine multiple field effects, indeed are critical in manyfield models of inflation.  Moreover, contrary to the na\"ive expectations, Fig.~\ref{fig:P(N)Expl} shows how the power of isocurvature relative to the adiabatic perturbations becomes increasingly suppressed during inflation. We will return to and explain this phenomenon in \S\ref{sec:isocurvature}. 

The power spectrum of the curvature perturbations (as evaluated both at the end of inflation and at horizon crossing) are plotted in the right panel of Fig.~\ref{fig:P(k)Expl}. Superhorizon evolution is typically most important for scales leaving the horizon long before the end of inflation, leading to a slight steepening of the curvature power spectrum. 
Of particular relevance for comparison with CMB experiments is the window of roughly 10 efolds around the pivot scale. Over these scales, the power spectrum is well fitted by a simple power law with spectral index
$n_s= 0.959$ and running $\alpha_s = -0.003$, clearly compatible with current observational bounds from Planck: $n_s= 0.965 \pm 0.005$ and $\alpha_s = - 0.006 \pm 0.007$ \cite{1502.02114}. This provides an explicit example of a non-trivial 100-field model that is compatible with CMB observations. 

Our model has sub-Planckian field displacements, and happens therefore at rather low energy scale with $H^2 \sim10^{-18} \Mpl^2$. The tensor to scalar ratio for this example is extremely low, $r=8 \times 10^{-12}$, which is a common feature across all of our ensembles. 
The `Lyth bound' \cite{Lyth:1996im}, which relates the field displacement during inflation to the tensor-to-scalar ratio in single-field models, states that,
\begin{equation}\label{eq:lythold}
r=16\epsilon_V<8\left(\frac{1}{N_{\rm exit}}\right)^{2}\left(\frac{\Delta\phi}{M_{\rm{Pl}}}\right)^{2} \, ,
\end{equation}
if $\epsilon$ is constant or monotonically increasing. For our model this bound is $r < 3 \times 10^{-5}$, which is clearly far from being saturated. There are two reasons for this: the first is related to the evolution of $\epsilon$, which remains small for most of the trajectory, only growing towards the end of inflation. The second reason is related to the superhorizon evolution of $\zeta$ --- as tensor modes are insensitive to isocurvature, superhorizon evolution necessarily decreases the value of $r$ compared to a single-field estimate. 

To summarise, this analysis of a single realisation highlights the importance of multifield effects in determining the phenomenology of our model. It also demonstrates how complex dynamics can result in Planck compatible inflation. We now turn to the study of ensembles of inflationary realisations and the resulting probability distributions for observables.

\begin{figure}[t]
\centering
\includegraphics[width=0.495\textwidth]{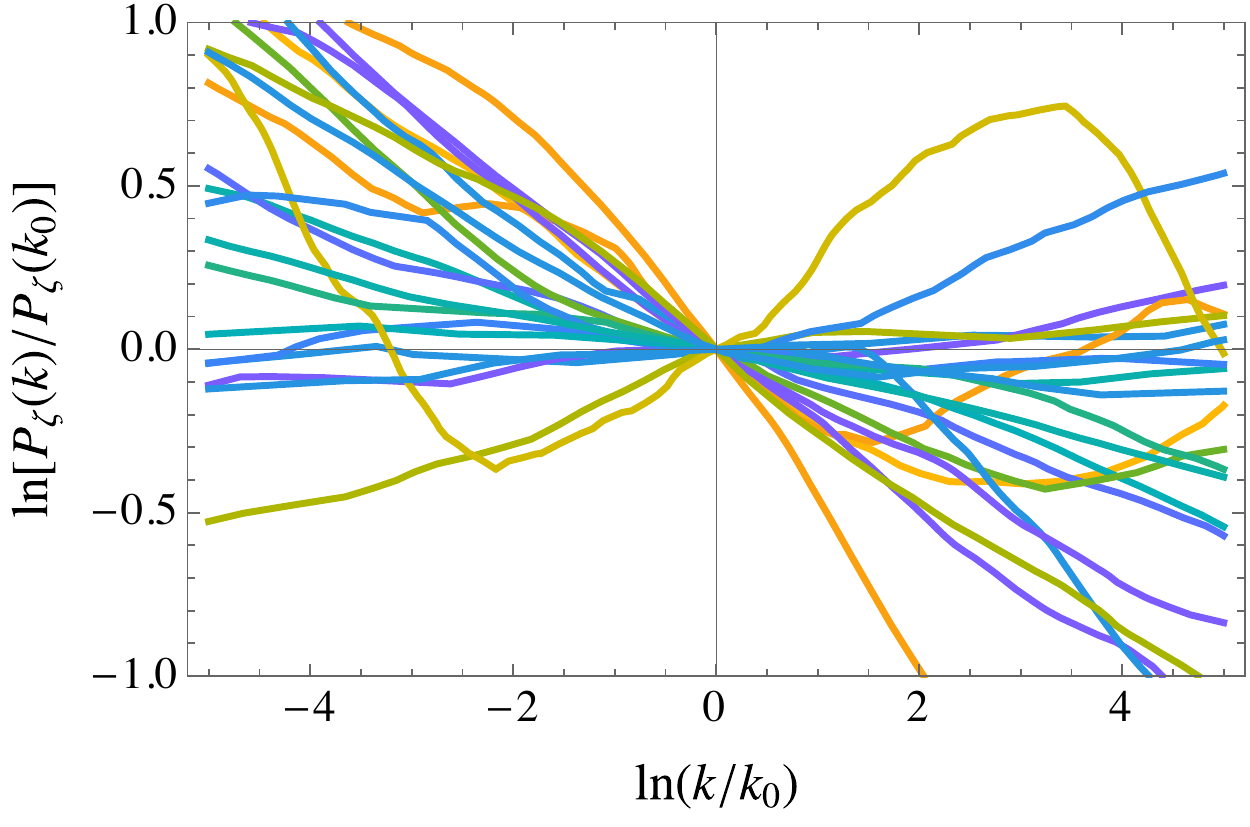}
\includegraphics[width=0.495\textwidth]{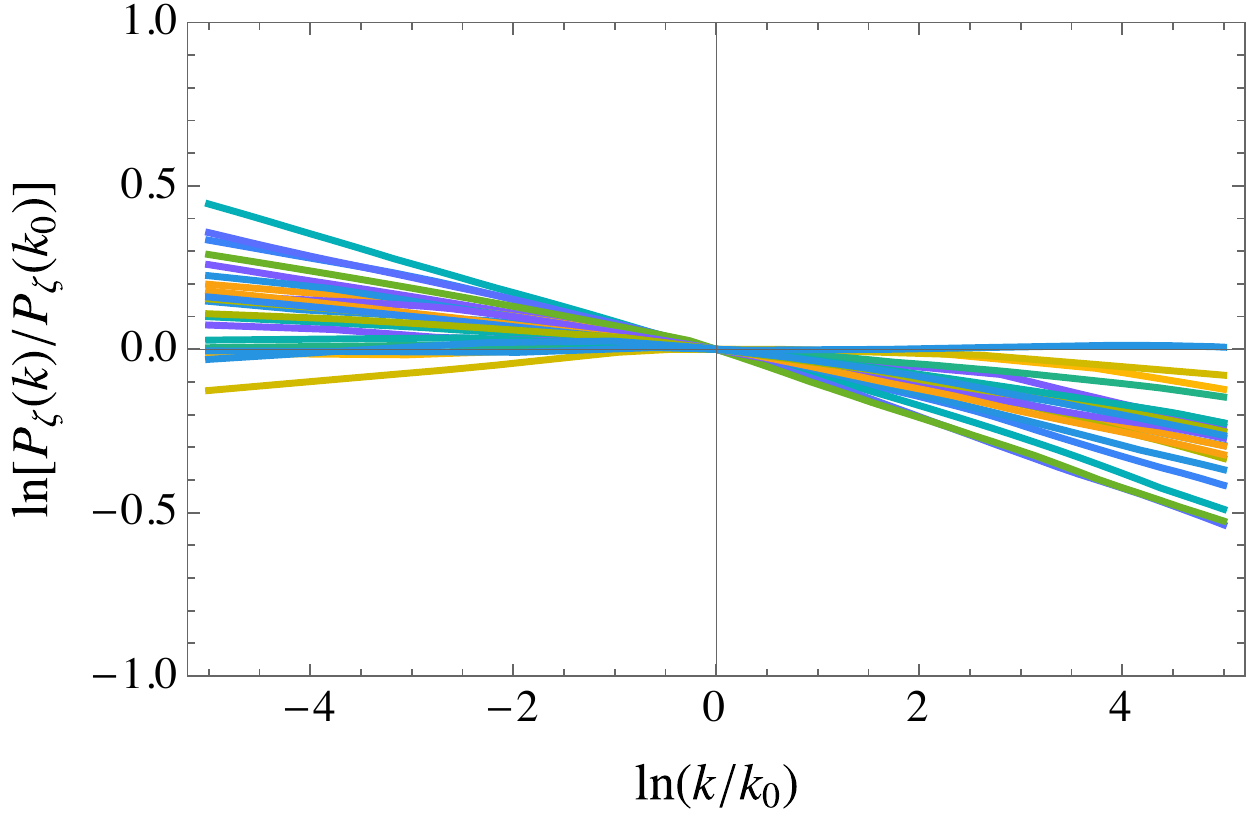}
  \caption{Example power spectra for the scales leaving the horizon between 50 and 60 efolds before the end of inflation for $N_f=2$ (left) and $N_f=100$ (right). Here $ \Lambda_{\rm h}=0.4 M_{\rm Pl}$, $\epsilon_{V0}=10^{-11}$ and $\eta_{V0}=-10^{-4}$. }
   \label{fig:Pzetaofkplots}
\end{figure}

\section{Lesson II:  The larger the number of fields, the simpler and sharper the predictions}
\label{sec:sharperpredswithN}
\label{sec:lesson2}

In this section we study  the statistical distributions 
of the primordial power spectra across our ensemble of  inflationary models. 
We first note  that the power spectra generated in the DBM potentials are not always simple power laws, requiring more than the amplitude and the spectral index to be described. Yet, we find that as the number of fields increases, the power spectra become simpler and more predictive. 


As mentioned, current  observations of the CMB constrain the primordial power spectrum over a roughly 
$10$ efold window which, depending on the details of reheating, left the horizon approximately 50--60 efolds before the end of inflation.\footnote{We do not explicitly model the reheating phase and assume for the rest of this paper, for concreteness, that the pivot scale left the horizon 55 efolds before the end of inflation.}
The two plots of Fig.~\ref{fig:Pzetaofkplots} show the primordial power spectra in this window for 20 randomly chosen realisations with, respectively, $N_f=2$ and $N_f=100$. 
When the number of fields is small, the power spectra are typically highly featured and scale dependent. 
%
When the number of fields increases, the power spectra become 
 less featured. 
For sufficiently many fields (in practice $N_f\gtrsim10$), 
%
the 
power spectra become  well approximated by the simple power law over the 10 efold window of scales observable through the CMB, cf.~Ref.~\cite{1604.05970}. 

\begin{figure}[t]
\centering
\includegraphics[width=0.48\textwidth]{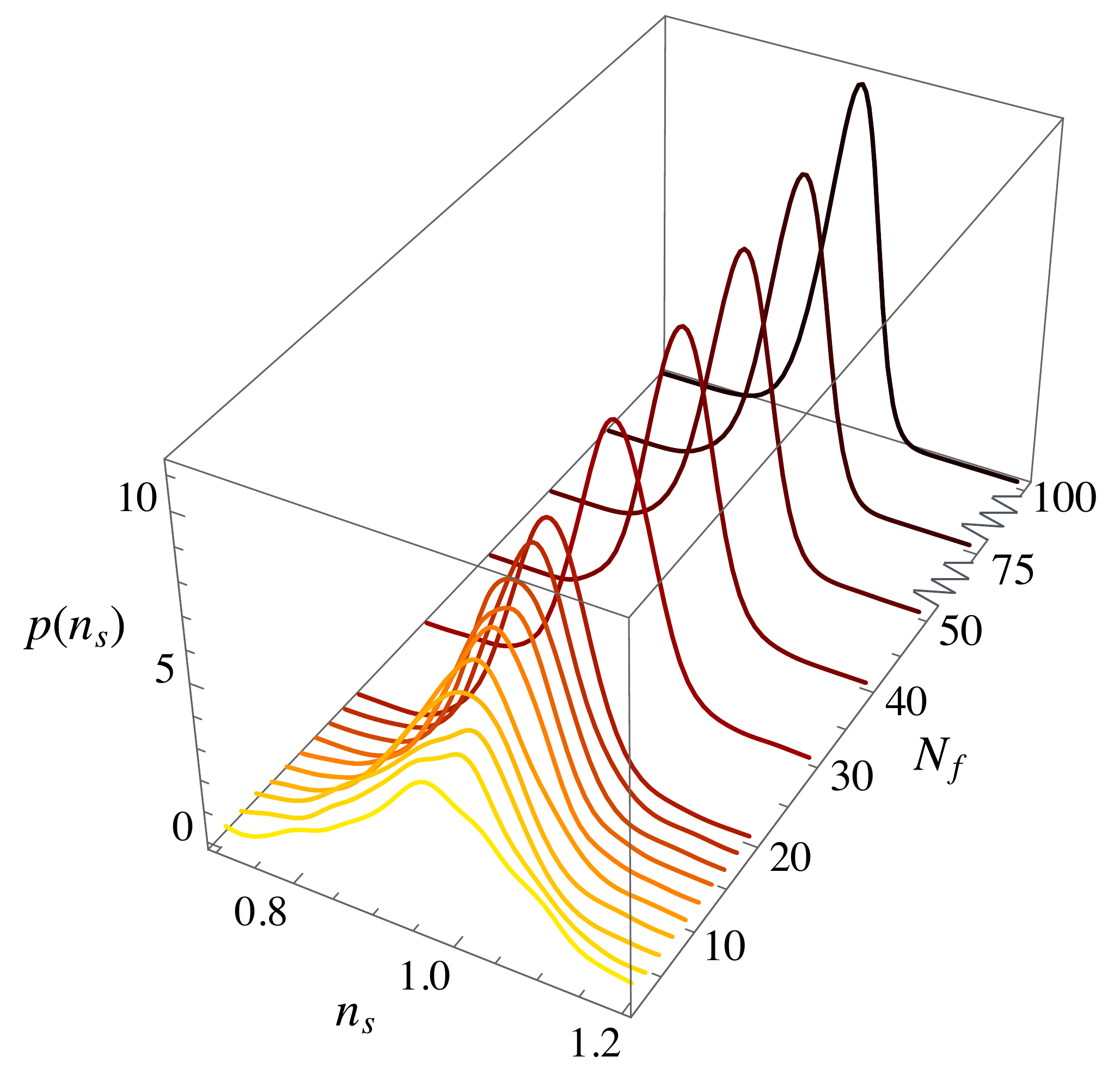}
\includegraphics[width=0.48\textwidth]{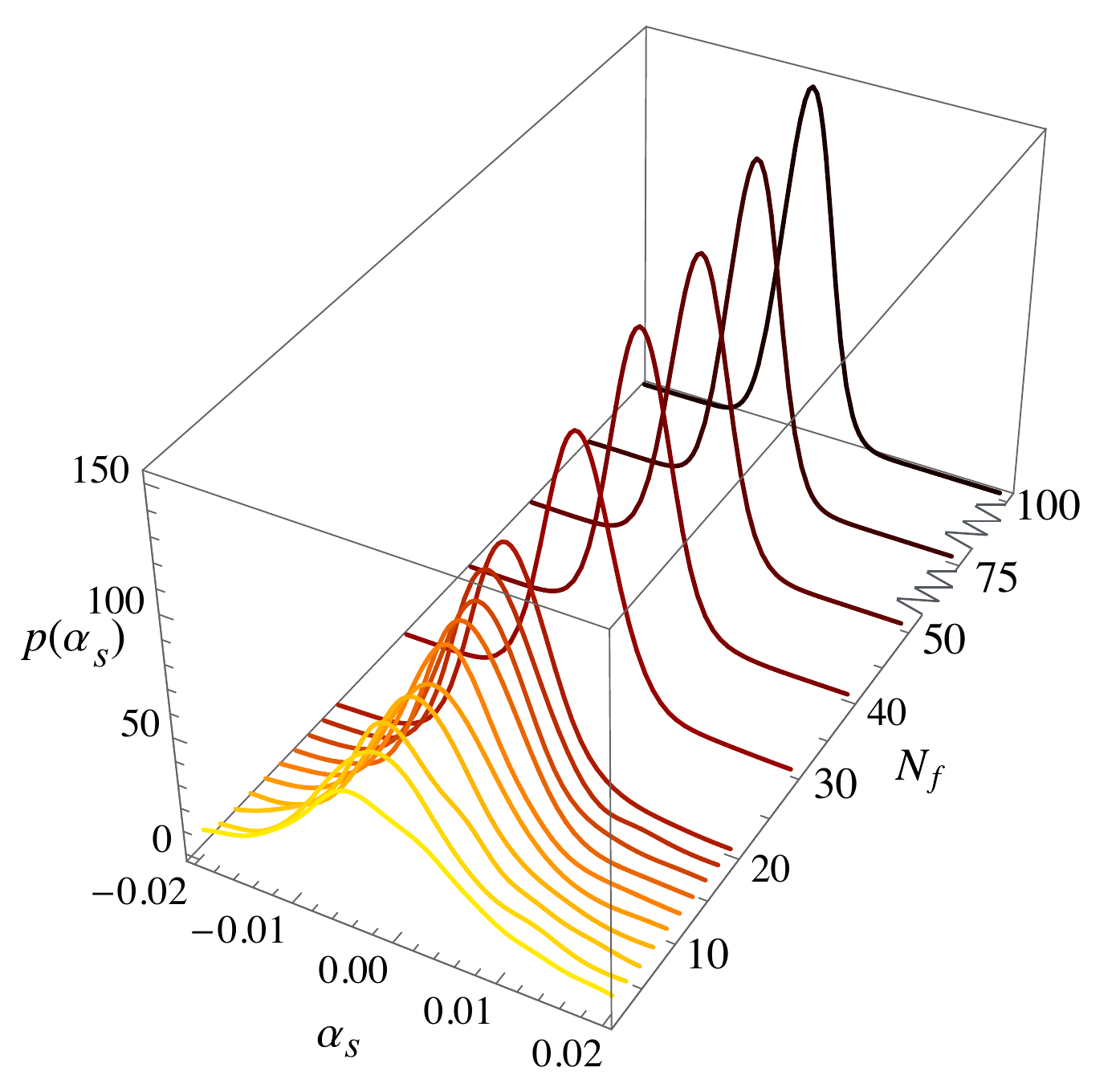}
  \caption{Empirical probability distributions for $n_s$ and $\alpha_s$ inferred from simulations of 1000 DBM models for each value of $N_f$. Here $ \Lambda_{\rm h}=0.4 M_{\rm Pl}$, $\epsilon_{V0}=10^{-11}$ and $\eta_{V0}=-10^{-4}$. }
   \label{fig:P(nsalpha)}
\end{figure}

A striking aspect of Fig.~\ref{fig:Pzetaofkplots} is that the bundles of generated power spectra not only become less featured at large $N_f$, but also more focussed. This decrease in the variation of the power spectra between different random realisations is further highlighted by Fig.~\ref{fig:P(nsalpha)}, which shows the empirical probability distributions of $n_s$ and the running, $\alpha_s = \d n_s/\d\ln k$, for $N_f$ from 2 to 100 computed from a numerical linear fit of  $\ln\left[ P_{\zeta}(k)\right]$  over the 10 efold range of Fig.~\ref{fig:Pzetaofkplots}.
At small $N_f$, the spectral index and 
the running are widely distributed, indicative of the complicated power spectra typical in random few-field models. 
As $N_f$  increases, the probability distributions  tend to rather sharp Gaussian distributions.
We  explain in \S\ref{sec:eigrep}  how this sharpening of the predictions can be understood as direct consequence of eigenvalue repulsion at large $N_{f}$.

\begin{figure}[t]
\centering
\includegraphics[width=0.56\textwidth]{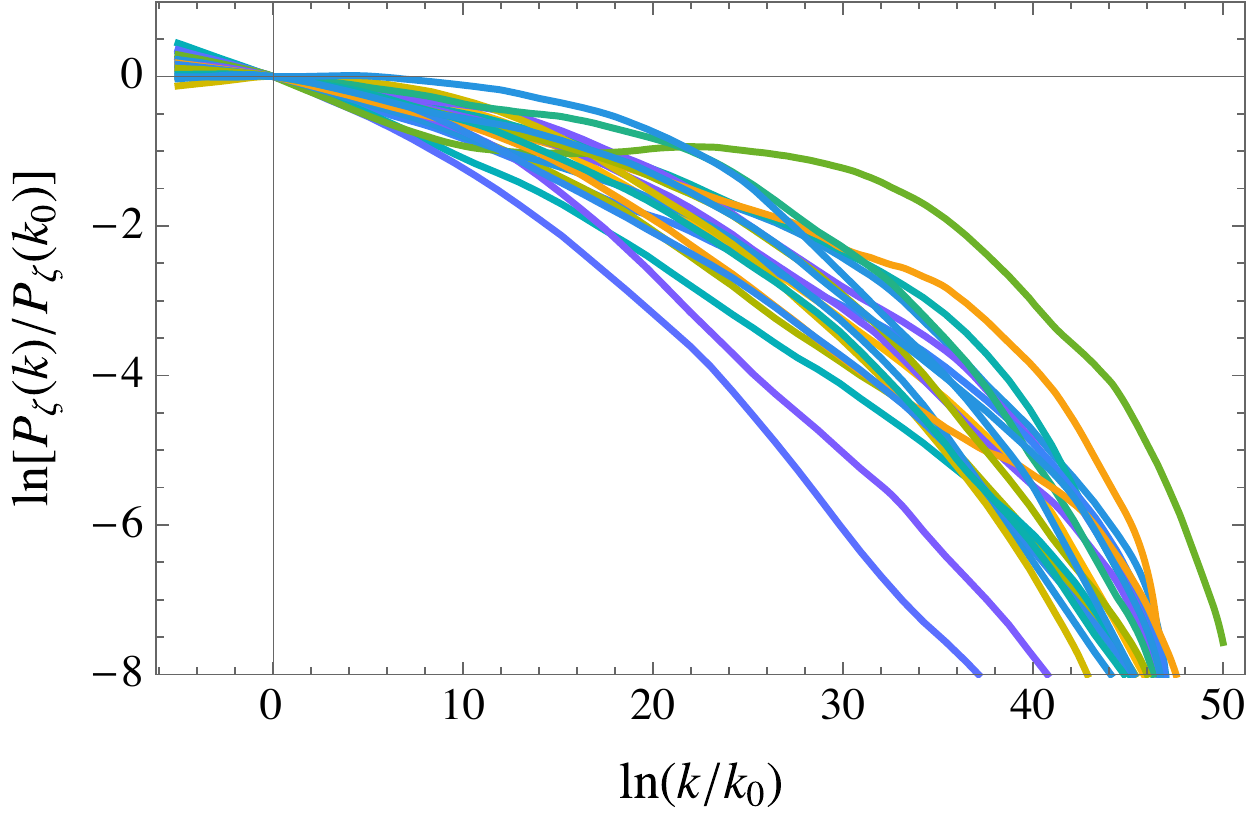}
\caption{The same $N_f=100$ examples shown on the lefthand plot of Fig.~\ref{fig:Pzetaofkplots}, but now showing the full range of scales exiting the horizon in the last 60 efolds of inflation. 
}
   \label{d100FullP_zeta}
\end{figure}

Even the very smooth power spectra generated  in the large $N_f$ regime may fail to be described by a simple power-law when studied over a larger range of scales. 
%
In Fig.~\ref{d100FullP_zeta} we show the same $N_f=100$ examples as in the right hand plot of Fig.~\ref{fig:Pzetaofkplots}, except that now we present the full range of scales exiting the horizon in the last 60 efolds of inflation. 
The power spectra show a clear downward trend indicative of a small negative running, but additional features on small scales are also common.

We note in closing that, even though in the DBM potentials sharper predictions can be understood through eigenvalue repulsion, such manifestation of universality at large-$N_f$ can arise in other situations. In Ref.~\cite{1312.4035, Price:2015qqb} a model of N-flation \cite{astro-ph/9804177, hep-th/0507205}, involving multiple free, non-interacting fields resulted in qualitatively similar behaviour. Despite the fact that these two examples are radically different, both models give rise to sharper predictions at large $N_f$. This suggests that there might be a more universal manifestation of the central limit theorem playing a role, and hence predictions becoming sharper at large $N_f$ may be a ubiquitous characteristic of manyfield inflation.

\section{Lesson III: Planck compatibility is not rare, but future experiments may rule out this class of models}
\label{sec:Planckcompatibility}
\label{sec:Planckcompat}
\label{sec:lesson3}

In this section we quantitatively study  the statistical distributions for the spectral index and the running inferred from our ensemble of DBM models. We show that a substantial fraction of the models are compatible with current observational constraints, but future experiments 
will provide stringent test and 
may even rule out this entire class of models.

As discussed in \S\ref{sec:lesson2}, randomly generated DBM models with only a few fields give rise to highly featured power spectra. 
In this case, the parametrisation of the primordial power spectrum  as a simple power law is not meaningful, and 
comparison with observations requires 
a dedicated analysis for each model, which falls beyond the scope of this paper. 
We note however that observations do not tend to favour highly scale-dependent power spectra, and we expect the vast majority of the highly featured models to be ruled out by current observations.

{\bf The spectral index:} 
For larger $N_f$, the power spectra are well-described by a power-law power spectrum. 
\begin{figure}[t]
\centering
\includegraphics[width=0.49\textwidth, height=4.8cm]{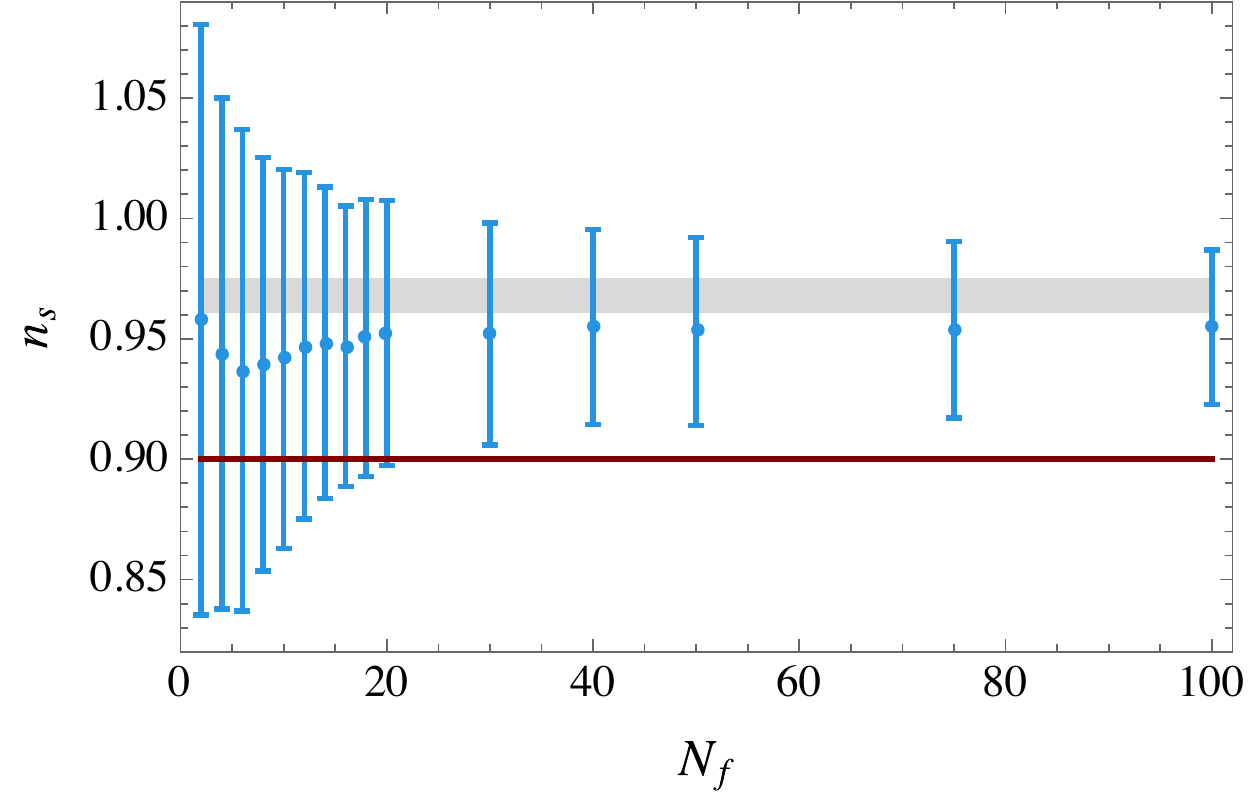}
\includegraphics[width=0.5\textwidth]{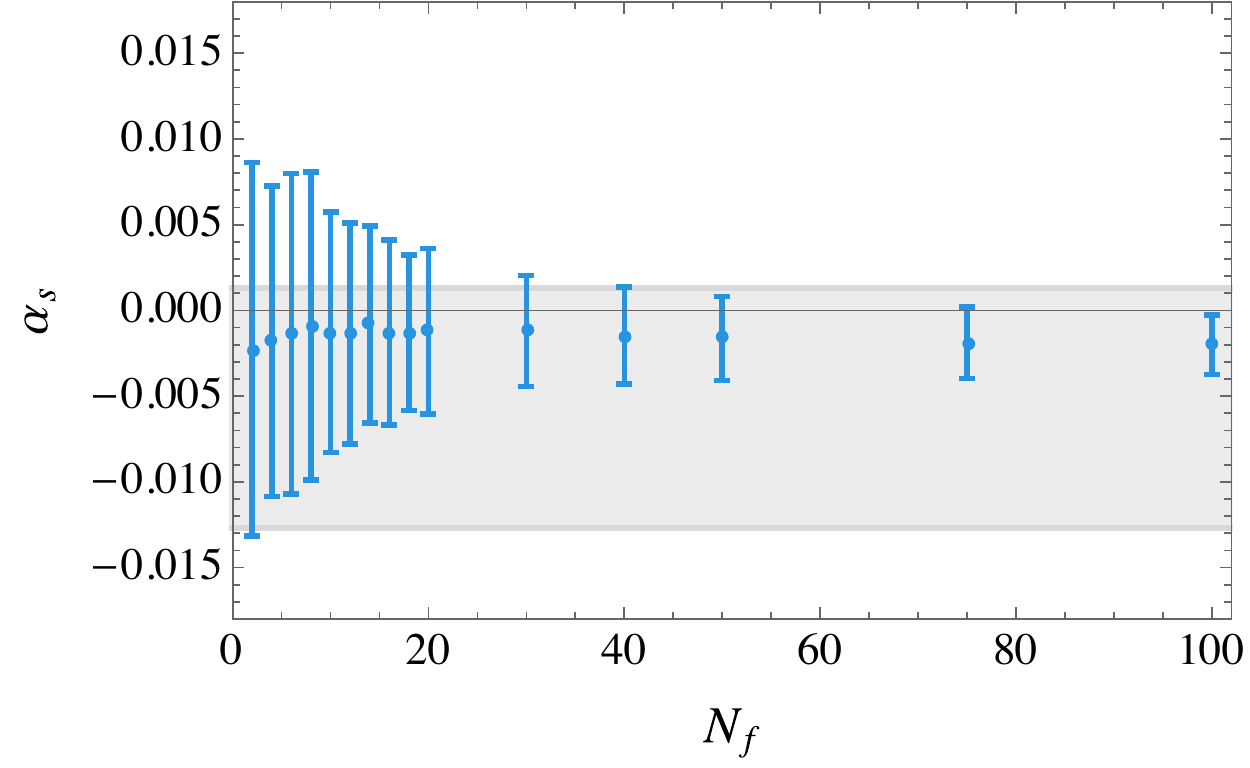}
\caption{
Left: Distribution of the spectral index. Right: Distribution of the running. 
The grey band indicates the current Planck 68\% confidence region \cite{1502.02114}; the red line indicates the single-field prediction of \cite{1608.00041}. 
 Here  $ \Lambda_{\rm h}=0.4 M_{\rm Pl}$, $\epsilon_{V0}=10^{-11}$ and $\eta_{V0}=-10^{-4}$.
%
}
   \label{ns_Nf_contours}
\end{figure}
Many of these ensembles of DBM models are compatible  
with current observational constraints from the Planck experiment.  
Figure~\ref{ns_Nf_contours} shows the
 distributions of spectral indices computed from   
 DBM realisations with $N_f$ ranging from 2 to 100. The distributions of 
 values of $n_s$ in DBM models tend to be broader than the observational constraints, and are typically not far off-set from the central value favoured by observations.\footnote{It should be noted that had we not seen any realisations exhibiting Planck compatibility, this would not necessarily imply that the model is incompatible with observation. One can imagine a situation where the variance of the model is vastly greater than the uncertainty in cosmological parameters given by the observed data. The question then would be if the observed data is typical in the model's distribution. Even if it was, it might be difficult to achieve sufficiently dense sampling to find a realisation compatible with observation. Only if the observed data were an outlier of the model could we regard the model as under pressure from observation.}

  Figure \ref{ns_Nf_contours} also shows the  stark contrast between the single-field model assumed to describe DBM potentials in Ref.~\cite{1608.00041}, and the actual distributions. This suggest that the very strong conclusions of  Ref.~\cite{1608.00041} should be regarded with caution.\footnote{We note however that the results of Fig.~\ref{ns_Nf_contours} are consistent with Ref.~\cite{1604.05970}, which also contained examples of DBM models with Planck-compatible power spectra. 
  }

{\bf The running:} 
Small deviations form a perfect power-law power spectrum are captured by the running of the spectral index, cf.~Eq.~\eqref{eq:alpha}. We compute $\alpha_s$ for each model by making a quadratic fit of $\ln[P_{\zeta}(k)]$ over the same 10 efold window as used above. 

This leads to  a striking result:
 not only do power spectra become well approximated by power laws for large number of fields, but they do so in such a dramatic way that the running is highly predicitve. 
Figure \ref{ns_Nf_contours}, to the right,  shows the 
ensemble average and standard deviation of the 
running as a function of the number of fields. 
For small $N_f$, the random DBM potentials provide 
highly featured power spectra and no 
reliable  predictions for the running. 
However, for sufficiently large $N_f$, the DBM potentials
are well-described by an approximate power-law and 
 make a sharp prediction pointing towards a preference for small negative running. 
In contrast to the case of $n_s$, the theoretical predictions for the running are much tighter than the current observational constraints. 
\begin{figure}[t]
\centering
\centering
\includegraphics[width=0.6\textwidth]{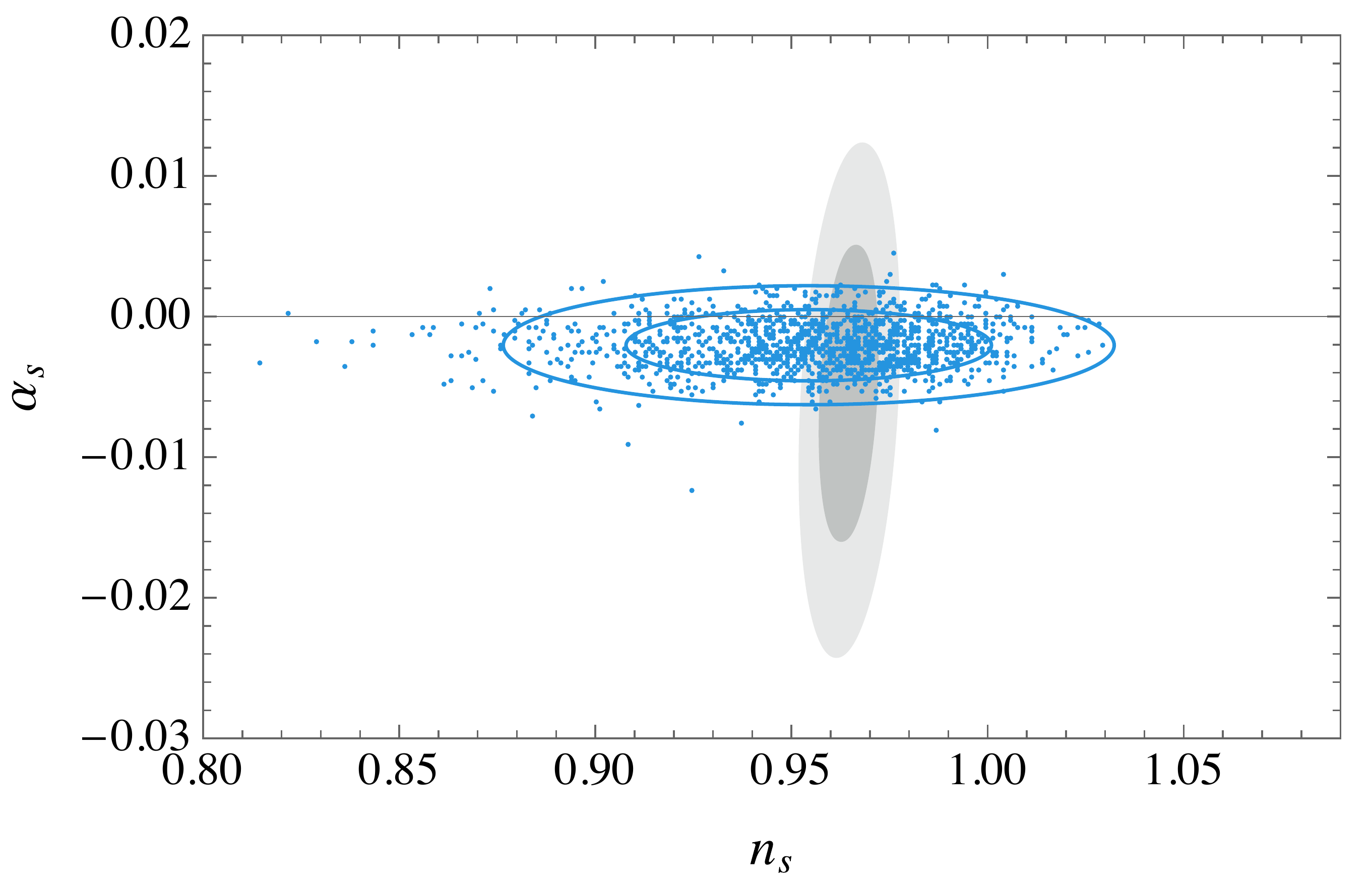}
\caption{The spectral index verses the running for 1000 models of 100-field inflation. Blue lines show $68\%$ and $95\%$ confidence limits for this ensemble. 
Planck $68\%$ and $95\%$ confidence limits are shown in grey. These contours use Planck high-$\ell$ polarisation, combined with temperature data \cite{1502.02114}.
Here  $ \Lambda_{\rm h}=0.4 M_{\rm Pl}$, $\epsilon_{V0}=10^{-11}$ and $\eta_{V0}=-10^{-4}$.
}
   \label{fig:scatter}
\end{figure}
To further highlight this point, 
Fig.~\ref{fig:scatter} shows a scatter plot of $n_s$ versus $\alpha_s$ for $1000$ DBM realisations with $N_f=100$ together with current constraints from  Planck \cite{1502.02114}. 

 Future  probes of the large-scale structure of the universe and the CMB spectrum and polarisation 
are expected to improve the observational constraints on the running: an optical galaxy survey of Euclid-type may reach a sensitivity of $\sigma(\alpha_s) = 0.002$ and a CoRe-like CMB experiment, in combination with other experiments, may be even more sensitive, obtaining $\sigma(\alpha_s) = 0.0011$--$0.0019$ \cite{1612.05138}. These experiments may rule out essentially all DBM models with a  large number of fields. This makes the running a promising observational test for manyfield inflation.\footnote{We note however that the DBM models typically produce a `running of the running' that is consistent with zero.}

\section{Lesson IV: The smoother the potentials, the sharper the predictions}
\label{sec:smootherV}
\label{sec:lesson4}
\label{sec:Lesson4}

\begin{figure}[t]
  \includegraphics[width=1\textwidth]{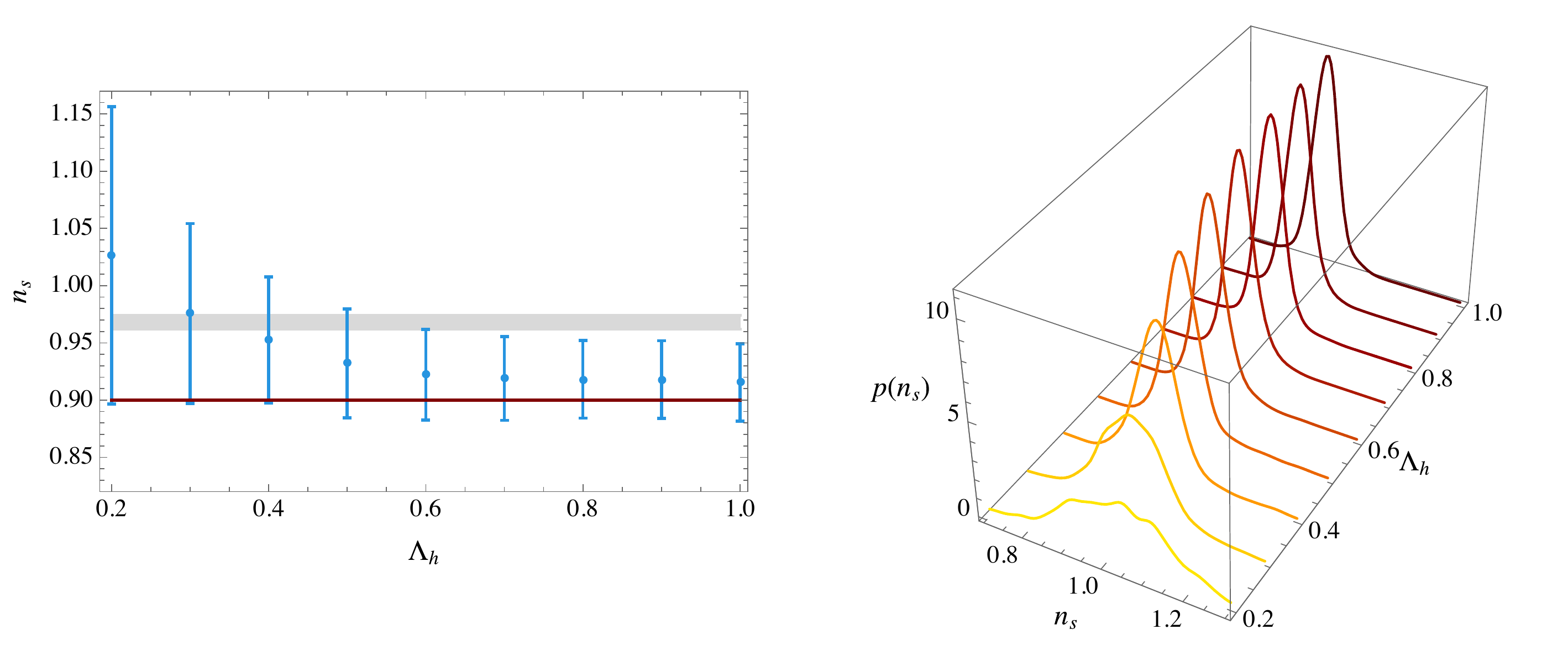}
    \centering
  \caption{
  The spectral index as function 
  of $\Lh$, in units of $\Mpl$. 
  The grey shaded region indicates the $68\%$ confidence limits from Planck \cite{1502.02114}. The red line shows prediction of Ref.~\cite{1608.00041}. Here
  $N_{f}=20,\, \epsilon_{V0}=10^{-11}$, and $\eta_{V0}=-10^{-4}$.
  }
   \label{fig:pnsvsLambdah}
\end{figure}

The intrinsic smoothness of the potentials affects the predictions of random multifield models. 
In the DBM construction, the `horizontal scale' $\Lh$ sets the distance scale in field space over which the eigenvalue spectrum of the Hessian matrix equilibrate. As mentioned in \S\ref{sec:DBMreview}, $\Lh$ can be given a  natural interpretation as a Wilsonian ultraviolet cut-off of the theory. DBM potentials with $\Lh \ll \Mpl$ tend to be highly featured on small scales and require excessive levels of fine-tuning of the initial $\epsilon_V$ parameter to support 60 efolds of inflation. We here focus on the interesting sub-Planckian range  of $0.2 \leq \Lh/\Mpl \leq 1$ and investigate the impact of the smoothness on the observables generated during inflation.  

In this section, we also study the influence of the initial $\epsilon$ parameter on the observables. While $\epsilon_{V0}$ only sets the magnitude of the gradient vector at a point in field space, the fields spend most of inflation very close to this point, and hence, a small $\epsilon_{V0}$ effectively smoothens the potential experienced by the fields during inflation.


Figure \ref{fig:pnsvsLambdah} shows the dependence of the spectral index on $\Lh$ for DBM models with $N_f=20$. 
%
For small $\Lh$,  models supporting at least 60 efolds of inflation are quite rare: we find 1134 successful examples out of 20,000 randomly generated DBM models with $N_f=20$, $\epsilon_{V0}=10^{-11}$ and $\eta_{V0} = -10^{-4}$.  The resulting power spectra for the successful models tend to be highly featured, leading to a large variance for the  spectral index computed over the 10 efold window as in \S\ref{sec:lesson3}. 
As $\Lh$ is increased, examples with a large number of efolds become  more common,\footnote{For $\Lh= 0.4 \Mpl$, we find 1066 examples with at least $60$ efolds of inflation from 2000 models. For $\Lh =\Mpl$, $70 \%$ of the examples were successful.}
and the 
 variance 
 of the spectral index decreases, as does its mean value. 
 Since $\Lh$ is a measure of the correlation length of the potential, we conclude that the smoother the potential, the sharper the prediction for the spectral index, and the redder the spectra.\footnote{Qualitatively similar behaviour was found in Ref.~\cite{1111.6646} for the case of 2-field Gaussian random potentials supporting large-field inflation. }
 
 From Fig.~\ref{fig:pnsvsLambdah}, we see that for large $\Lh$, most DBM models are too red to be consistent with the 
observationally favoured grey band from Planck \cite{1502.02114}.  
%
We conclude that small $\Lh$ is preferred by the data. 
Moreover, the red line in the left plot of Fig.~\ref{fig:pnsvsLambdah} shows the single-field prediction of  Ref.~\cite{1608.00041}, which is there assumed to be applicable for $\Lh < \Mpl$. We see that this prediction improves as  $\Lh \to \Mpl$, and is compatible with the result of the full simulations, at least for large $\Lh$ and moderate values of $N_f$. 
Again, as discussed, this single-field approximation should be regarded with caution. Given the compatibility of the DBM predictions with observational constraints for models with smaller $\Lh$, the strong conclusions of Ref.~\cite{1608.00041} appear harder to justify.

%

\begin{figure}[t]
  \includegraphics[width=0.56\textwidth]{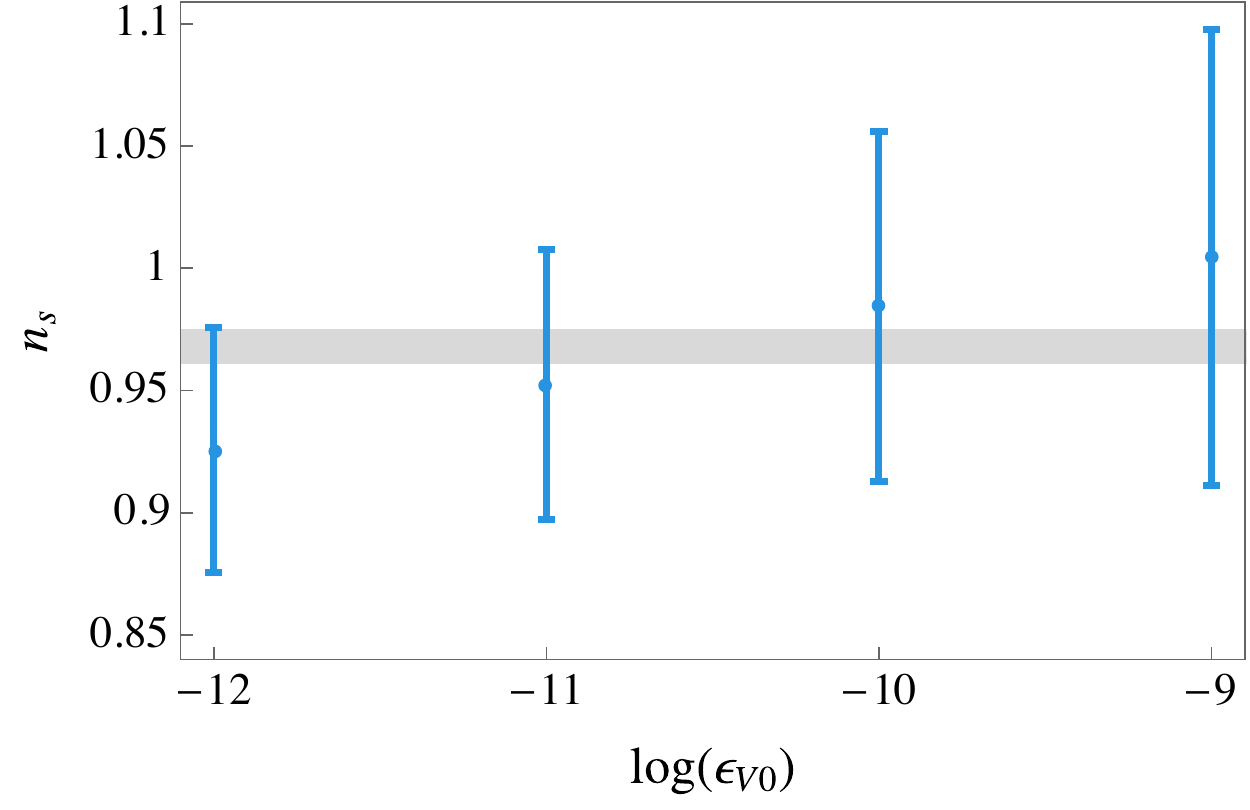}
    \centering
  \caption{The spectral index as a function of  $\epsilon_{V0}$.   The grey shaded region indicates the $68\%$ confidence limits from Planck \cite{1502.02114}. Here $N_{f}=20,\, \Lh=0.4 \Mpl$, and $\eta_{V0}=-10^{-4}$. }
   \label{fig:pnsvsepsilon}
\end{figure}

The initial value of $\epsilon_V$ is the hyperparameter that controls the  flatness of the potential near the approximate saddle-point. A smaller value of $\epsilon_{V0}$ leads to a more slowly rolling field and, generically, more efolds of inflation. 
The effect of 
a decrease in $\epsilon_{V0}$ 
on observables is
very similar to that of
an increase of $\Lh$.
Figure~\ref{fig:pnsvsepsilon} shows how, as $\epsilon_{V0}$ is decreased, the 
spectrum becomes more red and the 
 standard deviation 
 taken over 1000 random realisations decreases, in analogy with Fig.~\ref{fig:pnsvsLambdah} for increasing $\Lh$. Thus, also for the $\epsilon_{V0}$ hyperparameter, we see that a smoother potential leads to sharper predictions.

\section{Lesson V: 
Hyperparameters can transition from stiff to sloppy}
In any model of inflation, one generally seeks an understanding of how the model predictions depend on the model parameters. 
For the DBM models where the potentials are generated stochastically, we seek to understand the distribution of observables for a fixed choice of hyperparameters, and the dependence of these distributions to hyperparameter variations. We have already considered the dependence on $N_f$ in \S\ref{sec:lesson2} and the dependence on $\Lh$ and $\epsilon_{V0}$ in \S\ref{sec:Lesson4}. In this section we describe the dependence on the final hyperparameter of our DBM ensemble, the initial slow-roll parameter  $\eta_{V0}$. We also comment on 
 the problem of inference 
 for
this ensemble of models. 

\begin{figure}[t]
  \includegraphics[width=0.56\textwidth]{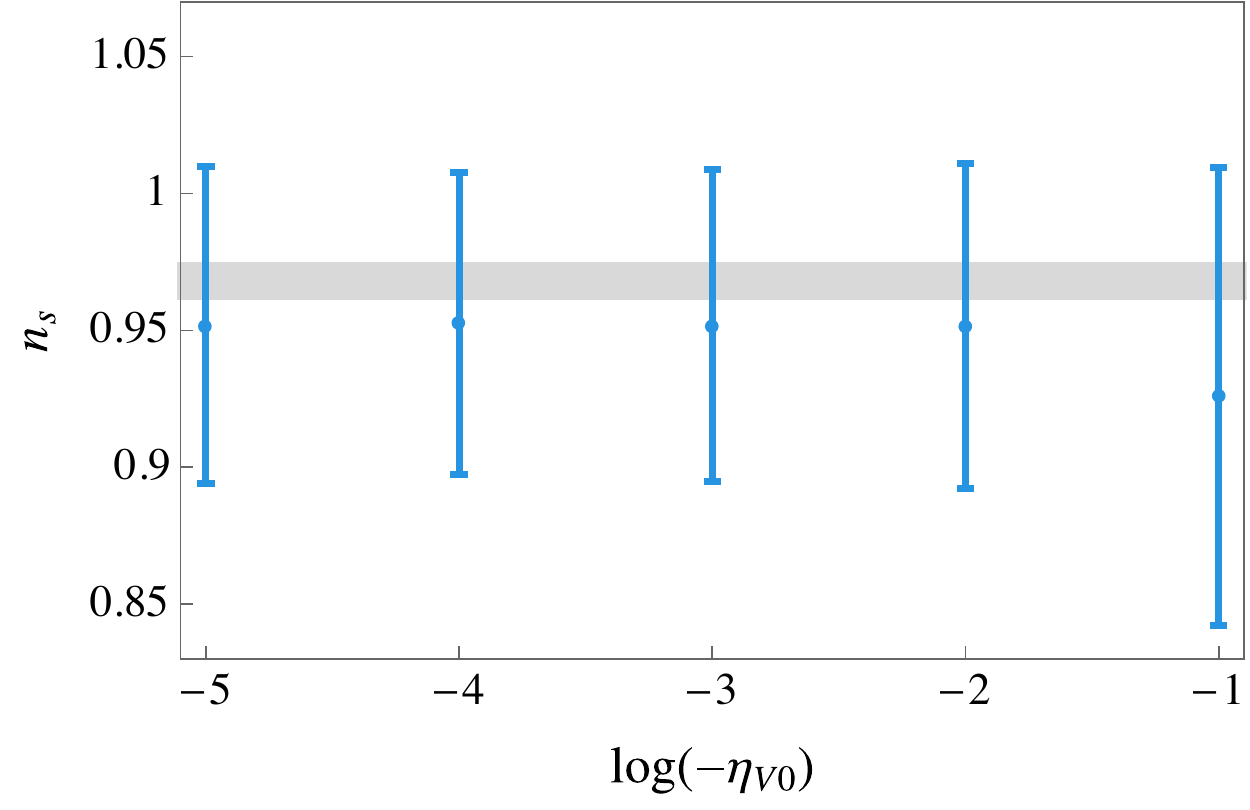}
    \centering
  \caption{Spectral index as a function of the initial slow-roll parameter $\eta_{V0}$. Here $N_{f}=20$, $\Lh=0.4 \Mpl$, and $\epsilon=10^{-11}$.   The grey shaded region indicates the $68\%$ confidence limits from Planck \cite{1502.02114}. }
   \label{fig:nseta}
\end{figure}

In Ref.~\cite{1307.3559}, it was found that the the Dyson Brownian motion of  the Hessian matrix 
quickly 
spoils any fine-tuning of the eigenvalue spectrum  made in the first patch: 
already a short distance ($\ll \Lh$) away from the 
approximate critical point, the Hessian matrix develops small off-diagonal terms which 
cause `eigenvalue relaxation' and a prompt erasure of any tuning of the curvature of the potential. 
As a direct consequence, 
for 
$|\eta_{V0}|$ not too large, we expect the predictions of the model to be independent of $|\eta_{V0}|$. This should be contrasted with 
the case common to 
 many  single-field models, in which 
 $N_e \sim 1/|\eta_{V0}|$, and $n_s-1 \sim \eta_{V*} \propto \eta_{V0}$.
Reference \cite{1307.3559} found that the distribution of the number of efolds generated by DBM models become independent of $\eta_{V0}$ as $|\eta_{V0}| \lesssim 0.1$. 
 Here, we 
 compute the effects of $\eta_{V0}$ on observables, such as the spectral index. 
 
 Figure \ref{fig:nseta} shows the ensemble averages and standard deviations for five sets of 1000 randomly generated DBM models with $N_f=20$, $\Lh=0.4 \Mpl$ and $\epsilon_{V0}=10^{-11}$, and with $\eta_{V0}$ varying from $-10^{-5}$ to $-10^{-1}$. For $\eta_{V0} = -0.1$, examples supporting at least 60 efolds of inflation are quiet rare (success rate: 1/15.6), and the predictions, just as in the case of small $\Lh$, become unsharp.   For $|\eta_{V0}|<0.1$, the spectral index becomes independent of $\eta_{V0}$, consistent with   the picture of eigenvalue relaxation proposed in~\cite{1307.3559}.

 \subsection{Stiffness, sloppiness and inference} 

The appearance of parameters that have little or no effect on observables is  
 a ubiquitous feature of complex multi-parameter systems. 
For instance, in systems biology, nuclear physics and statistical physics 
such parameters are sometimes referred to as `sloppy' or `ill-conditioned', to distinguish them from `stiff'
parameters 
for which a small change has a direct effect 
on observables \cite{1478-3975-1-3-006, brown2003statistical, waterfall2006sloppy}.
`Sloppiness', i.e.~the separation of the model parameters into a few stiff and multiple sloppy parameters, has been argued to be a natural emergent phenomenon in complex systems.  

We have already found several examples of sloppy parameters in the DBM models: 
for sufficiently large $N_f$, the  distribution of $n_s$ and $\alpha_s$ are given by Gaussians that are insensitive  to the number of fields, cf.~Fig's.~\ref{fig:P(nsalpha)} and \ref{ns_Nf_contours}; for sufficiently large $\Lh$, the observables are independent of $\Lh$, cf.~Fig.~\ref{fig:pnsvsLambdah}; and finally, we have seen that the predictions become independent of $\eta_{V0}$ for small enough absolute values, cf.~Fig.~\ref{fig:nseta}.
As the sloppy parameters no longer affect the observables, these parameter regimes exhibit `universal' predictions that are rather insensitive to the details of the system. By contrast, in the complementary parameter regimes these parameters are stiff and affect the predictions. In our construction, the $\epsilon_V$ parameter remains stiff over the entire sampled parameter space.\footnote{The number of efolds of inflation is expected to diverge as $\epsilon_V\to 0$, and we do not expect this parameter to become sloppy. However, for extremely flat potentials that lead to $\gg 60$ efolds of inflation, we expect the eigenvalues of the Hessian matrix to have relaxed to the equilibrium configuration at horizon crossing (55 efolds before the end of inflation). This will lead to a sloppiness of the spectral index and its running in the small $\epsilon_V$ limit. }



An intriguing feature of  the hyperparameters of the DBM model is that parameters can transition from being stiff to being sloppy.  This behaviour is 
evident in e.g.~Fig.~\ref{fig:P(nsalpha)}:
for  small numbers of fields, the distribution changes quite dramatically as $N_{f}$ is increased but when the number of field is larger, a similar change in the number of fields seems to have a much less dramatic effect. Similar transitions occur also for the hyperparameters $\Lh$ and $\eta_{V0}$. A
stiff-to-sloppy transition for $N_{f}$ was also observed in the context of N-flation in Ref.~\cite{Price:2015qqb}. In \S\ref{sec:eigrep} we will relate large-$N_f$ sloppiness to  the notion of universality in random matrix theory.

Clearly, there are two sides to sloppiness. Sloppy parameters enable one to make robust, `universal' predictions for observables, but precisely because these parameters are sloppy, they are `ill-conditioned' and hard or impossible to determine from  observations. For example, the well-defined large-$N_f$ limit of DBM inflation limits our ability to, even in principle, determine the precise number of fields present in the early universe. 
Inference in sloppy systems is sometimes discussed using
the notion of `model complexity', or `Bayesian complexity' \cite{spieglhalter2002bayesian, kunz2006measuring} which seeks to compute the effective number of parameters of a model which can be inferred from data.

\section{Lesson VI: Despite tachyons, isocurvature can decay}\label{sec:isocurvature}
In multifield models of inflation, the curvature perturbation evolves on superhorizon scales if sourced by the isocurvature perturbations, cf.~Eq.~\eqref{eq:Pzetaprime}. This can make it challenging to extract reliable predictions from these models as
one may need to follow the perturbations through reheating to extract the observational predictions for the spectrum of the primordial perturbations.  
A common approach 
to this problem is to ensure (or assume) the existence of an `adiabatic limit'  in which entropic perturbations become massive and hence suppressed, leaving only the then constant curvature perturbation. However, an adiabatic limit in the sense of only a single remaining light field is not a general feature of multifield models of inflation, and it is 
certainly not a property of the models of inflation that we study here. 
In this section, we address the critical question of the  evolution of the curvature and isocurvature perturbations in the absence of an adiabatic limit. 

\begin{figure}[t]
  \includegraphics[width=0.6\textwidth]{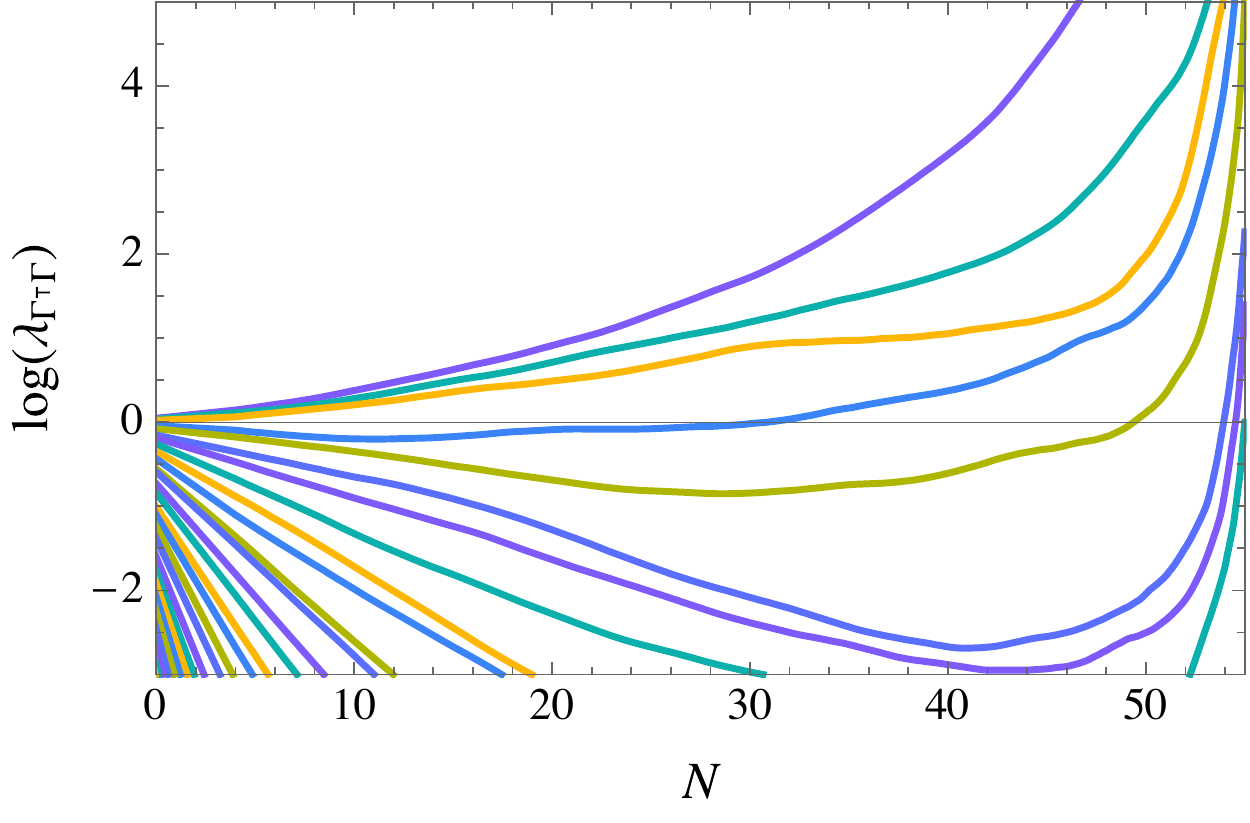}
    \centering
  \caption{Evolution of  the eigenvalues of $\left(\Gamma^{\top}\Gamma\right)$ during inflation in the 100-field model of \S\ref{sec:casestudy}. 
  }
%
   \label{fig:lambdaplot}
\end{figure}

\subsection{Isocurvature suppression in the case study}
In slow-roll inflation, the field perturbations in the flat gauge evolve according to Eq.~\eqref{eq:JocobiEoM}. Entropic perturbations that correspond to eigendirections of the Hessian matrix with negative eigenvalues grow in magnitude. In other words, multiple tachyons in the spectrum may signal the growth of entropic modes. 
In the transport formalism, field perturbations evolve with the propagator $\Gamma_{ab}(N, N_0)$, as discussed in \S\ref{sec:perts}. An estimate of the growth of the modes can then be obtained by computing 
%
the eigenvalues of $\left(\Gamma^{\top}\Gamma\right)_{ac} = \Gamma_{ab}(N, N_0) \Gamma_{cb}(N, N_0)$.
Eigenvalues $\lambda_{\Gamma^{\top}\Gamma}>1$ correspond to modes that have grown since horizon crossing, while modes with $\lambda_{\Gamma^{\top}\Gamma} <1$ have become suppressed.\footnote{
If at horizon crossing all fields are comparatively light and have $M^2< 9/4 H^2$, 
%
$\Sigma_{ab}(N_0) \propto \delta_{ab}$ and  the eigenvalues of $\Gamma^{\top}\Gamma$ precisely measure the growth of the field perturbations. For 
the more general situation with both massive and light modes, the eigenvalues of $\Gamma^{\top}\Gamma$ overestimates the importance of modes with $M^2> 9/4 H^2$. } An adiabatic limit corresponds to having a single non-vanishing eigenvalue, and all others negligibly small.

Figure \ref{fig:lambdaplot} shows the evolution of the eigenvalues of $\left(\Gamma^{\top}\Gamma\right)$ during inflation in the 100-field case study of \S\ref{sec:casestudy}. A large fraction of the eigenvalues quickly become suppressed: these correspond to modes that during the  slow passage of the approximate critical point have $m^2 > 0$, yielding  exponentially suppressed perturbations. A few modes are nearly massless or tachyonic during this period, and remain unsuppressed or grow during inflation: these modes are the main source of superhorizon evolution of $\zeta$. 

Towards the end of inflation, as the eigenvalue spectrum of the Hessian matrix relaxes back to the Wigner semi-circle, more modes become tachyonic. This results in an `upturn' in the spectrum of $\left(\Gamma^{\top}\Gamma\right)$ towards the end of inflation and an increase in the number of modes with $\lambda_{\Gamma^{\top}\Gamma}>1$.
 Clearly, the eigenvalue spectrum of Fig.~\ref{fig:lambdaplot} does not correspond to only one non-vanishing eigenvalue at the end of inflation, and hence, an adiabatic limit is not reached in this model. 
  \begin{figure}[t]
  \includegraphics[width=0.6\textwidth]{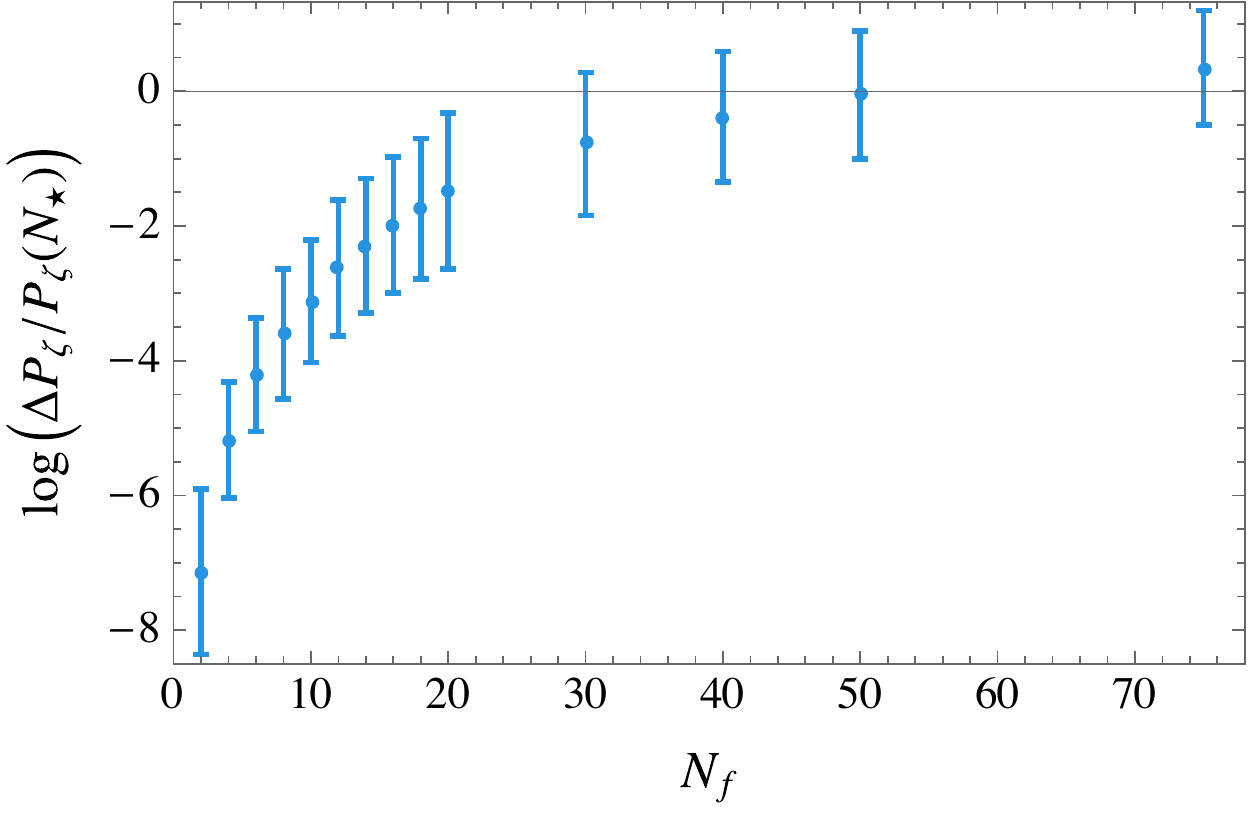}
    \centering
  \caption{More fields lead to more superhorizon evolution. Here $\Delta P_{\zeta} = P_{\zeta}(N_{\rm end}) - P_{\zeta}(N_{\star})$. 
  }
%
   \label{fig:DeltaP}
\end{figure}

An important question is then: with multiple modes growing after horizon crossing,  can a model like the 100-field case study be at all predictive? Exactly how relic isocurvature power impacts observables through post-inflationary physics is highly model dependent. As a first step to answering this question we now show how, despite growing field perturbations, the relative magnitude of isocurvature-to-curvature  perturbations can decay during inflation, and in many cases becomes radically suppressed. 

In the case study of \S\ref{sec:casestudy},  Fig.~\ref{fig:P(N)Expl} shows the magnitudes of $P_{\zeta}(N, k_*)$, $P_{\rm iso}(N, k_*)$ and $P_{\rm cross}(N, k_*)$ during inflation. At the horizon crossing,
each light mode fluctuates with an amplitude set by $H_*$, and since there is only one adiabatic mode but multiple light entropic modes, 
 $P_{\rm iso}(N_*, k_*) > P_{\zeta}(N_*, k_*)$. On superhorizon scales, the power spectra evolve non-trivially:  
the isocurvature perturbation shows a decaying trend, while the curvature perturbation increases in a step-like fashion. 
At the end of inflation, $P_{\rm iso}/P_{\zeta} =0.0067$. Thus, despite the numerous positive eigenvalues of $\Gamma^{\rm \top} \Gamma$, the ratio of isocurvature-to-curvature perturbations decreases substantially during inflation.  

\subsection{Isocurvature suppression: generalities}
\begin{figure}[t]
  \includegraphics[width=0.48\textwidth]{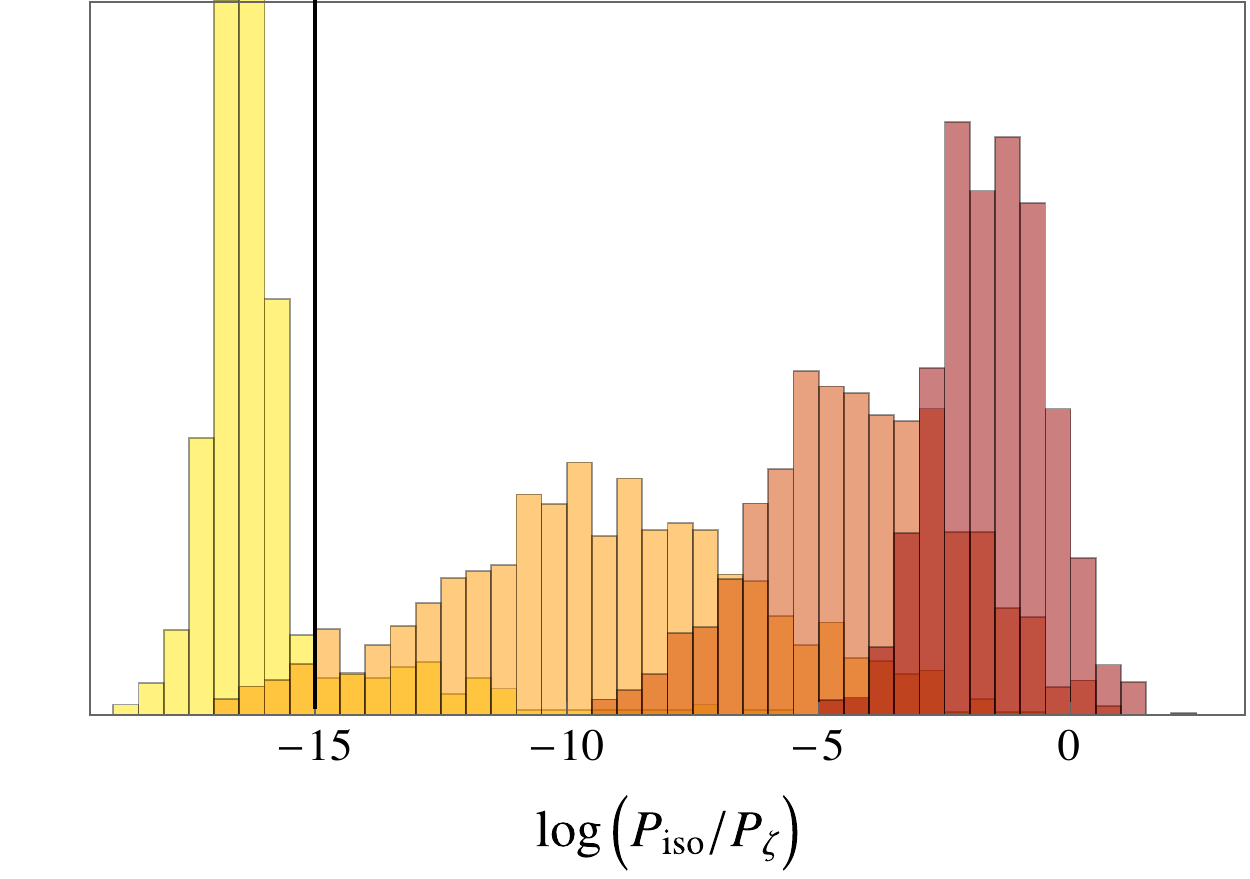}
  \includegraphics[width=0.48\textwidth]{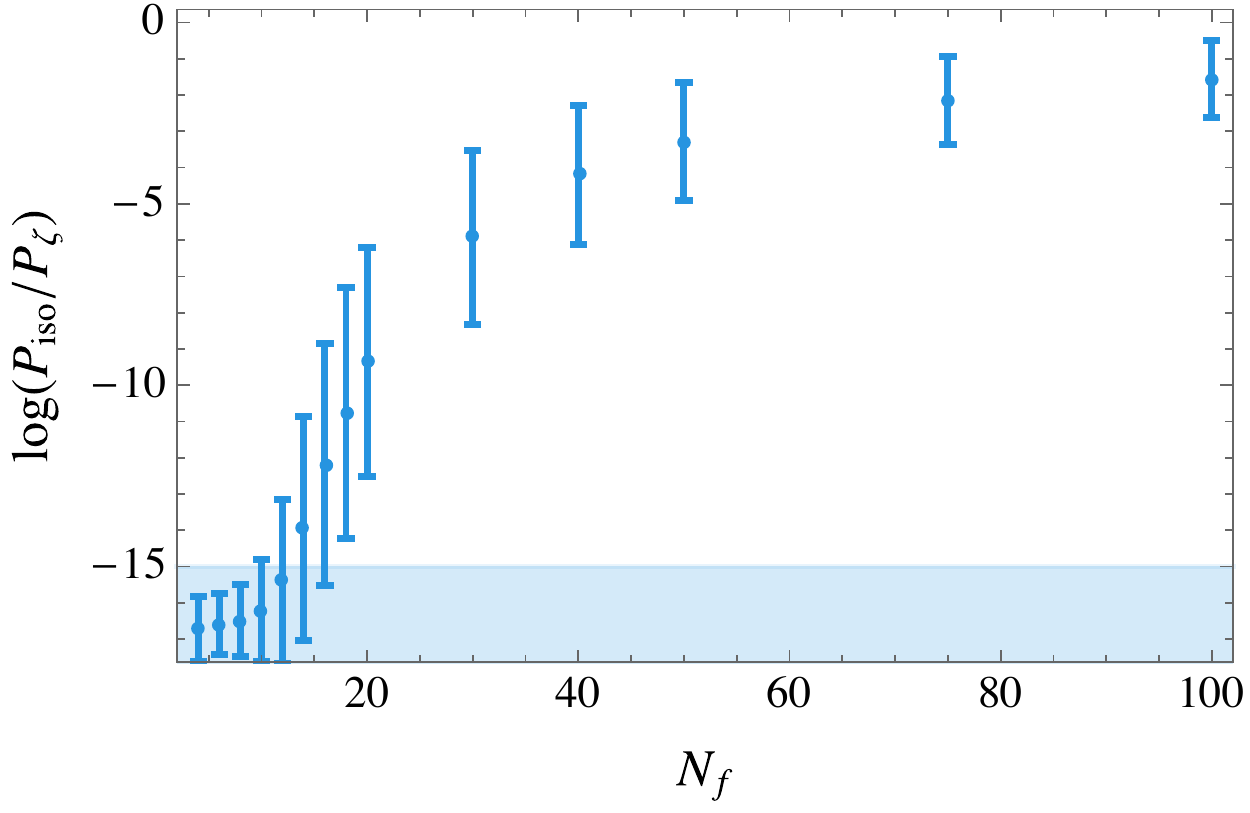}
    \centering
  \caption{Distribution of isocurvature-to-curvature ratio at the end of inflation. Left: Histograms for, from left to right, $N_f=10, 20,  40, 100$; the black vertical line indicates the numerical accuracy. Right: Dependence with $N_f$;  the shaded blue area indicates the limit of numerical accuracy. For $N_f = 10$, most examples are consistent with vanishing isocurvature at the end of inflation.   
  }
%
   \label{fig:isocurvH}
\end{figure}

The suppression of the isocurvature-to-curvature ratio at the end of inflation is not unique to the case study of \S\ref{sec:casestudy}. On the contrary, we find even smaller ratios in the vast majority of 
inflationary realisations that we study. 
Figure \ref{fig:isocurvH} shows the distributions of $\log(P_{\rm iso}/P_{\zeta})$ as evaluated at the end of inflation for various numbers of fields. For $N_f\leq10$, most models give ratios smaller than the  numerical accuracy our our simulation:  $P_{\rm iso}/P_{\zeta} < 10^{-15}$.  For larger $N_f$, the levels of isocurvature  increase slowly, but remain small until $N_f =100$, for which a moderate suppression of $P_{\rm iso}/P_{\zeta} \approx 10^{-2}$ is typical.

Thus, despite numerous tachyons and high levels of isocurvature at horizon crossing, isocurvature can decay and become negligibly small at the end of inflation. 
 To gain some intuition for these results, we first consider the very simple (non-random) case in which the Hessian matrix is constant during inflation and the field velocity is aligned with the eigenvector of the smallest eigenvalue of the Hessian matrix. With these assumptions,  the field makes no turns during inflation and, as is evident from
 \eqref{eq:Pzetaprime},  $P_{\zeta}'=0$. 
However,
the Hessian matrix may have several tachyonic directions leading to 
growing field perturbations with $\lambda_{\Gamma^{\top}\Gamma}>1$. 
Since $n^a$ is aligned with the eigenvector of the smallest eigenvalue, $\delta \phi_{\parallel}$ is the fastest growing mode. 
Since $\zeta$ is constant, the growth of this field perturbation has to be exactly matched by a  corresponding growth of the slow-roll parameter $\epsilon_V$, cf.~Eq.~\eqref{eq:zeta}. The remaining, entropic modes grow more slowly than $\delta \phi_{\parallel}$ and lead to isocurvature perturbations, cf.~Eq.~\eqref{eq:iso}, that become increasingly suppressed as $\epsilon_V$ grows.  
%
%
This illustrates how the existence of multiple large eigenvalues of   $\Gamma^{\top} \Gamma$ can be  consistent with both a constant curvature perturbation and decreasing levels of isocurvature. 
 For more details on this example, see Appendix~\ref{appendix}.
As we now discuss, these arguments 
directly generalise to the 
case of manyfield inflation in random potentials.


 The superhorizon evolution of the power spectra
 in a general model of multifield slow-roll inflation are governed by Eqns.~\eqref{eq:Pzetaprime}--\eqref{eq:Pcrossprime}. 
As discussed in \S\ref{sec:perts}, we do not solve these equations directly, but rather use the transport method to evolve the correlators. 
We here use 
Eqns.~\eqref{eq:Pzetaprime}--\eqref{eq:Pcrossprime}
to obtain a conceptual understanding of the evolution of isocurvature in manyfield inflation.  Figure  \ref{fig:P(N)Expl} provides a clear illustration of these general considerations. 

First,
Eq.~\eqref{eq:Pzetaprime} indicates that the curvature power spectrum is only sourced by the isocurvature-curvature cross-correlation, and the sourcing only occurs when $n^a$ is misaligned with an eigenvector of $V_{ab}$, i.e.~when the slow-roll trajectory turns. 
At horizon crossing, 
the isocurvature-curvature cross-correlation is zero, as is the cross-correlation between different isocurvature modes. 
 The initial suppression of $P_{\rm cross}$ makes $P_{\zeta}$ evolve very slowly immediately after horizon crossing. 
 
 The isocurvature-curvature cross-correlation is sourced directly by the isocurvature modes, cf.~Eq.~\eqref{eq:Pisoprime}. This  leads to a rapid growth of the cross-correlation after horizon crossing, and substantial  levels of superhorizon evolution are possible once $P_{\rm cross} \approx P_{\zeta}$. 
However, both the $P_{\rm cross}$ and $P_{\rm iso}$ tend to decay during inflation. This decay is related to the terms, 
\bea
%
(P_{\rm iso}^{ii})' &=& 2 \left(n^a  \frac{V_{ab}}{V}  n^b - v_i^a  \frac{V_{ab}}{V}  v^b_i - 2 \epsilon_V \right) P_{\rm iso}^{ii} + \ldots  \, , \nonumber \\
(P_{\rm cross}^i)' &=& \left(n^a  \frac{V_{ab}}{V}  n^b
- v^a_i\,  \frac{V_{ab}}{V} \, v^b_i\, 
 - 2 \epsilon_V \right) P_{\rm cross}^i + \ldots  \, .
 \label{eq:Pprimapprox}
\eea
The effects of these contributions can be understood as follows: during inflation $n^a$ tends  to align with the eigenvector of the smallest eigenvalue of the Hessian matrix. The combination $n^a  \frac{V_{ab}}{V}  n^b - v_i^a  \frac{V_{ab}}{V}  v^b_i$ will then be negative and contribute to a suppression of $P_{\rm iso}^{ii}$ and $P_{\rm cross}^i$. In the small-field models that we study, the factor of $\epsilon_V$ is often too small to be important during most of inflation, but  towards the end of inflation it grows substantially and causes additional suppression of the isocurvature-curvature cross-correlation and the isocurvature power spectrum. 
%
 %
 %
%
Once the isocurvature decays and becomes negligible,  $P_{\zeta}$ again effectively become constant. This explains how, even in the absence of an adiabatic limit, $P_{\zeta}$ often ceases to evolve well before the end of inflation. 

Equation \eqref{eq:Pprimapprox} also indicates that the eigenvalue spacing of the Hessian matrix is a key factor in determining the suppression of isocurvature.
As the number of fields increases, the separation of the squared masses is squeezed and the suppression of isocurvature is less substantial. 
%

To summarise, the models we study typically have several tachyons, and multiple components of the flat-gauge field perturbations grow during inflation. However, despite the absence of an adiabatic limit,  isocurvature tends to become highly suppressed, forcing the evolution of $P_{\zeta}$ to peter out. 


We note in closing that observations constrain isocurvature left in the post-inflationary, post-reheating universe. We have here only analysed the evolution of the curvature and isocurvature perturbations during inflation. To our knowledge, there exists no systematic understanding of how isocurvature at the end of inflation translates into isocurvature post-reheating, and thus, the levels of isocurvature suppression computed here do not lend themselves to direct comparison with observational constraints.

\section{Lesson VII: Eigenvalue repulsion drives the predictions}
\label{sec:eigrep}
\label{sec:lesson7}
In \S\ref{sec:lesson2}, we showed that 
the primordial power spectra generated by  the DBM models  become both simpler and sharper as the number of fields is increased. 
In this section we show that these results can be understood as being a direct consequence of eigenvalue repulsion, and we suggest that this may indicate  broad applicability of the results, extending beyond the details of the DBM construction.

\subsection{Why  predictions become simpler and sharper at large $N_{f}$ }
As measures of how predictive and simple  the models are, we consider the ensemble variance of the spectral index and its running. A small variance of the model prediction for the spectral index  indicates a highly predictive model, while small values of the running indicate that the power spectra are simple and not very featured.

In the small-field models that we study in this paper, $\epsilon_*$ tends to be highly suppressed with respect to $\eta_{V*}$, and the dominant contribution to the spectral index is given by, 
\be
n_{s}-1\approx -2\, e^{a}\left(\frac{V_{ab}}{V} \right)_* e^b
\, .
\label{eq:nsapprox}
\ee
Equation \eqref{eq:newalpha} for the running similarly simplifies to,
\begin{equation}
\alpha_s  \approx 
 4 e^a\left( \frac{V_{ac} V_{cb}}{V^{2}\Lambda_{\rm h}^{4}}\right)_*  e^b
 - 4 \left(e^a \left ( \frac{ V_{ab}}{V \Lambda_{\rm h}^{2}} \right)_* e^b \right)^2 
 + 2 e^a \left(\frac{V'_{ab}}{V\Lambda_{\rm h}^{2}} \right)_*e^b
  \, .
  \label{eq:alphasf}
\end{equation}
Here, as in Eq.~\eqref{eq:e_a}, the unit vector $e^{a}$ is proportional to $N_{b}\Gamma^{b}_{~a}(N_{\rm end}, N_*)$ and
 captures the entire superhorizon evolution of the spectral index and its running. Unfortunately, $e^a$ is 
hard  to compute analytically for a general multifield model. 
Here, we use 
our numerical simulations to gain some insight into the properties of $e^a$, as evaluated at the end of inflation. 

\begin{figure*}
\centering
\includegraphics[width=1\textwidth]{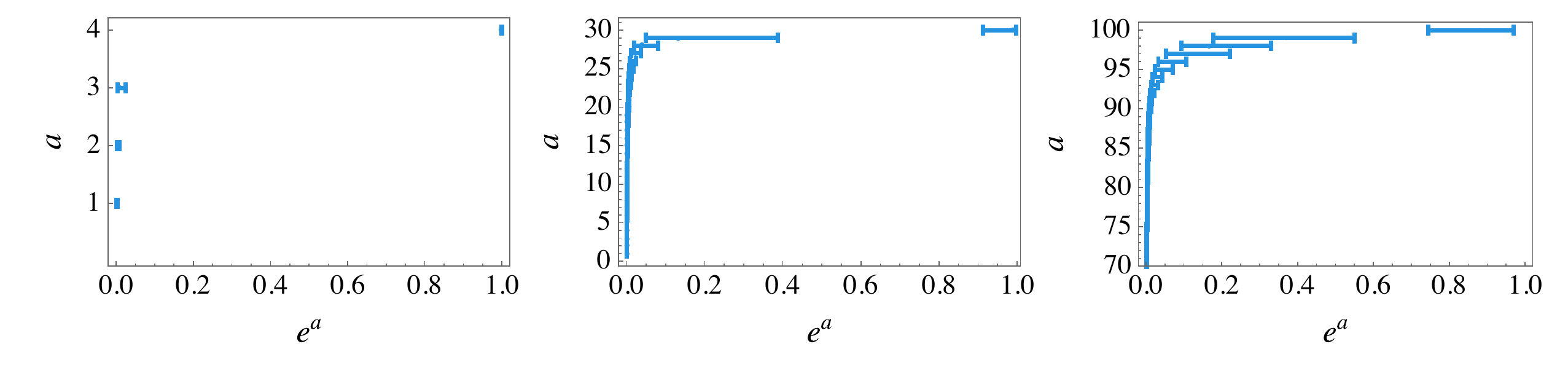}
\caption{
Ensemble median and standard deviation of the components of the vector $e_{a}$, 
defined with respect to an ordered eigenbasis  of $(V_{ab})_*$,
for ensembles with, from left to right, $N_f= 4,\, 30$ and 100.
Here $\Lh=0.4 \Mpl$, $\epsilon_{V0}=10^{-11}$ and  $\eta_{V0}=-10^{-4}$. }
  \label{fig:e_a}
\end{figure*}

Figure \ref{fig:e_a} shows the ensemble medians and standard deviations of the components of $e^{a}$ in the eigenbasis of $(V_{ab})_{*}$, ordered by descending squared mass: $(m^2_{a})_*>(m^2_{b})_*$ for $a<b$. 
%
%
For a small number of fields we see almost perfect alignment of $e^a$ with the
eigenvector of the
 smallest eigenvalue, $(m^2_{\rm min})_*$. For a large number of fields, $e^a$ 
 is still 
 well aligned with this
eigenvector but develops a few non-negligible components. The increased misalignment of $e^a$  is consistent with the increased levels of superhorizon evolution as $N_f$ is increased (cf.~Fig.~\ref{fig:DeltaP}).
 
  
Using Eq.~\eqref{eq:nsapprox}, we see that   $n_{s}$ is primarily determined by the weighted sum of the smallest few eigenvalues of $(V_{ab})_{*}$, and the dominant  contribution corresponds to  the most negative squared mass. 
%
%
Hence, we can gain a qualitative understanding of 
the distribution of $n_s$
by considering the behaviour of the smallest eigenvalue, $\lambda_{\rm min}$, of $v_{ab}$. 

\begin{figure}[t]
  \includegraphics[width=0.8\textwidth]{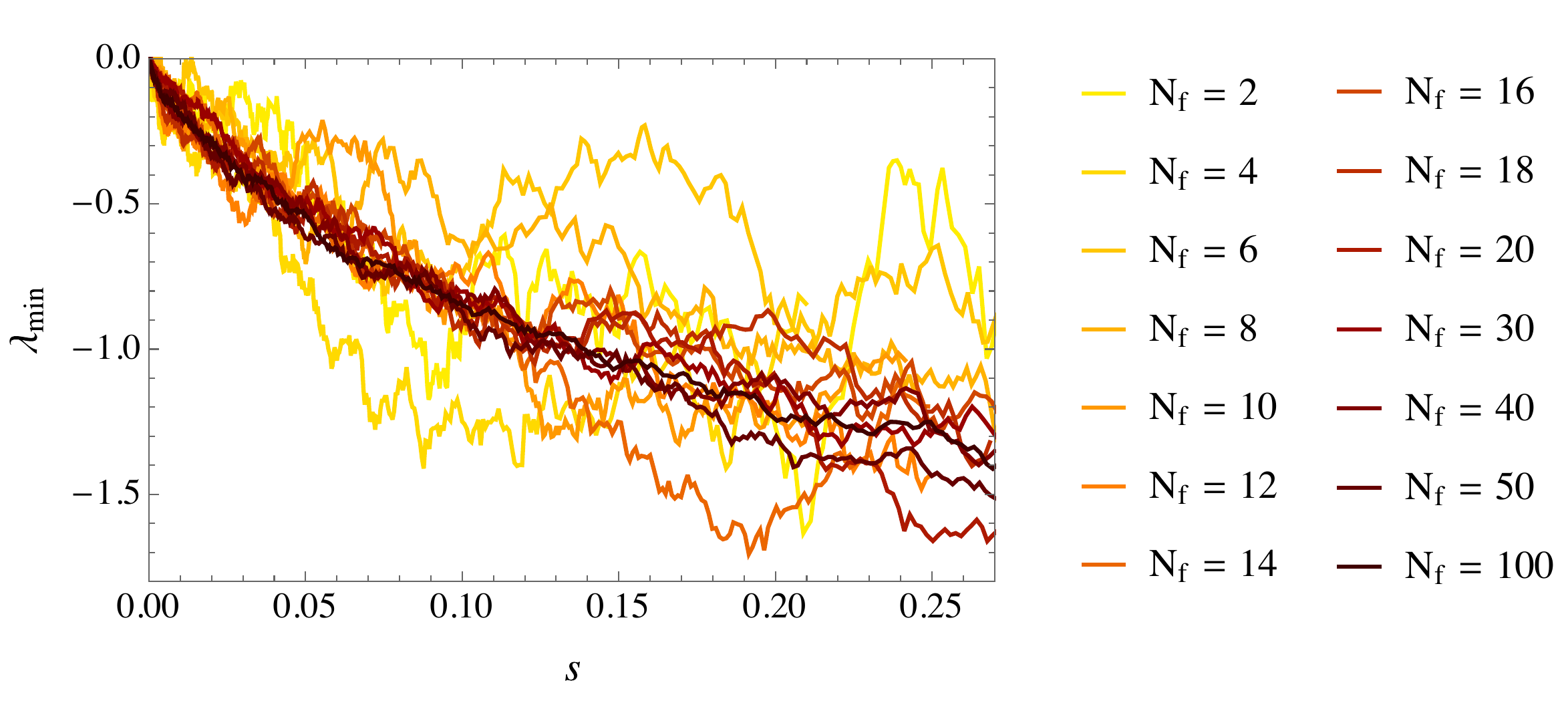}
    \centering
  \caption{Evolution of the smallest eigenvalue for example trajectories as a function of the path length. }
   \label{fig:min_lambda_vs_s}
\end{figure}

Figure \ref{fig:min_lambda_vs_s} shows the evolution of $\lambda_{\rm min}$ over the course of example trajectories for models with $N_f$ from 2 to 100. All examples start with a highly fluctuated spectrum,
cf.~Fig.~\eqref{fig:wigner},  but the subsequent evolution looks very different depending on the number of fields. In particular the trajectories are much more varied when the number of fields is small, resulting in a large variance in the smallest mass at horizon crossing. In contrast,
 at large $N_{f}$
 the eigenvalues of $v_{ab}$ become more dense, and  
  eigenvalue repulsion tames the variability of trajectories considerably.
  This leads to a sharp reduction in  the variance in the smallest squared mass at horizon crossing. Hence we can ascribe the predictive nature of the DBM construction at large $N_{f}$ to eigenvalue repulsion.

Figures \ref{fig:e_a} and  \ref{fig:min_lambda_vs_s}  also indicate why the power spectra become smoother and less featured as the number of fields is increased. 
Using Eq.~\eqref{eq:alphasf} for the running, we see that for  $e^{a}$ exactly aligned with the eigendirection of the smallest eigenvalue,  the first two terms   cancel. For an approximate alignment, this cancelation is approximate and the  dominant term is  the final one, which depends on the volatility of $V_{ab}$ at horizon crossing. 
Figure \ref{fig:min_lambda_vs_s} demonstrates how this  volatility becomes  heavily suppressed   at large $N_{f}$. 
As the number of fields increases, the spacing between eigenvalues decrease (cf.~Fig.~\ref{fig:Expl1}). Eigenvalue repulsion then suppresses the random fluctuations, resulting in smoother evolution of the smallest eigenvalue.

Eigenvalue repulsion is not unique to the DBM construction. For example, constructions of random scalar potentials using Gaussian Random Fields (GRFs) take the distribution of the Hessian matrix to be slightly different from that of the GOE \cite{1307.3559, Bachlechner:2014rqa, 1612.03960}. Still, eigenvalue repulsion is a manifestation of the presence of a Vandermonde determinant appearing in the change of variables from the randomly distributed entries of the Hessian to its eigenvalues,
\be
\prod_{a<b} dV_{ab} = \prod_{a<b} |m^2_a - m^2_b|\, \prod_a dm^2_a \, ,
\ee
which is common to  DBM, GRFs and many other possible constructions. This suggests that the results found in this paper may extend to much broader classes of models than those considered here. 

\section{Conclusions}
We have used Dyson Brownian Motion to construct random scalar potentials for multiple fields, and we have adapted the transport method  to  study the generation of observables during inflation.  These methods are numerically very efficient and have allowed us, for the first time, to explicitly study models of inflation with a very large number of interacting fields. We have statistically generated an ensemble of inflationary models giving at least 60 efolds of inflation, and we have used these to determine the statistical properties of manyfield inflation. We draw seven lessons from these studies:  
\begin{enumerate}
\item {\bf Manyfield inflation is not single-field inflation.} In particular, superhorizon evolution of the curvature perturbation is common in manyfield models, and cannot be captured by single-field models.  
\item {\bf Planck compatibility is not rare.} It has been speculated, based on single-field toy models designed to mimic the DBM potentials, that manyfield inflation typically gives a too red spectrum compared to observations, and can be ruled out by current observations \cite{1608.00041}. We find, consistently with   \cite{1604.05970}, that  this is not true: many of the manyfield models that we study are compatible with Planck constraints on $n_s$. Interestingly however,  the DBM potentials lead to rather sharp predictions for a small negative  running of the spectral index, with a typical spread of the predictions much below the current observational sensitivity. 
\item {\bf The larger the number of fields, the sharper the predictions.} As the number of fields increases, the number of couplings and the complexity of the model increases. The predictions, however, become sharper, and the power spectra become smoother and less featured.
\item {\bf The smoother the potentials, the sharper the predictions.} One of the hyperparameters of the DBM models is the horizontal scale, $\Lh$, which can be interpreted as an ultraviolet cut-off of the theory. We find that as  $\Lh/ \Mpl$ increases towards unity, the predictions become sharper. Conversely, for $\Lh \leq 0.2 \Mpl$, the distributions of observables generated during inflation become exceedingly broad. 
\item {\bf Hyperparameters can make stiff to sloppy transitions.} We have shown that 
there exist regions of the hyperparameter space in which 
%
 the predictions for observables are universal and largely insensitive to the values of these parameters. Such `sloppy' regions
 exist for large $N_f$, large  $\Lh$ and small $|\eta_{V0}|$.

\item {\bf Despite tachyons, isocurvature can decay.} With many light fields, the cosmological perturbations, in addition to the adiabatic mode, include multiple entropic modes which give rise to isocurvature. We show that while isocurvature typically dominates the curvature perturbation at horizon crossing, inflation leads to a dynamical suppression of the power in isocurvature modes, which in many cases leads to drastically suppressed levels of isocurvature at the end of inflation, despite many tachyonic directions in field space.  
\item {\bf Eigenvalue repulsion drives the predictions.} Several of the predictions of the DBM models can be understood from the evolution of a few of the smallest  squared masses, which due to eigenvalue repulsion `repel' when closely spaced. As the number of fields increases, the spacing between eigenvalues decreases and random fluctuations decrease in amplitude. We show that this `eigenvalue repulsion'  drives the predictions of the DBM models. We note that as eigenvalue repulsion is common to systems with interacting fields, our results may extend well beyond the particular setup of DBM potentials.   
\end{enumerate}

It is often said that the generic predictions of  single-field inflation include adiabatic perturbations and an approximately scale invariant spectrum.  
Our results emphasise the strength of these generic predictions: 
despite the complicated nature of the randomly interacting manyfield models that we study, 
isocurvature is commonly 
highly suppressed, and spectra are approximately scale invariant. 
%
Reversely, our results illustrate how many microscopic models map to the same observables, which makes it impossible from observations of the power spectrum of curvature and isocurvature alone to `invert' the map and  find the underlying microphysical model. 

There are a number of interesting future directions to this work. Some models of multiple-field inflation can generate local non-Gaussianity with an amplitude of $f_{\rm NL} \sim {\cal O}(1)$, while single-field inflation predicts $|f_{\rm NL}| \ll 1$. As the next generation of cosmological experiments aim for a sensitivity of $f_{\rm NL} \sim {\cal O}(1)$, it is  a pressing question to determine what we can learn from these observations: {\it Should a future observational constraint of $|f_{\rm NL}| \ll 1$ be interpreted as evidence for single-field inflation?} Moreover, at the level of the 3-point function, additional effects such as particle production can lead to distinctive signatures, which would be interesting to study in these (or related) models. 

Interesting theoretical extensions of this work include a comparison between the predictions of models constructed using random matrix theory with those constructed using Gaussian Random Fields (GRFs). Thus far, due to the numerical complexity of this construction, GRFs have only been used to explicitly study inflation with a small number of fields \cite{1111.6646,Battefeld:2012qx,Bachlechner:2014rqa, 1612.03960,astro-ph/0410281}. It would be interesting to extend this work to the manyfield case. Moreover, we have here worked with a flat field space metric  and we 
expect our results to be applicable to models 
which are not too strongly curved around the inflationary approximate saddle-point. Strong curvature can lead to interesting `flattening' effects (cf.~Ref.~\cite{1612.04505}), which would be interesting to incorporate. 


More generally, we have demonstrated that the use of stochastic methods can be effective in overcoming critical challenges in early universe cosmology. A natural question to ask is where else these techniques may be useful. Very little is known about reheating after multifield inflation, in large part due to the computational complexity of the problem. It is tempting to think that stochastic methods could be useful in making progress on this front. Finally, while string theory strongly motivates
the study of low-energy effective theories with a large number of fields, 
most explicit models of inflation in string theory have focussed on more tractable single-field case.
It would be interesting to see if random matrix theory and random function theory techniques can 
be adapted to shed light on more complicated multifield string theory compactifications. 

\pagebreak

\section*{Acknowledgements}
We are very grateful to Thomas Bachlechner, Daniel Baumann, Theodor Bj\"orkmo, Alan Guth, Andrei Linde, Liam McAllister, Sonia Paban, Enrico Pajer, Eva Silverstein, Kepa Sousa, and Alexander Westphal for stimulating conversations. 
We thank Juha J\"aykk\"a and Kacper Kornet for excellent computational support. 
 MD acknowledges funding from the German Science Foundation (DFG) within the Collaborative Research Centre 676 \emph{Particles, Strings and the Early Universe} and by the ERC Consolidator Grant STRINGFLATION under the HORIZON 2020 contract no.~647995. 
DM is supported by a Stephen Hawking Advanced Fellowship at the Centre for Theoretical Cosmology at the University of Cambridge. 
JF acknowledges funding from the ERC Consolidator Grant STRINGFLATION under the HORIZON 2020 contract no.~647995.
This work used the COSMOS Shared Memory system at DAMTP, University of Cambridge operated on behalf of the STFC DiRAC HPC Facility. This equipment is funded by BIS National E-infrastructure capital grant ST/J005673/1 and STFC grants ST/H008586/1, ST/K00333X/1.
This work was completed at the Aspen Center for Physics, which is supported by National Science Foundation grant PHY-1066293.

\bibliographystyle{modifiedJHEP}
\bibliography{references}

\appendix
\section{A very simple model of inflating off a steep saddle point}\label{appendix}
Qualitatively we can reproduce the behaviour observed in \S\ref{sec:isocurvature} with a simple model of saddle point inflation where all quantities of interest can be computed analytically. Consider a potential of the form,
\be
V=\Lambda^{4}\left[v_{0}+\frac{1}{2}\lambda_{a}(\phi^{a})^{2}\right] \, ,
\ee
where $\lambda_{a}$ are ordered such that $\lambda_{a}<\lambda_{b}$ for $a<b$. We assume that inflation evolves purely along $\phi^{1}$ with all the other fields stationary at 0, such that,
   \be
   u_{ab} = 
   \begin{pmatrix} 
      -\frac{\lambda_{1}}{v_{0}+\frac{1}{2}\lambda_{1}\phi^{2}} + \frac{\lambda_{1}^{2}\phi^{2}}{(v_{0}+\frac{1}{2}\lambda_{1}\phi^{2})^{2}} \\
       & \ddots \\
      & & -\frac{\lambda_{\alpha}}{v_{0}+\frac{1}{2}\lambda_{1}\phi^{2}}\\
      &&&\ddots
   \end{pmatrix} \, .
   \ee
Here we drop the label of field $1$, $\phi\equiv\phi^{1}$, since all other fields are stationary. Furthermore, all off-diagonal terms are zero. With this field configuration, Eq.~\eqref{eq:Gammaeq} is a bit fiddly to evaluate. Instead we solve Eq.~\eqref{eq:GammaEoM} directly,
\be
\Gamma_{ab} = \PathOrder\exp\left\{-\int\d\phi\frac{v_{0}+\frac{1}{2}\lambda_{1}\phi^{2}}{\lambda_{1}\phi}u_{ab} \right\} \, ,
\ee
where we have used $\d N = 1/\phi' \d\phi$ and $\PathOrder$ denotes path ordering. We find that,
   \be\label{eq:Gamma_saddle}
   \Gamma_{ab} = 
   \begin{pmatrix} 
      \frac{\phi_{f}}{\phi_{i}}\frac{v_{0}+\frac{1}{2}\lambda_{1}\phi^{2}_{i}}{v_{0}+\frac{1}{2}\lambda_{1}\phi^{2}_{f}} \\
       & \ddots \\
      & & \left(\frac{\phi_{f}}{\phi_{i}}\right)^{\frac{\lambda_{a}}{\lambda_{1}}}\\
      &&&\ddots
   \end{pmatrix}\, .
   \ee
If $\lambda_{1}<0$ the field $\phi$ evolves to larger values, and hence $\Gamma_{11}$ grows. If $\lambda_{a}>0$ then $\Gamma_{aa}$ decreases, as expected --- this is simply the usual result that inflation along the bottom of a valley will suppress isocurvature, resulting in single field like behaviour. If $\lambda_{a}<0$ then $\Gamma_{aa}$ grows, but slower than $\Gamma_{11}$. This reproduces the behaviour that was seen in \S\ref{sec:isocurvature}. Here it is clear that $\Gamma_{11}$ determines the evolution of fluctuations along the trajectory. Note, that $\Gamma_{11}$ evolves at a rate proportional to $\sqrt{\epsilon_V}$, hence upon substituting Eq.~\eqref{eq:Gamma_saddle} in to Eq.~\eqref{eq:2p}, we see that $P_{\zeta}$ is independent of $\phi_{f}$ and so conserved, as expected. Substituting Eq.~\eqref{eq:Gamma_saddle} in to Eq.~\eqref{eq:Piso}, we see that each tachyonic direction makes a contribution $P_{\rm iso}^{aa}$ to $P_{\rm iso}$ of the form,
\be
P_{\rm iso}^{aa} \propto \frac{1}{2 \epsilon_V}\left(\frac{\phi_{f}}{\phi_{i}}\right)^{\frac{2\lambda_{a}}{\lambda_{1}}}\, , 
\ee
which, since $1/\epsilon_V$ decreases faster than the second factor, decays as
 observed in \S\ref{sec:isocurvature}. We therefore see $P_{\rm iso}$ will always decrease as long as inflation takes place along the most tachyonic direction.

\end{document}